\documentclass[11pt,a4paper]{article}

\usepackage{jheppub}
\usepackage{amsmath,amssymb,mathtools}
\usepackage{afterpage}
\usepackage{tikz}

\usetikzlibrary{decorations.pathmorphing,decorations.markings,arrows.meta,calc}
\usepackage[compat=1.1.0]{tikz-feynman}
\tikzfeynmanset{warn luatex=false}
\allowdisplaybreaks[3]
\makeatletter
\gdef\@fpheader{\phantom{Prepared for submission to JHEP}}
\renewcommand\subsection{\@startsection{subsection}{2}{\z@}%
                                   {-3.0ex\@plus -1ex \@minus -.5ex}%
                                   {1.2ex \@plus .3ex \@minus .3ex}%
                                   {\normalfont\normalsize\bfseries}}
\makeatother

\newcommand{\dd}{\mathop{}\!\mathrm{d}}
\newcommand{\Tr}{\operatorname{Tr}}
\newcommand{\Ree}{\operatorname{Re}}
\newcommand{\Imm}{\operatorname{Im}}
\newcommand{\IR}{\mathrm{IR}}
\newcommand{\e}{\mathrm{e}}
\newcommand{\ii}{\mathrm{i}}
\newcommand{\calM}{\mathcal{M}}

\allowdisplaybreaks

\title{Decoherence as a Detector-Resolution Flow}

\author[a]{Yao Ji}
\author[a]{and Yuqi Wang}
\affiliation[a]{School of Science and Engineering, The Chinese University of Hong Kong, Shenzhen, Guangdong, 518172, China}
\emailAdd{yaoji@cuhk.edu.cn}
\emailAdd{yuqiwang5@link.cuhk.edu.cn}

\abstract{%
We consider the QED final-state radiation modeled by
$\gamma\gamma^\ast\to\pi^+\pi^-+X$, including its finite-temperature
generalization, as an open quantum system, and study the evolution of the
pion-pair momentum density matrix as the detector resolution scale is varied.  We derive an explicit relation between the Lindblad jump operator and the Wilson line associated with soft photon radiation. We find that real emission and virtual correction serve as two separate contributions to momentum decoherence, and together they give rise to the real--virtual cancellation at the density-matrix level.  The external finite-temperature photon bath modifies the decoherence rate by a Bose factor consistent with the Feynman--Vernon influence-functional approach. We emphasize that our conclusions also apply to hard fermion production thanks to the universality of the soft radiation. }


\begin{document}

\maketitle

\section{Introduction}
\label{pion:sec:introduction}

The finite detector resolution in inclusive measurements with final-state radiation serves as a primary parameter suitable for an open quantum system description~\cite{Sterman:1986aj,Collins:1988ig,Catani:1989ne}, in the sense that the observed hard particles form the subsystem, while soft radiation below the resolution scale is left unobserved and can be viewed as the environmental degrees of freedom to be integrated/traced out~\cite{Bauer:2000ew,Bauer:2001yt,Feige:2014wja,Agarwal:2021ais}. Varying the detector resolution therefore changes how much final-state information is discarded, and hence changes the reduced density matrix of the hard particles.

The reduced density matrix contains the usual inclusive cross sections as its
diagonal entries, whose infrared finiteness follows from real--virtual
cancellation~\cite{Bloch:1937pw,Yennie:1961ad,Kinoshita:1962ur,Lee:1964is,Grammer:1973db,Frye:2018xjj,Jauch:1954rnc,Rohrlich:1975np,Nakanishi:1983jpi,Kapec:2017tkm},
which can also be formulated at the amplitude level via dressed states~\cite{Dollard:1964,Chung:1965zza,Kibble:1968pya,Kibble:1968oug,Kibble:1968npb,Kibble:1968lka,Kulish:1970ut,Frohlich:1979uu,Jakob:1990zi,Morchio:1982ym,Lavelle:2005bt,Gabai:2016kuf,Furugori:2020vdl,Balachandran:2014hra,Dybalski:2019jwj}
and finite $S$-matrix constructions~\cite{Hannesdottir:2019rqq,Hannesdottir:2019opa}.
Its off-diagonal entries, on the other hand, encode the quantum coherence between
momentum eigenstates.
The suppression of these quantum correlations by unresolved final-state radiation is the scattering analog of
bremsstrahlung decoherence~\cite{Breuer:2000ha,Calucci:2003sx},
environment-induced decoherence~\cite{Feynman:1963fq,Caldeira:1982iu,Caldeira:1984,Leggett:1987,Zeh:1970,Joos:1985,Zurek:2003zz,Schwinger:1960qe,Hu:1991di,Polarski:1995jg,Grabert:1988yt,Alicki:1979wxy,Unruh:1989dd,Zurek:1991vd,Anglin:1997,Paz:2001,Schlosshauer:2019,Schlosshauer:2003zy,Brune:1996zz},
and collisional decoherence~\cite{Diosi:1995,Gallis:1990,Hornberger:2003,Hornberger:2003wik,Adler:2006gt,Vacchini:2003nmq,Busse:2009kyq}.
It is also related to studies of infrared information and soft sectors~\cite{Lysov:2014csa,Campiglia:2016hvg,Carney:2017jut,Campiglia:2015qka,Kapec:2014zla,He:2014laa,Cachazo:2014fwa,Campiglia:2014yka,Strominger:2015bla,Conde:2016csj,Sahoo:2020ryf,Strominger:2021mtt},
scattering with partial information~\cite{Carney:2016fbk},
soft dressings~\cite{Carney:2017oxp,Carney:2018ygh},
infrared coherence and scales~\cite{Gomez:2018ylx,Gomez:2018szy},
hard--soft entanglement~\cite{Tomaras:2019wvo,Irakleous:2021nbn},
and collider quantum-information observables~\cite{Afik:2021,Afik:2022kwm,Fabbrichesi:2021npl,Severi:2021cnj,AguilarSaavedra:2022wam,Afik:2022dgh,Barr:2021zcp,Barr:2024,ATLAS:2024,Aoude:2025,Gu:2025}.

Very recently, Ref.~\cite{Rey:2026yaa} showed that the reduced density matrix of the hard particles evolves with the detector resolution according to a Lindblad equation.  In this paper, we first motivate this Lindblad description from first principles via an amplitude-level calculation based on the leading-power (LP) soft theorem, and then reveal its field-theoretic origin by making the connection between the Wilson-line operator and the Lindblad jump operator manifest through the LP factorization of the $S$-matrix.  This correspondence also allows us to generalize this open quantum system framework to the finite-temperature case.  We use \(\gamma\gamma^\ast\to\pi^+\pi^-\) as a concrete example, while the
construction applies directly to any process in which hard charged particles
are observed and soft photons below the detector resolution are traced over due to the universality of the eikonal approximation.
Section~\ref{sec:amplitude-level-calculation} derives the reduced hard density
matrix from the final-state Fock expansion and shows how real emission and
virtual correction combine to give both real--virtual cancellation and
momentum decoherence.  Section~\ref{sec:soft-smatrix-wilson-line} derives the
factorization of the scalar-QED \(S\)-matrix into momentum-dependent Wilson
lines and establishes necessary identities for later use.  Section~\ref{sec:kraus-representation} starts from the factorized
$S$-matrix to reveal the connection between the Wilson line and the
Lindblad jump operator, both with and without a nonzero-temperature photon bath. As a consistency check,
Section~\ref{sec:influence-functional} reproduces
the same results of previous sections from the Feynman--Vernon influence-functional approach.

\section{Amplitude-level calculation}
\label{sec:amplitude-level-calculation}

In this section, we apply the amplitude-level soft theorem of scalar QED to
demonstrate the structure of the open quantum system induced by unresolved photon
radiation. The central quantity is the reduced density matrix: its diagonal elements reproduce the inclusive cross section, while its
off-diagonal elements encode quantum coherence between different final states. It therefore
contains richer quantum information than the cross section alone.  The
real--virtual cancellation familiar from the
Bloch--Nordsieck/Kinoshita–Lee–Nauenberg (KLN) theorem translates into the finiteness of the diagonal
density-matrix elements. In addition, the real and virtual contributions
together determine the decoherence between the hard charged particles, whose
resolution-scale evolution takes a Lindblad form. We subsequently derive the corresponding decoherence rate in the pion production process.

\subsection{Reduced density matrix and soft exponentiation}
\label{pion:sec:reduced-density-soft-exponentiation}

We consider the semi-inclusive pion-pair production process in the center-of-mass frame,
\begin{equation}
  \gamma\gamma^\ast \to \pi^+\pi^-
  +X \, .
\end{equation}
The initial photon state removes the initial-initial and initial-final state interactions at the leading-power (LP) approximation.
We identify the charged pions of the same mass $m$ as the hard final state, while $X$
denotes the
photon radiation, part of which corresponds to the unresolved photons
to be traced over at a fixed detector
resolution.
An $n$-photon state is written as
$X_n=\{(q_a,\lambda_a)\}_{a=1}^n$, where $q_a$ is the momentum of the on-shell photon with
energy $\omega_{q_a}$, and $\lambda_a$ is its polarization.  For
the two outgoing pions, we denote the momentum configuration by
\(p=(p_+,p_-)\), with \(|p\rangle=|p_+,p_-\rangle\), and write
\(\calM_n(p;X_n)\) for the amplitude that produces the hard pion state
together with the photon radiation \(X_n\).
Before the detector performs any measurement, the full outgoing state is a
superposition in Fock space over the pion momentum configurations and over all photon occupation numbers:
\begin{equation}\label{eq:Psi-state}
  |\Psi\rangle =
  \sum_{n=0}^{\infty}\frac{1}{n!}
  \prod_{a=1}^n\left[
    \sum_{\lambda_a}\int \dd\Phi(q_a)
  \right]
  \int\dd\Pi(p)\,\calM_n(p;X_n)\,|p\rangle\otimes |X_n\rangle .
\end{equation}
The phase-space measures appearing in this expansion are
\begin{equation}
  \dd\Phi(q)=\frac{\dd^3 q}{(2\pi)^3 2\omega_q},
  \qquad
  \dd\Pi(k)=\frac{\dd^3 k}{(2\pi)^3 2E_{\mathbf k}}.
\end{equation}
Here $\omega_q\equiv|\mathbf q|$ due to on-shell condition, and $\dd\Pi(k)$ denotes the one-pion measure over its momentum, while for the composite label \(p=(p_+,p_-)\) we use
\(\dd\Pi(p)=\dd\Pi(p_+)\dd\Pi(p_-)\).
At a fixed detector resolution scale \(\lambda_{\rm det}\), we use the dimensionless resolution parameter
\(\xi=\ln(\lambda_{\rm det}/\lambda_{\IR})\), where \(\lambda_{\IR}\) is an infrared cutoff, to describe the $\lambda_{\rm det}$ dependence of various quantities.
This IR scale should be understood as the scale at which the soft photon
cloud is included in the infrared dressing of the charged pion states; photons
softer than \(\lambda_{\IR}\) are therefore indistinguishable from the charged pions and thus not counted as environmental modes
to be traced over.  The environmental trace instead runs over photon modes in
the detector-unresolved window between \(\lambda_{\IR}\) and \(\lambda_{\rm det}\).  Tracing
over these unresolved modes defines the reduced hard density matrix
\begin{align}
  \widehat\rho(\xi)
  &=
  \Tr_{X,\xi}(|\Psi\rangle\langle\Psi|)=
  \sum_{n=0}^{\infty}\frac{1}{n!}
  \prod_{a=1}^{n}\left[
    \sum_{\lambda_a}\int\dd\Phi_\xi(q_a)
  \right]
  \langle X_n|\Psi\rangle\langle\Psi|X_n\rangle \,, \label{eq:reduced-trace}
\end{align}
where to restrict the photon trace to this detector-unresolved window, we replace the full
one-photon phase-space measure by
\begin{equation}
  \dd\Phi_\xi(q)=\dd\Phi(q)\Theta_\xi(q),
  \qquad
  \Theta_\xi(q)
  \equiv
  \Theta\!\left(\ln\frac{\omega_q}{\lambda_{\IR}}\right)
  \Theta\!\left(\xi-\ln\frac{\omega_q}{\lambda_{\IR}}\right).
  \label{pion:eq:unresolved-one-photon-measure}
\end{equation}
Here \(\Theta_\xi(q)\) is one for photon modes that the detector does not
resolve and zero otherwise; increasing \(\xi\) therefore integrates over a larger photon
phase space corresponding to leaving more photons unresolved.
The logarithms reflect the common knowledge on the form of scale dependence.
Since we are only interested in the soft photons in the unresolved window, we will regard $|X_n\rangle$ as the unresolved $n$-soft-photon states hereafter.
Equation~\eqref{eq:reduced-trace} is the partial trace written in the unresolved photon Fock
basis. The remaining task is to evaluate the projected state
\(\langle X_n|\Psi\rangle\), which is the inner product between the unresolved photon Fock state and the complete outgoing state,
\begin{align}
  \langle X'_n|\Psi\rangle
  &=
  \sum_{m=0}^{\infty}\frac{1}{m!}
  \prod_{b=1}^{m}\left[
    \sum_{\lambda_b}\int\dd\Phi(q_b)
  \right]\int\dd\Pi(p)\,\calM_m(p;X_m)|p\rangle\langle X'_n|X_m\rangle
  \notag\\
  &=
  \resizebox{0.88\textwidth}{!}{$\displaystyle
  \frac{1}{n!}
  \prod_{b=1}^{n}\left[
    \sum_{\lambda_b}\int\dd\Phi(q_b)
  \right]
  \sum_{\sigma\in S_n}\prod_{a=1}^{n}
  (2\pi)^3 2\omega_{q'_a}\,\delta^{(3)}(q'_a-q_{\sigma(a)})
  \delta_{\lambda'_a\lambda_{\sigma(a)}}
  \int\dd\Pi(p)\,\calM_n(p;X_n)|p\rangle$}
  \notag\\
  &=
  \frac{1}{n!}\sum_{\sigma\in S_n}
  \int\dd\Pi(p)\,
  \calM_n\big(p;q_{\sigma(1)},\lambda_{\sigma(1)},\ldots,
  q_{\sigma(n)},\lambda_{\sigma(n)}\big)|p\rangle
  \notag\\
  &=
  \int\dd\Pi(p)\,\calM_n(p;X'_n)|p\rangle .
\end{align}
To go from the first to the second line, we have used the identities for the inner products between different photon modes in Fock space,
\begin{align}
\langle X'_n|X_m\rangle=\delta^{nm}\sum_{\sigma\in S_n}\prod_{a=1}^{n}
(2\pi)^3 2\omega_{q'_a}\,\delta^{(3)}(q'_a-q_{\sigma(a)})
\delta_{\lambda'_a\lambda_{\sigma(a)}}\, ,
\end{align}
where $S_n$ denotes all permutations of $\{1,2,\cdots ,n\}$ and $\sigma(a)$ is the $a$-th element in a particular permutation $\sigma$.
Integrating over $\prod_{b=1}^n {\dd}\Phi(q_b)$ using the delta functions together with the sum $\sum_{\sigma\in S_n}$ then leads to the cancellation of $1/n!$ as a manifestation of Bose symmetry.
Relabeling the projected state \(X'_n\) as \(X_n\) and
substituting it into the Fock-basis trace gives
\begin{align}
  \widehat\rho(\xi)
  &=
  \sum_{n=0}^{\infty}\frac{1}{n!}
  \prod_{a=1}^{n}\left[
    \sum_{\lambda_a}\int\dd\Phi_\xi(q_a)
  \right]
  \int\dd\Pi(p)\dd\Pi(p')\,
  \calM_n(p;X_n)\calM^\ast_n(p';X_n)
  |p\rangle\langle p'| .\label{eq:Xtrace}
\end{align}

\begin{figure}[!t]
  \centering
  \begin{tikzpicture}[
    scale=0.96,
    pion/.style={draw=black!86,line width=0.9pt,postaction={decorate,
      decoration={markings,mark=at position 0.58 with
      {\arrow{Stealth[length=2.25mm,width=1.55mm]}}}}},
    pionp/.style={draw=black!50,line width=0.82pt,postaction={decorate,
      decoration={markings,mark=at position 0.58 with
      {\arrow{Stealth[length=2.15mm,width=1.45mm]}}}}},
    hardphoton/.style={draw=black!75,line width=0.75pt,decorate,
      decoration={snake,amplitude=1.15pt,segment length=6.2pt}},
    wilson/.style={draw=black!86,double=white,double distance=1.45pt,
      line width=0.62pt,postaction={decorate,
      decoration={markings,mark=at position 0.60 with
      {\arrow{Stealth[length=2.15mm,width=1.45mm]}}}}},
    wilsonp/.style={draw=black!50,double=white,double distance=1.35pt,
      line width=0.58pt,postaction={decorate,
      decoration={markings,mark=at position 0.60 with
      {\arrow{Stealth[length=2.05mm,width=1.35mm]}}}}},
    softtext/.style={font=\small\itshape,red!45!black},
    lab/.style={font=\small}
  ]
    \begin{scope}[xshift=-3.80cm]
      \coordinate (h) at (0,0);
      \draw[hardphoton] (-2.38,1.04) -- (-0.48,0.32);
      \draw[hardphoton] (-2.38,-1.04) -- (-0.48,-0.32);
      \draw[pion] (0.48,0.32) -- (2.38,1.04);
      \draw[pion] (0.48,-0.32) -- (2.38,-1.04);
      \filldraw[fill=black!8,draw=black!72,line width=0.7pt] (h) circle (0.62);
      \node[font=\footnotesize] at (h) {$\calM_0(p)$};
      \node[lab,left=2pt] at (-2.38,1.04) {$\gamma$};
      \node[lab,left=2pt] at (-2.38,-1.04) {$\gamma^\ast$};
      \node[lab,right=2pt] at (2.38,1.04) {$\pi^+$};
      \node[lab,right=2pt] at (2.38,-1.04) {$\pi^-$};
      \node[lab] at (0,-1.55) {(a) hard production};
    \end{scope}

    \begin{scope}[xshift=3.95cm]
      \coordinate (o) at (0,0);
      \coordinate (rp) at (2.34,1.02);
      \coordinate (lm) at (-2.34,-1.02);
      \coordinate (rpp) at (2.34,-1.02);
      \coordinate (lmp) at (-2.34,1.02);

      \shade[inner color=red!18,outer color=red!3,opacity=0.95]
        (o) ellipse [x radius=2.18cm,y radius=1.26cm];

      \draw[wilson] (o) -- (rp);
      \draw[wilson] (o) -- (lm);
      \draw[wilsonp] (o) -- (rpp);
      \draw[wilsonp] (o) -- (lmp);

      \fill[black] (o) circle (1.1pt);
      \node[lab,right=2pt] at (rp) {$\pi^+(p)$};
      \node[lab,left=2pt] at (lm) {$\pi^-(p)$};
      \node[lab,right=2pt,black!55] at (rpp) {$\pi^+(p')$};
      \node[lab,left=2pt,black!55] at (lmp) {$\pi^-(p')$};
      \node[softtext] at (0,0.88) {$\exp S_{pp'}(\xi)$};
      \node[lab] at (0,-1.55) {(b) soft overlap};
    \end{scope}
  \end{tikzpicture}
  \caption{
  Hard production and unresolved soft overlap.
  (a) The short-distance process produces the charged pion pair with hard
  amplitude $\calM_0(p)$.
  (b) A schematic picture of the coherence between two hard momentum states,
  where the double lines denote the eikonal Wilson-line dressing of the outgoing
  pions.
  }
  \label{pion:fig:hard-production-soft-overlap}
\end{figure}
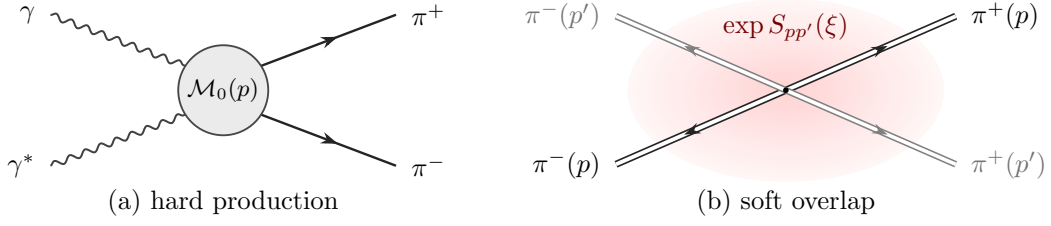

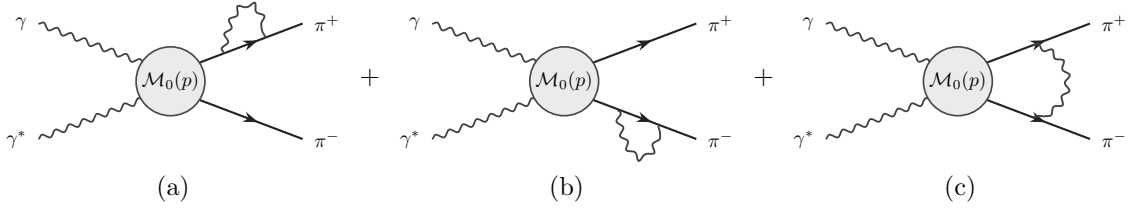
\begin{figure}[!t]
  \centering
  \tikzset{
    vpion/.style={draw=black!86,line width=0.9pt,postaction={decorate,
      decoration={markings,mark=at position 0.58 with
      {\arrow{Stealth[length=2.25mm,width=1.55mm]}}}}},
    vphoton/.style={draw=black!75,line width=0.75pt,decorate,
      decoration={snake,amplitude=1.15pt,segment length=6.2pt}},
    vsoft/.style={draw=black!75,line width=0.78pt,decorate,
      decoration={snake,amplitude=1.15pt,segment length=6.2pt}},
    vlab/.style={font=\small},
    vblob/.style={circle,draw=black!72,fill=black!8,line width=0.7pt}
  }
  \resizebox{\textwidth}{!}{%
  \begin{tabular}{@{}c@{\hspace{0.5em}}c@{\hspace{0.5em}}c@{\hspace{0.5em}}c@{\hspace{0.5em}}c@{}}
    \begin{tikzpicture}[scale=0.82, transform shape, baseline=(h.center)]
      \coordinate (h) at (0,0);
      \draw[vphoton] (-2.38,1.04) node[vlab,left=2pt] {$\gamma$} -- (-0.48,0.32);
      \draw[vphoton] (-2.38,-1.04) node[vlab,left=2pt] {$\gamma^\ast$} -- (-0.48,-0.32);
      \draw[vpion] (0.48,0.32) -- (2.38,1.04) node[vlab,right=2pt] {$\pi^+$};
      \draw[vpion] (0.48,-0.32) -- (2.38,-1.04) node[vlab,right=2pt] {$\pi^-$};
      \filldraw[vblob] (h) circle (0.62);
      \node[font=\small] at (h) {$\calM_0(p)$};
      \coordinate (a) at (0.96,0.50);
      \coordinate (b) at (1.70,0.78);
      \draw[vsoft] (a) .. controls (1.08,1.50) and (1.62,1.58) .. (b);
    \end{tikzpicture}
    & {\raisebox{-0.2ex}{$+$}} &
    \begin{tikzpicture}[scale=0.82, transform shape, baseline=(h.center)]
      \coordinate (h) at (0,0);
      \draw[vphoton] (-2.38,1.04) node[vlab,left=2pt] {$\gamma$} -- (-0.48,0.32);
      \draw[vphoton] (-2.38,-1.04) node[vlab,left=2pt] {$\gamma^\ast$} -- (-0.48,-0.32);
      \draw[vpion] (0.48,0.32) -- (2.38,1.04) node[vlab,right=2pt] {$\pi^+$};
      \draw[vpion] (0.48,-0.32) -- (2.38,-1.04) node[vlab,right=2pt] {$\pi^-$};
      \filldraw[vblob] (h) circle (0.62);
      \node[font=\small] at (h) {$\calM_0(p)$};
      \coordinate (a) at (0.96,-0.50);
      \coordinate (b) at (1.70,-0.78);
      \draw[vsoft] (a) .. controls (1.08,-1.50) and (1.62,-1.58) .. (b);
    \end{tikzpicture}
    & {\raisebox{-0.2ex}{$+$}} &
    \begin{tikzpicture}[scale=0.82, transform shape, baseline=(h.center)]
      \coordinate (h) at (0,0);
      \draw[vphoton] (-2.38,1.04) node[vlab,left=2pt] {$\gamma$} -- (-0.48,0.32);
      \draw[vphoton] (-2.38,-1.04) node[vlab,left=2pt] {$\gamma^\ast$} -- (-0.48,-0.32);
      \draw[vpion] (0.48,0.32) -- (2.38,1.04) node[vlab,right=2pt] {$\pi^+$};
      \draw[vpion] (0.48,-0.32) -- (2.38,-1.04) node[vlab,right=2pt] {$\pi^-$};
      \filldraw[vblob] (h) circle (0.62);
      \node[font=\small] at (h) {$\calM_0(p)$};
      \coordinate (a) at (1.35,0.65);
      \coordinate (b) at (1.35,-0.65);
      \draw[vsoft] (a) .. controls (2.10,0.48) and (2.10,-0.48) .. (b);
    \end{tikzpicture}
    \\[1.4ex]
    (a) & & (b) & & (c)
  \end{tabular}
  }
  \caption{One-soft-photon virtual correction in the eikonal approximation. The
  gray blob denotes the hard amplitude \(\calM_0(p)\), while the unresolved
  virtual photon attaches only to the outgoing charged eikonal lines. Panels
  (a) and (b) are the two same-line attachments, and panel (c) is the exchange
  attachment between the \(\pi^+\) and \(\pi^-\) lines; their sum gives the
  soft virtual factor \(V_p(\xi)\) proportional to \(J_p(q)\cdot J_p(-q)\).}
  \label{pion:fig:virtual-eikonal-graphs}
\end{figure}

\noindent
To proceed, we apply the LP soft theorem:
\begin{equation}
  \calM_n(p;X_n)\Big|_{\rm LP} =
  \calM_0(p)\calM_{\rm vir}(p)\prod_{a=1}^{n}S_p(q_a,\lambda_a).
  \label{pion:eq:leading-soft-amplitude-factorization}
\end{equation}
The one-photon soft factor $S_p(q,\lambda)$ is the product of the outgoing photon
polarization and the classical eikonal current of the charged pion pair, which reads,
\begin{align}
    S_p(q,\lambda)=\epsilon^\ast(q,\lambda)\cdot J_p(q)\, ,\label{eq:Sp}
\end{align}
with
\begin{align}
 J_p^\mu(q)=J^\mu_+(q)+J^\mu_-(q)\, ,\qquad\quad J_\pm^\mu(q)=\pm \frac{e p_\pm^\mu}{p_\pm\cdot q+\ii0}
 \label{eq:Jp-current}
\end{align}
in scalar QED~\cite{Low:1958sn,Burnett:1967km,Weinberg:1965nx,DelDuca:1990gz,Bohm:1986fg,Schonherr:2008av,Elvang:2016qvq}.
The factor \(\calM_{\rm vir}(p)\) is the virtual correction associated with the
hard pion-pair in momentum eigenstate $|p\rangle$.  At leading power,
the virtual corrections can be written as products of the one-photon
exchange contribution \(V_p(\xi)\) shown in
Fig.~\ref{pion:fig:virtual-eikonal-graphs}.
Summing over any number of disconnected copies gives the exponential
\(\calM_{\rm vir}(p)=\exp[V_p(\xi)]\), the usual Yennie–Frautschi–Suura (YFS)/Bloch--Nordsieck (BN)
exponentiation of soft photons~\cite{Bloch:1937pw,Yennie:1961ad}.  We adopt
the shorthand $\displaystyle\int_q\equiv\int\dd^4q/(2\pi)^4$, and restrict the soft virtual loop to the same resolution window by
$\displaystyle\int_{q,\xi}\equiv\int_q\Theta_\xi(q)$ to extract the relevant effects of virtual contributions to the reduced density matrix.  The eikonal Feynman rule for a single photon attachment to an outgoing charged pion line is
\begin{alignat}{3}
  \pi^\pm&:\quad&
  (\mp \ii e)(2p_\pm^\mu+q^\mu)\,
  \frac{\ii}{(p_\pm +q)^2-m^2+\ii0}
  &\simeq&
  {\pm}\,\frac{e p_\pm^\mu}{p_\pm\cdot q+\ii0} = J^\mu_\pm(q)\, ,\label{pion:eq:eikonal-rule}
\end{alignat}

\noindent
which directly determines the contribution from panel (c) of
Fig.~\ref{pion:fig:virtual-eikonal-graphs}. Panels (a) and (b) are self-energy
diagrams that contribute to the soft amplitude only through the wave-function
renormalization factor \(\sqrt{Z_\pm}=1+\frac12\Sigma'_\pm(m^2)\), where
\(\Sigma_\pm(p_\pm^2)\) is the \(\pi^\pm\) QED self-energy amplitude and the prime denotes
\(\partial/\partial p_\pm^2\). Here we are expanding the self-energy correction around $p^2=m^2$ to extract the IR contribution relevant to our  consideration.
At leading power, the contribution of panel (a) in Feynman gauge is
\begin{align}
  V_p^{(a)}(\xi)
  &=
  \frac12\,\Sigma'_+(m^2)\Big|_{\rm soft}
  \notag\\
  &=
  \frac12\,\frac{\partial}{\partial p_+^2}
  \bigg[-4\ii e^2\int_{q,\xi}
  \frac{p_+^2}{(q^2+\ii0)(p_+^2-m^2+2p_+\cdot q+\ii 0)}\bigg]_{p_+^2=m^2}
  \notag\\
  &\simeq
  \frac12\,(-4\ii e^2)\int_{q,\xi}\frac{p_+^2}{q^2+\ii0}\,
  \frac{-1}{(2p_+\cdot q+\ii0)^2}
  \notag\\
  &=
  \frac12\int_{q,\xi}\frac{-\ii}{q^2+\ii0}\,
  \frac{e^2p_+^2}{(p_+\cdot q+\ii 0)(-p_+\cdot q-\ii 0)} ,
  \label{pion:eq:Va-wave-function}
\end{align}
where the second line isolates the leading-soft contribution from the one-loop self-energy of panel (a),
and in the third line,
we have neglected the subleading contribution from $\frac{\partial}{\partial p_+^2}$ acting on $p_+^2$ in the numerator with $p_\pm^2\sim m^2\gg |p\cdot q|$.
Panel (b) follows with \(p_+\) replaced by \(p_-\), and adding panel (c)
gives the leading-power soft part of the one-loop virtual correction,
\begin{align}
  V_p(\xi)
  &= V_p^{(a)}(\xi)+V_p^{(b)}(\xi)+V_p^{(c)}(\xi)
  \notag\\
  &=
  \frac{1}{2}\int_{q,\xi}
  \frac{-\ii}{q^2+\ii0}
  \frac{e^2 p_+^2}
  {(p_+\cdot q+\ii0)(-p_+\cdot q-\ii0)}
  \notag\\
  &+\frac{1}{2}\int_{q,\xi}
  \frac{-\ii}{q^2+\ii0}
  \frac{e^2 p_-^2}
  {(p_-\cdot q+\ii0)(-p_-\cdot q-\ii0)}
  \notag\\
  &-\frac{1}{2}\int_{q,\xi}
  \frac{-\ii}{q^2+\ii0}
  \left[
    \frac{e^2 p_+\cdot p_-}
    {(p_+\cdot q+\ii0)(-p_-\cdot q+\ii0)}
    +
    \frac{e^2 p_-\cdot p_+}
    {(p_-\cdot q+\ii0)(-p_+\cdot q+\ii0)}
  \right]
  \notag\\
  &=
  \frac{1}{2}\int_{q,\xi} \frac{\ii}{q^2+\ii0}\left(\frac{e p_+}{p_+\cdot q+\ii0}-\frac{e p_-}{p_-\cdot q-\ii0}\right)^2
  \, .
  \label{pion:eq:virtual-eikonal-current}
\end{align}
The real part of \(V_p(\xi)\) calculated below involves only on-shell photons, for which $p_\pm\cdot q\neq0$. This makes the $\ii0$ prescription in $1/(p_\pm\cdot q)$ negligible, so in this case the \(-\ii0\) of the \(\pi^-\) eikonal factor can be flipped, turning the square into \(J_p(q)\cdot J_p(q)\) with \(J_p\) of eq.~\eqref{eq:Sp}.  The \(\ii0\) prescription therefore affects only the imaginary part of \(V_p(\xi)\).
We note that while the integrands of the leading-soft contribution for the massive and massless cases are identical, the latter, which we do not consider in this work, requires additional regularization for the collinear divergences from the phase integral.
The real part of the virtual exponent is therefore given by
\begin{align}
  2\Ree V_p(\xi)
  &=V_p(\xi)+V^\ast_p(\xi)
  \notag\\
  &=
  \frac{1}{2}\int_{q,\xi}
  \left[
    \frac{-\ii}{q^2+\ii0}
    +
    \frac{\ii}{q^2-\ii0}
  \right]
  \sum_\lambda
  \left|
    \epsilon^\ast(q,\lambda)\cdot J_p(q)
  \right|^2
  \notag\\
  &=
  -\pi\int_{q,\xi}
  \delta(q^2)
  \sum_\lambda
  \left|
    \epsilon^\ast(q,\lambda)\cdot J_p(q)
  \right|^2
  \notag\\
  &=
  -\sum_\lambda\int\dd\Phi_\xi(q)\,|S_p(q,\lambda)|^2
  \notag\\
  &=
  -S_{pp}(\xi).
  \label{pion:eq:virtual-loop-soft-factor-form}
\end{align}
In the last equality we have introduced the integrated soft overlap
\begin{equation}
  S_{pp'}(\xi)
  \equiv
  \sum_\lambda\int\dd\Phi_\xi(q)\,
  S_p(q,\lambda)S^\ast_{p'}(q,\lambda)\, .
  \label{pion:eq:integrated-soft-overlap}
\end{equation}
As expected, computing the real part corresponds to cutting the soft virtual photon
loop, which puts the photon on shell and hence gives the same phase-space measure as the single real-photon emission.
The minus sign in \(2\Ree V_p(\xi)=-S_{pp}(\xi)\) is
necessary for canceling out the $\lambda_{\rm IR}$
dependence between real and virtual corrections in
the diagonal entries of the density matrix, a manifestation of the infrared finiteness of the inclusive cross section.  The remaining
imaginary part of the loop integral contains a momentum-dependent eikonal phase, which we keep by decomposing the same virtual factor as
\begin{equation}
  V_p(\xi)
  =
  -\frac12 S_{pp}(\xi)-\ii\chi_p(\xi),
  \qquad
  \chi_p(\xi)\equiv -\Imm V_p(\xi).
  \label{pion:eq:virtual-phase-decomposition}
\end{equation}
Returning to the Fock-space trace, the hard density matrix reads,
\begin{align}
  \rho_{pp'}(0)\equiv \calM_0(p)\calM^\ast_0(p') \,.
\end{align}
Substituting the factorized amplitude eq.~\eqref{pion:eq:leading-soft-amplitude-factorization} into the trace~\eqref{eq:Xtrace} and then projecting onto the momentum eigenstates $(p,p')$ of the pion pair leads to the full LP density matrix,
\begin{align}
  \rho_{pp'}(\xi)
  &=
  \sum_{n=0}^{\infty}\frac{1}{n!}
  \prod_{a=1}^{n}\left[
    \sum_{\lambda_a}\int\dd\Phi_\xi(q_a)
  \right]
  \calM_n(p;X_n)\calM^\ast_n(p';X_n)
  \notag\\
  &\simeq
  \rho_{pp'}(0)\,
  \exp[V_p(\xi)+V_{p'}^\ast(\xi)]
  \sum_{n=0}^{\infty}\frac{1}{n!}
  \prod_{a=1}^{n}\left[
    \sum_{\lambda_a}\int\dd\Phi_\xi(q_a)\,
    S_p(q_a,\lambda_a)S^\ast_{p'}(q_a,\lambda_a)
  \right]
  \notag\\
  &=
  \rho_{pp'}(0)\,
  \exp[V_p(\xi)+V_{p'}^\ast(\xi)+S_{pp'}(\xi)]
  \notag\\
  &\equiv
  \rho_{pp'}(0)\,
  \exp[-\Gamma_{pp'}(\xi)-\ii\chi_{pp'}(\xi)].
  \label{pion:eq:combined-coherence-factor}
\end{align}
The decoherence exponent and the phase can therefore be identified,
\begin{align}
  \Gamma_{pp'}(\xi)
  &=
  \frac{1}{2}S_{pp}(\xi)+\frac{1}{2}S_{p'p'}(\xi)-\Ree S_{pp'}(\xi)
  \notag\\
  &=
  \frac{1}{2}\int\dd\Phi_\xi(q)\sum_\lambda
  |S_p(q,\lambda)-S_{p'}(q,\lambda)|^2\, ,
  \label{pion:eq:gamma-square}
  \\
  \chi_{pp'}(\xi)
  &\equiv
  \chi_p(\xi)-\chi_{p'}(\xi)-\Imm S_{pp'}(\xi)\, .
  \label{pion:eq:phase-difference}
\end{align}
For \(p=p'\), the diagonal matrix element shows the real--virtual cancellation
directly following $\Gamma_{pp}(\xi)=\chi_{pp}(\xi)= 0$, resulting in
$\rho_{pp}(\xi)=%
\rho_{pp}(0)$.
For the off-diagonal elements where \(p\neq p'\), the
decoherence exponent $\Gamma_{pp'}(\xi)$ is always positive as shown in the second line of eq.~\eqref{pion:eq:gamma-square}.
Thus the unresolved soft radiation preserves inclusive probabilities but
suppresses hard momentum coherence.

\subsection{Lindblad flow in resolution space}
\label{pion:sec:lindblad-flow}

We now turn the resolution dependence found in
Section~\ref{pion:sec:reduced-density-soft-exponentiation} into an evolution equation
for the reduced hard density matrix with the dimensionless resolution parameter \(\xi\) as the flow
parameter.  Differentiating equation
\eqref{pion:eq:combined-coherence-factor} with respect to \(\xi\) yields,
\begin{equation}
  \frac{\dd\rho_{pp'}}{\dd\xi}
  =
  \left[-\dot\Gamma_{pp'}(\xi)-\ii\dot\chi_{pp'}(\xi)\right]\rho_{pp'}(\xi),
  \label{pion:eq:matrix-flow-full-exponent}
\end{equation}
where overdots denote derivatives with respect to the resolution parameter \(\xi\).  The physical
interpretation of this evolution is illustrated in
Fig.~\ref{pion:fig:resolution-flow}.
To write compactly the shell contribution from the newly resolved/unresolved photon modes, we define the shell integration
operation following eq.~\eqref{pion:eq:unresolved-one-photon-measure}, including the summation over polarization states as,
\begin{align}
  \int\!\dd\nu_\xi\,F(q,\lambda)
  &\equiv
  \sum_\lambda\int\dd\Phi(q)\,
  \partial_\xi\Theta_\xi(q)\,F(q,\lambda)
  \notag\\
  &=
  \sum_\lambda\int\dd\Phi(q)\,
  \Theta\!\left(\ln\frac{\omega_q}{\lambda_{\IR}}\right)
  \delta\!\left(\xi-\ln\frac{\omega_q}{\lambda_{\IR}}\right)
  F(q,\lambda).
  \label{pion:eq:shell-integration-operation}
\end{align}
\afterpage{%
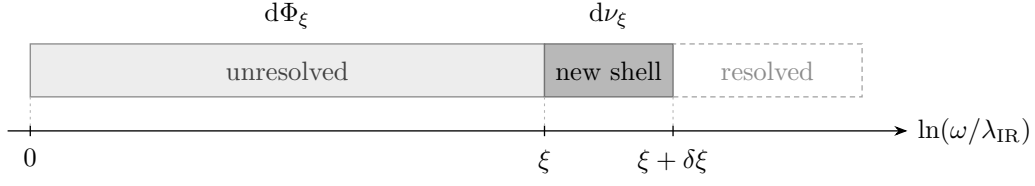
\begin{figure}[!t]
  \centering
  \begin{tikzpicture}[font=\small]
    \def\xs{6.8}\def\xd{8.5}\def\xe{11.0}\def\h{0.7}
    \fill[black!7] (0,0) rectangle (\xs,\h);
    \draw[black!45, line width=0.5pt] (0,0) rectangle (\xs,\h);
    \fill[black!28] (\xs,0) rectangle (\xd,\h);
    \draw[black!55, line width=0.5pt] (\xs,0) rectangle (\xd,\h);
    \draw[black!35, line width=0.5pt, dash pattern=on 2pt off 1.5pt] (\xd,0) -- (\xe,0) -- (\xe,\h) -- (\xd,\h);
    \node[black!75] at ({\xs/2},{\h/2}) {unresolved};
    \node[black] at ({(\xs+\xd)/2},{\h/2}) {new shell};
    \node[black!45] at ({(\xd+\xe)/2},{\h/2}) {resolved};
    \node[above=2pt] at ({\xs/2},\h) {$\dd\Phi_\xi$};
    \node[above=2pt] at ({(\xs+\xd)/2},\h) {$\dd\nu_\xi$};
    \draw[-{Stealth[length=5pt]}, line width=0.5pt] (-0.3,-0.45) -- (\xe+0.6,-0.45) node[right] {$\ln(\omega/\lambda_\IR)$};
    \foreach \x/\l in {0/0, \xs/\xi, \xd/\xi+\delta\xi}
      \draw[line width=0.5pt] (\x,-0.52) -- (\x,-0.38) node[pos=0, below=1pt] {$\l$};
    \draw[black!30, line width=0.4pt, dash pattern=on 1pt off 1.5pt] (0,0) -- (0,-0.38) (\xs,0) -- (\xs,-0.38) (\xd,0) -- (\xd,-0.38);
  \end{tikzpicture}
  \caption{Resolution flow in soft-photon phase space. The measure \(\dd\Phi_\xi\) integrates over the unresolved photons with \(0<\ln(\omega_q/\lambda_{\IR})<\xi\). Increasing \(\xi\) by \(\delta\xi\) turns the shell \(\xi<\ln(\omega_q/\lambda_{\IR})<\xi+\delta\xi\) from resolved to unresolved, and the operation \(\int\dd\nu_\xi\) of eq.~\eqref{pion:eq:shell-integration-operation} restricts the photon-mode integral to this shell, which supplies the resolution rates \(\dot\Gamma_{pp'}(\xi)\) and \(\dot\chi_{pp'}(\xi)\).}
  \label{pion:fig:resolution-flow}
\end{figure}
}
The matrix-element-level equation~\eqref{pion:eq:matrix-flow-full-exponent} can be cast into the following
operator form,
\begin{equation}
  \frac{\dd\widehat\rho}{\dd\xi}
  =
  -\ii[\widehat H_{\rm eff}(\xi),\widehat\rho(\xi)]
  +
  \int\!\dd\nu_\xi\,
  \left[
    \widehat L_1(\xi)\widehat\rho(\xi)\widehat L_1^\dagger(\xi)
    -
    \frac{1}{2}
    \left\{
      \widehat L_1^\dagger(\xi)\widehat L_1(\xi),
      \widehat\rho(\xi)
    \right\}
  \right],
  \label{pion:eq:lindblad-flow}
\end{equation}
where we identify the Lindblad jump operator and the effective Hamiltonian as
\begin{equation}
  \widehat L_1(\xi)
  =
  \sum_r |r\rangle S_r(q,\lambda)\langle r|,
  \qquad
  \widehat H_{\rm eff}(\xi)
  =
  \sum_r |r\rangle
  \dot\chi_r(\xi)
  \langle r|
  \label{pion:eq:L1-and-H-identification}
\end{equation}
with $|r\rangle$ running over the momentum eigenstates of the hard particles and \(\sum_r\equiv\int\dd\Pi(r)\).
Physically, \(\widehat L_1(\xi)\) describes the leading-power emission of an
unresolved soft photon. The emission does not change the hard particle
momentum, but carries a weight of the soft factor \(S_r(q,\lambda)\) entering the flow equation~\eqref{pion:eq:lindblad-flow} and the soft factor can also be projected out by $\langle r|\widehat L_1(\xi) | r\rangle$.  The operator \(\widehat H_{\rm eff}(\xi)\) is the effective Hamiltonian associated
with the phase of the virtual correction~\eqref{pion:eq:virtual-phase-decomposition} and generates the corresponding
closed-system evolution.

We now verify directly that \eqref{pion:eq:lindblad-flow} reproduces exactly the matrix-element flow~\eqref{pion:eq:matrix-flow-full-exponent} by sandwiching eq.~\eqref{pion:eq:lindblad-flow} between momentum eigenstates. Using the definition of $\widehat H_{\rm eff}$,
we find the Hamiltonian part gives,
\begin{equation}
  \langle p|-\ii[\widehat H_{\rm eff}(\xi),\widehat\rho(\xi)]|p'\rangle
  =
  -\ii\left[
    \dot\chi_p(\xi)
    -
    \dot\chi_{p'}(\xi)
  \right]\rho_{pp'}(\xi)\, ,
  \label{pion:eq:H-matrix-element}
\end{equation}
with $\rho_{pp'}\equiv\langle p|\widehat\rho|p'\rangle$. Similarly, the jump and anticommutator terms are
\begin{align}
  \langle p|\widehat L_1(\xi)\widehat\rho(\xi)\widehat L_1^\dagger(\xi)|p'\rangle
  &=
  \sum_{r,s}
  S_r(q,\lambda)S^\ast_s(q,\lambda)
  \langle p|r\rangle
  \langle r|\widehat\rho(\xi)|s\rangle
  \langle s|p'\rangle
  =
  S_p(q,\lambda)S^\ast_{p'}(q,\lambda)\rho_{pp'}(\xi),
  \notag\\
  \langle p|\widehat L_1^\dagger(\xi)\widehat L_1(\xi)\widehat\rho(\xi)|p'\rangle
  &=
  \sum_r
  |S_r(q,\lambda)|^2
  \langle p|r\rangle
  \langle r|\widehat\rho(\xi)|p'\rangle
 =
  |S_p(q,\lambda)|^2\rho_{pp'}(\xi),
  \notag\\
  \langle p|\widehat\rho(\xi)\widehat L_1^\dagger(\xi)\widehat L_1(\xi)|p'\rangle
  &=
  \sum_r
  |S_r(q,\lambda)|^2
  \langle p|\widehat\rho(\xi)|r\rangle
  \langle r|p'\rangle
 =
  |S_{p'}(q,\lambda)|^2\rho_{pp'}(\xi).
  \label{pion:eq:L1-matrix-elements-section-two}
\end{align}
Combining these pieces, we find,
\begin{align}
  \frac{\dd\rho_{pp'}}{\dd\xi}
  &=
  -\ii\left[
    \dot\chi_p(\xi)
    -
    \dot\chi_{p'}(\xi)
    -
    \int\!\dd\nu_\xi\,
    \Imm\!\left(S_p(q,\lambda)S_{p'}^\ast(q,\lambda)\right)
  \right]\rho_{pp'}(\xi)
  \notag\\
  &-
  \frac12
  \int\!\dd\nu_\xi\,
  |S_p(q,\lambda)-S_{p'}(q,\lambda)|^2
  \rho_{pp'}(\xi)
  \notag\\
  &=
  \left[-\dot\Gamma_{pp'}(\xi)-\ii\dot\chi_{pp'}(\xi)\right]\rho_{pp'}(\xi)\, ,
  \label{pion:eq:matrix-flow-with-H}
\end{align}
which reproduces eq.~\eqref{pion:eq:matrix-flow-full-exponent} as advocated.
We further show in Appendix~\ref{pion:app:cptp} that the dissipative dephasing part $\dot\Gamma_{pp'}(\xi)$ is completely positive trace-preserving
(CPTP) and completely positive (CP)
divisible.

\subsection{Angular decoherence between hard scalar charged particles}
\label{pion:sec:angular-decoherence}

We now evaluate the decoherence exponent in the center-of-mass frame,
where the \(\pi^\pm\) in the two hard configurations $(p_\pm,p'_\pm)$ have the same energy \(E\) and speed $v$,
but travel in different directions, given by the unit vectors \(\mathbf n\) and \(\mathbf n'\):
\begin{equation}
  p_\pm=E(1,\pm v\mathbf n),\qquad
  p'_\pm=E(1,\pm v\mathbf n'),\qquad
  \mathbf n\cdot\mathbf n'=\cos\delta\theta .
  \label{pion:eq:nearby-pion-kinematics}
\end{equation}
For small angular separation,
\(|\mathbf n-\mathbf n'|^2\simeq(\delta\theta)^2\). Writing the soft photon momentum
as \(q=\omega(1,\mathbf n_q)\), with \(\mathbf n_q\) the unit vector along the photon momentum, and using the transverse gauge with
\(\epsilon^0(q,\lambda)=0\) and \(\mathbf n_q\cdot\epsilon(q,\lambda)=0\), we isolate the simplest angular scaling in the non-relativistic limit
\(v^2\ll1\).  For relativistic pions only the overall coefficient changes, and
the scaling with \(\xi\) and \(\delta\theta\) stays the same. The soft factor then reduces to,
\begin{equation}
  S_p(q,\lambda)=-
  \frac{2ev\,\mathbf n\cdot\epsilon^\ast(q,\lambda)}
  {\omega\left(1-v^2(\mathbf n\cdot\mathbf n_q)^2\right)}
  \simeq-
  \frac{2ev\,\mathbf n\cdot\epsilon^\ast(q,\lambda)}{\omega}.
\end{equation}
For the energy-window defined in~\eqref{pion:eq:unresolved-one-photon-measure}, substituting this approximation
into \eqref{pion:eq:gamma-square} gives the leading-power decoherence exponent as a function of $\delta\theta\ll1$,
\begin{align}
  \Gamma_{pp'}(\xi)
  &=
  \frac{1}{2}\int\dd\Phi_\xi(q)\sum_\lambda
  |S_p(q,\lambda)-S_{p'}(q,\lambda)|^2
  \notag\\
  &\simeq
  \frac{1}{2}\int\dd\Phi_\xi(q)\,
  \frac{4e^2v^2}{\omega^2}
  \sum_\lambda |(\mathbf n-\mathbf n')\cdot\epsilon^\ast(q,\lambda)|^2
  \notag\\
  &=
  \frac{1}{2}\int\dd\Phi_\xi(q)\,
  \frac{4e^2v^2}{\omega^2}
  \sum_{ij}(\mathbf n-\mathbf n')_i(\mathbf n-\mathbf n')_j
  (\delta_{ij}-n_{q,i}n_{q,j})
  \notag\\
  &=
  \frac{e^2v^2}{(2\pi)^3}
  \int_{\lambda_{\IR}}^{\lambda_{\rm det}}\frac{\dd\omega}{\omega}
  \int\dd\Omega\,
  \left(|\mathbf n-\mathbf n'|^2-(\mathbf n_q\cdot(\mathbf n-\mathbf n'))^2\right)
  \notag\\
  &=
  \frac{e^2v^2}{(2\pi)^3}\frac{8\pi}{3}
  |\mathbf n-\mathbf n'|^2\ln\frac{\lambda_{\rm det}}{\lambda_{\IR}}
  \notag\\
  &=
  \frac{4\alpha}{3\pi}v^2(\delta\theta)^2 \xi ,
  \label{pion:eq:angular-decoherence-main}
\end{align}
where we have used \(\int\dd\Omega\,n_{q,i}n_{q,j}=(4\pi/3)\delta_{ij}\),
\(|\mathbf n-\mathbf n'|^2\simeq(\delta\theta)^2\), and the polarization sum in transverse gauge,
\(\sum_\lambda\epsilon_i(q,\lambda)\epsilon_j^\ast(q,\lambda)
=\delta_{ij}-n_{q,i}n_{q,j}\). The radial integral gives
\(\ln(\lambda_{\rm det}/\lambda_{\IR})=\xi\), yielding the coefficient quoted in
\eqref{pion:eq:angular-decoherence-main} with \(\alpha\equiv e^2/(4\pi)\).

Clearly, the matrix element obeys
\begin{equation}
  \frac{\dd\rho_{pp'}}{\dd\xi}
  =
  -\frac{4\alpha}{3\pi}v^2(\delta\theta)^2\rho_{pp'}(\xi).
\end{equation}
This is the resolution-flow equation in the pion example at small angles $\delta\theta\ll1$ with $v^2\ll1$: identical directions
have \(\delta\theta=0\) corresponding to the diagonal elements of the density matrix, so the unresolved photons cannot tell the two hard
states apart and the same-direction coherence is not suppressed, while correlations between nearby
but distinct directions are damped at a rate proportional to their squared
angular separation. The factor \(v^2\) is the usual velocity suppression of
soft radiation from slow-moving charges.

The same result can be interpreted as an angular coherence scale, obtained by
defining \(\delta\theta_{\mathrm{coh}}(\xi)\) through
\(\Gamma_{pp'}(\xi)\sim 1\) above which coherence between different momentum eigenstates becomes negligible,
\begin{equation}
  \delta\theta_{\mathrm{coh}}(\xi)\sim
  \sqrt{\frac{3\pi}{4\alpha v^2 \xi}} .
  \label{pion:eq:angular-coherence-scale}
\end{equation}
Angles much smaller than this scale are not efficiently distinguished by the
unresolved photons, whereas larger angles give strongly suppressed off-diagonal
entries; increasing the detector resolution interval therefore narrows the
angular coherence region as more soft photons are classified as environmental degrees of freedom.

\section{\texorpdfstring{\(S\)-matrix-level soft theorem and the Wilson-line operators}{S-matrix-level soft theorem}}
\label{sec:soft-smatrix-wilson-line}

The amplitude-level calculation above identified the Lindblad evolution of the reduced density matrix a posteriori, but left the origin of this Lindblad structure unclear. To reveal its origin from first principles, we start from the scalar-QED \(S\)-matrix that governs the final-state dynamics of the spin-zero charged pion pair. In this section, we show that its leading-power soft contribution factorizes into Wilson-line operators conditioned on the hard particle momenta. This factorized form is shown in Section~\ref{sec:kraus-representation} to directly result in the Lindblad structure, which makes it manifest that the Wilson-line operator serves as the jump operator. The same construction extends straightforwardly to the finite-temperature case, and reproduces these results within the influence-functional approach of Section~\ref{sec:influence-functional}.

\subsection{\texorpdfstring{\(S\)-matrix factorization in scalar QED}{S-matrix factorization in scalar QED}}
\label{wl:sec:soft-smatrix-factorization}

At leading power, the soft \(S\)-matrix takes the form,
\begin{align}
  \widehat S_{\rm soft}
  &=
  \mathcal T
  \exp\!\left[
  -\ii\int\dd^4x
  \sum_p
  |p\rangle\langle p|\otimes
  J_p^\mu(x)\widehat A_\mu(x)
  \right]\, ,
  \label{wl:eq:Ssoft-before-factorization}
\end{align}
where $|p\rangle$ is the momentum eigenstate of the hard particle pair, $\widehat A_\mu$ is the field operator of the soft photon, and
\begin{equation}
  J_p^\mu(x)
  =
  e\int_0^\infty\dd s\,
  p_+^\mu\delta^{(4)}(x-z_{p_+}(s))
  -
  e\int_0^\infty\dd s\,
  p_-^\mu\delta^{(4)}(x-z_{p_-}(s)) \label{eq:eikonal-EM-current}
\end{equation}
is the scalar EM eikonal current induced by a pair of particles with opposite charge ($e$ and $-e$) traveling with momenta $p_\pm$ along the worldlines parameterized by $z_{p_\pm}$ under the leading-power approximation. We refer to Appendix~\ref {app:scalarQED} for detailed derivations.
Expanding
eq.~\eqref{wl:eq:Ssoft-before-factorization} as a Dyson series gives
\begin{align}
  \widehat S_{\rm soft}
  &=
  \sum_{n=0}^{\infty}
  \frac{(-\ii)^n}{n!}
  \prod_{a=1}^n
  \left[
  \int\dd^4x_a
  \sum_{p_a}
  \right]
  \left(
  |p_1\rangle\langle p_1|
  \cdots
  |p_n\rangle\langle p_n|
  \right)
  \notag\\
  &\otimes
  \mathcal T
  \left\{
  J_{p_1}^{\mu_1}(x_1)\widehat A_{\mu_1}(x_1)
  \cdots
  J_{p_n}^{\mu_n}(x_n)\widehat A_{\mu_n}(x_n)
  \right\}.
  \label{wl:eq:Ssoft-dyson-expanded}
\end{align}
The momentum eigenstate projectors $|p_a\rangle\langle p_a |$ are time independent and commute with the photon operators, so
the time-ordering operation acts only on the photon fields. With the common normalization condition
$\langle p|k\rangle =1$ for the eigenstates,  the projector chain collapses under
the product momentum integrals for an arbitrary function $F(p_1,\ldots,p_n)$,
\begin{align}
  &\prod_{a=1}^n
  \sum_{p_a}
  |p_1\rangle\langle p_1|
  \cdots
  |p_n\rangle\langle p_n|\,
  F(p_1,\ldots,p_n)
 =
  \sum_p
  |p\rangle\langle p|\,
  F(p,\ldots,p).
  \label{wl:eq:projector-chain-collapse}
\end{align}
Applying eq.~\eqref{wl:eq:projector-chain-collapse} to
eq.~\eqref{wl:eq:Ssoft-dyson-expanded}
immediately yields,
\begin{align}
  \widehat S_{\rm soft}
  &=
  \sum_p
  |p\rangle\langle p|
  \otimes
  \sum_{n=0}^{\infty}
  \frac{(-\ii)^n}{n!}
  \prod_{a=1}^n
  \int\dd^4x_a\,
  \mathcal T
  \left\{
  J_p^{\mu_1}(x_1)\widehat A_{\mu_1}(x_1)
  \cdots
  J_p^{\mu_n}(x_n)\widehat A_{\mu_n}(x_n)
  \right\}
  \notag\\
  &=
  \sum_p |p\rangle\langle p|
  \otimes
  \mathcal T
  \exp\!\left[
  -\ii\int\dd^4x\,
  J_p^\mu(x)\widehat A_\mu(x)
  \right].
  \label{wl:eq:Ssoft-factorized}
\end{align}
Inserting the worldline representation of
\(J_p^\mu\) in eq.~\eqref{eq:eikonal-EM-current}, which is effectively the classical EM current, yields,
\begin{align}
  \widehat S_{\rm soft}
  &=
  \sum_p |p\rangle\langle p|
  \otimes
  \mathcal T
  \exp\!\left[
  -\ii e\int_0^\infty\dd s\,
  p_+^\mu\widehat A_\mu(z_{p_+}(s))
  +\ii e\int_0^\infty\dd s\,
  p_-^\mu\widehat A_\mu(z_{p_-}(s))
  \right]
  \notag\\
  &=
  \sum_p |p\rangle\langle p|
  \otimes
  \mathcal P
  \exp\!\left[
  -\ii e\int_{C_{p_+}}\!\dd z^\mu\,
  \widehat A_\mu(z)
  +\ii e\int_{C_{p_-}}\!\dd z^\mu\,
  \widehat A_\mu(z)
  \right]
  \notag\\
  &=
  \sum_p |p\rangle\langle p|
  \otimes
  \mathcal P\Bigg\{
  \exp\!\left[
  -\ii e\int_{C_{p_+}}\!\dd z^\mu\,\widehat A_\mu(z)
  \right]
  \exp\!\left[
  +\ii e\int_{C_{p_-}}\!\dd z^\mu\,\widehat A_\mu(z)
  \right]
  \Bigg\}
  \notag\\
  &\equiv
  \sum_p |p\rangle\langle p|
  \otimes
  \mathcal P\left\{
  \widehat W_{p_+}\widehat W_{p_-}
  \right\}
  \notag\\
  &\equiv
  \sum_p |p\rangle\langle p|\otimes \widehat W_p.
  \label{wl:eq:Ssoft-wilson-factorization}
\end{align}
Here $C_{p_\pm}$ denotes the path of the charged pion with momentum $p_\pm$. \(\widehat W_p\) is the pion-pair Wilson line built from the individual Wilson lines \(\widehat W_{p_\pm}\) induced by \(\pi^\pm\) with path ordering ${\cal P}$.

The full \(\gamma\gamma^\ast\to\pi^+\pi^-+X\) scattering process can be
viewed as an \(S\)-matrix evolution from \(t=-\infty\) to
\(t=+\infty\), split at the production time \(t=0\): the hard \(S\)-matrix \(\widehat S_0\)
evolves the initial state from \(-\infty\) to \(0\) and produces the hard
pion pair, while the soft \(S\)-matrix evolves this pion pair from \(0\) to
\(+\infty\) through the soft-photon dynamics described above. With
this separation, the full leading-power \(S\)-matrix takes the form,
\begin{equation}
  \widehat S
  \simeq
  \left(
    \sum_p |p\rangle\langle p|\otimes\widehat W_p
  \right)
  (\widehat S_0\otimes\widehat I),
  \label{wl:eq:general-S-factorization}
\end{equation}
where
\(\widehat I\) is the identity operator on the photon Fock space.
Let us first show that eq.~\eqref{wl:eq:general-S-factorization} reproduces the usual soft theorem at the amplitude level. We take an initial hard state \(|i\rangle\) with the photon vacuum \(|0\rangle\), and a final state with hard momentum \(p\) and soft photon configuration \(X\). With \(|p;X\rangle\equiv |p\rangle\otimes |X\rangle\), \(|i;0\rangle\equiv |i\rangle\otimes |0\rangle\), and \(\calM_0(p)\equiv\langle p|\widehat S_0|i\rangle\) denoting the initial hard-production $S$-matrix projected onto the momentum eigenstate $\langle p|$, the amplitude becomes
\begin{align}
  \calM(p;X)
  &\equiv
  \langle p;X|\widehat S|i;0\rangle
  \notag\\
  &\simeq
  \langle p;X|
  \left(
    \sum_{p'} |p'\rangle\langle p'|\otimes\widehat W_{p'}
  \right)
  (\widehat S_0\otimes\widehat I)
  |i;0\rangle
  \notag\\
  &=
  \sum_{p'}
  \langle p|p'\rangle
  \langle p'|\widehat S_0|i\rangle
  \langle X|\widehat W_{p'}|0\rangle
  \notag\\
  &=
  \calM_0(p)
  \langle X|\widehat W_p|0\rangle .
  \label{wl:eq:amplitude-soft-factorization}
\end{align}
Thus the \(S\)-matrix relation at the amplitude level adheres to the physical picture that the hard process produces the charged particles first, then the Wilson-line matrix element supplies the soft-radiation factor associated with the hard momentum \(p\).

\subsection{Physical content of the Wilson line}
\label{wl:sec:wilson-line-physical-content}

The factorization above isolates the leading-power photon dynamics in the
conditional Wilson-line operator $\widehat W_p$.  We therefore evaluate its
matrix elements to make the virtual-correction and real-emission
content encoded in it explicit.

Let \(\mathcal C_p\equiv C_{p_+}\cup C_{p_-}\), and let
\(\Theta_{\mathcal C_p}(x,y)\) denote the step function that is one when
\(x\) is later in time than \(y\) on \(\mathcal C_p\).  Using the eikonal current in
eq.~\eqref{eq:eikonal-EM-current}, the Wilson-line operator yields,
\begin{align}
  \widehat W_p
  &=
  \exp\!\bigg[
  -\ii\int\dd^4x\,
  J_p^\mu(x)\widehat A_\mu(x)
  -
  \frac12
  \int\dd^4x\,\dd^4y\,
  \Theta_{\mathcal C_p}(x,y)
  \left[
  J_p^\mu(x)\widehat A_\mu(x),
  J_p^\nu(y)\widehat A_\nu(y)
  \right]
  \bigg]
  \notag\\
  &=
  \exp\!\left[
  -\ii\int\dd^4x\,
  J_p^\mu(x)\widehat A_\mu(x)
  -
  \ii\chi_p
  \right]
  \notag\\
  &=
  \exp\!\left[
  -\ii e
  \int_0^\infty\!\dd s\,
  p_+^\mu\widehat A_\mu(z_{p_+}(s))
  +\ii e
  \int_0^\infty\!\dd s\,
  p_-^\mu\widehat A_\mu(z_{p_-}(s))
  -
  \ii\chi_p
  \right]
  \notag\\
  &=
  \exp\!\bigg[
  \sum_\lambda\int\dd\Phi(q)
  \int_0^\infty\!\dd s\,\e^{-\varepsilon s}
  \bigg(
  -\ii e
  \left(
  p_+\cdot\epsilon^\ast(q,\lambda)\e^{\ii s p_+\cdot q}
  \widehat a^\dagger_{q,\lambda}
  +
  p_+\cdot\epsilon(q,\lambda)\e^{-\ii s p_+\cdot q}
  \widehat a_{q,\lambda}
  \right)
  \notag\\
  &\qquad+
  \ii e
  \left(
  p_-\cdot\epsilon^\ast(q,\lambda)\e^{\ii s p_-\cdot q}
  \widehat a^\dagger_{q,\lambda}
  +
  p_-\cdot\epsilon(q,\lambda)\e^{-\ii s p_-\cdot q}
  \widehat a_{q,\lambda}
  \right)
  \bigg)
  -
  \ii\chi_p
  \bigg]
  \notag\\
  &=
  \exp[-\ii\chi_p]
  \exp\!\Bigg[
  \sum_\lambda\int\dd\Phi(q)
  \left(
  S_p(q,\lambda)\widehat a^\dagger_{q,\lambda}
  -
  S_p^\ast(q,\lambda)\widehat a_{q,\lambda}
  \right)
  \Bigg].
  \label{wl:eq:Wilson-line-soft-factor-form}
\end{align}
The photon field $\widehat A_\mu$ is tacitly restricted to the modes with $0 \leq \ln (\omega_q/\lambda_{\rm IR})\leq\xi$, corresponding to unresolved photons.  In the first line, we have employed the Abelian path-ordering identity proved in Appendix~\ref{wl:app:path-ordering-commutator} to rewrite the path-ordered expression in eq.~\eqref{wl:eq:Ssoft-wilson-factorization}.  Since the photon-field commutator is a $c$-number, the
current-current commutator term, together with the phase of the mass counterterm from each pion propagator, gives the eikonal phase \(\chi_p\) of eq.~\eqref{pion:eq:virtual-phase-decomposition} (see Appendix~\ref{app:wilson-matrix-element}).  The convergence factor \(\varepsilon>0\) regularizes the
endpoint behavior of the integral and produces the standard \(\ii0\) prescriptions after
\(\varepsilon\to0^+\).  The last line is again the
soft factor
in eq.~\eqref{eq:Sp}.
Suppressing the photon-mode labels and the corresponding sum and phase-space
integral, we write
\begin{equation}
  \widehat W_p
  =
  \e^{-\ii\chi_p}
  \exp\!\left(S_p\widehat a^\dagger-S_p^\ast\widehat a\right)\, ,
  \label{if:eq:compact-wilson-line}
\end{equation}
with
\begin{align}
  S_p\widehat a^\dagger
  &\equiv
  \sum_\lambda\int\dd\Phi(q)\,
  S_p(q,\lambda)\widehat a^\dagger_{q,\lambda},&
  S_p^\ast\widehat a
  &\equiv
  \sum_\lambda\int\dd\Phi(q)\,
  S_p(q,\lambda)^\ast\widehat a_{q,\lambda},
  \label{if:eq:compact-notation}
  \\
  |S_p|^2
  &\equiv
  \sum_\lambda\int\dd\Phi(q)\,
  |S_p(q,\lambda)|^2,&
  |S_p-S_{p'}|^2
  &\equiv
  \sum_\lambda\int\dd\Phi(q)\,
  |S_p(q,\lambda)-S_{p'}(q,\lambda)|^2 .\notag
\end{align}

\noindent
It is convenient to use the Baker--Campbell--Hausdorff formula to first normal order the Wilson-line operator before computing its matrix element,
\begin{align}
  \widehat W_p
  &=
  \e^{-\ii\chi_p}
  \e^{S_p\widehat a^\dagger-S_p^\ast\widehat a}
  =
  \e^{-\ii\chi_p}
  \e^{-\frac12 |S_p|^2}
  \e^{S_p\widehat a^\dagger}
  \e^{-S_p^\ast\widehat a},
  \label{if:eq:normal-ordered-soft-evolution}
\end{align}
where we have used
\([S_p\widehat a^\dagger,-S_p^\ast\widehat a]=|S_p|^2\). The vacuum expectation
value can then be obtained straightforwardly as
\begin{align}
  \langle0|\widehat W_p|0\rangle
  &=
  \e^{-\ii\chi_p}
  \e^{-\frac12 |S_p|^2}
  \langle0|
  \e^{S_p\widehat a^\dagger}
  \e^{-S_p^\ast\widehat a}
  |0\rangle
  =
  \e^{-\ii\chi_p}
  \e^{-\frac12 |S_p|^2},
  \label{if:eq:vacuum-soft-evolution-factor}
\end{align}
using \(\e^{-S_p^\ast\widehat a}|0\rangle=|0\rangle\) and
\(\langle0|\e^{S_p\widehat a^\dagger}=\langle0|\).
The vacuum matrix element~\eqref{if:eq:vacuum-soft-evolution-factor} agrees with the
virtual correction \(\calM_{\rm vir}(p)=\exp[V_p(\xi)]\) of
Section~\ref{pion:sec:reduced-density-soft-exponentiation} as expected (see
Appendix~\ref{app:wilson-matrix-element}).
Expanding the Wilson-line operator makes this correspondence explicit at the diagrammatic level,
\begin{align}
  \e^{-\frac12 |S_p|^2}
  &=
  \e^{\ii\chi_p}\langle0|\widehat W_p|0\rangle
  =
  \langle0|
  \e^{S_p\widehat a^\dagger-S_p^\ast\widehat a}
  |0\rangle
  \notag\\
  &=
  1
  -
  \frac{|S_p|^2}{2!}\langle0|\widehat a\widehat a^\dagger|0\rangle
  +
  \frac{|S_p|^4}{4!}
  \langle0|
  \widehat a\widehat a\widehat a^\dagger\widehat a^\dagger
  +
  \widehat a\widehat a^\dagger\widehat a\widehat a^\dagger
  |0\rangle
  +\cdots \, ,
  \label{if:eq:virtual-series-from-soft-evolution}
\end{align}
where the creation and annihilation operators for the soft photons coupling to the hard particles must always be paired up to yield a nonzero matrix element.
For the two pion lines, \(S_p=S_++S_-\) with $S_\pm(q,\lambda)\equiv\epsilon^\ast(q,\lambda)\cdot J_\pm(q)$, so
\(|S_p|^2=|S_+|^2+|S_-|^2+S_+S_-^\ast+S_-S_+^\ast\). The first two terms $|S_\pm|^2$ are
the loop corrections attached to the individual \(\pi^+\) or \(\pi^-\) propagators,
corresponding to panels (a) and (b) of
Fig.~\ref{pion:fig:virtual-eikonal-graphs}, while the cross terms attach the
soft photon to different pion propagators, as in the exchange graph in panel
(c).

After identifying the factor $\e^{-\frac12 |S_p|^2}$ as the virtual corrections using the vacuum-to-vacuum matrix element, the
real soft-emission amplitudes are obtained by projecting the same
Wilson-line operator onto the $n$-soft-photon final state
\(\langle X_n|=\langle0|\widehat a_{q_n,\lambda_n}\cdots\widehat a_{q_1,\lambda_1}\) with kinematics $X_n=\{(q_a,\lambda_a)\}_{a=1}^n$. Then, from eq.~\eqref{if:eq:normal-ordered-soft-evolution}, we find,
\begin{align}
  \langle X_n|\widehat W_p|0\rangle
  &=
  \e^{-\ii\chi_p}
  \e^{-\frac12 |S_p|^2}
  \langle X_n|
  \e^{S_p\widehat a^\dagger}
  \e^{-S_p^\ast\widehat a}
  |0\rangle
  \notag\\
  &=
  \e^{-\ii\chi_p}
  \e^{-\frac12 |S_p|^2}
  \langle X_n|
  \e^{S_p\widehat a^\dagger}
  |0\rangle
  \notag\\
  &=
  \e^{-\ii\chi_p}
  \e^{-\frac12 |S_p|^2}
  \frac{1}{n!}
  \langle0|
  \widehat a_{q_n,\lambda_n}\cdots\widehat a_{q_1,\lambda_1}
  (S_p\widehat a^\dagger)^n
  |0\rangle
  \notag\\
  &=
  \e^{-\ii\chi_p}
  \e^{-\frac12 |S_p|^2}
  \prod_{a=1}^{n}S_p(q_a,\lambda_a)\, .
  \label{if:eq:n-real-photon-soft-evolution}
\end{align}
Thus the same Wilson-line operator encodes both the
\(n\)-photon LP emission soft factor and the exponential
\(\e^{-\ii\chi_p}\e^{-|S_p|^2/2}\) appearing in the vacuum-to-vacuum matrix
element~\eqref{if:eq:vacuum-soft-evolution-factor}.

The annihilation factor in \eqref{if:eq:normal-ordered-soft-evolution} does not
act on the vacuum, but it becomes visible once the initial photon state is not
\(|0\rangle\), as occurs in a thermal soft-photon bath. These preexisting environmental soft photons interact with the charged pions in the same way as those from self-emission at zero temperature.
For a single occupied photon mode,
\begin{align}
  \langle0|\widehat W_p|1\rangle
  &=
  \e^{-\ii\chi_p}
  \e^{-\frac12 |S_p|^2}
  \langle0|
  \e^{S_p\widehat a^\dagger}
  \e^{-S_p^\ast\widehat a}
  |1\rangle
  \notag\\
  &=
  \e^{-\ii\chi_p}
  \e^{-\frac12 |S_p|^2}
  \langle0|
  \e^{S_p\widehat a^\dagger}
  (|1\rangle-S_p^\ast|0\rangle)
  \notag\\
  &=
  -S_p^\ast\,\e^{-\ii\chi_p}\e^{-\frac12 |S_p|^2},
  \label{if:eq:one-photon-absorption}
\end{align}
where we have used \(\widehat a|1\rangle=|0\rangle\) corresponding to the absorption of one soft
photon by the eikonal current.
The stimulated emission also becomes visible. As a simple example, consider the $1\to2$ transition with normalization \(\widehat a^\dagger|1\rangle=\sqrt2\,|2\rangle\),
\begin{align}
  \langle2|\widehat W_p|1\rangle
  &=
  \e^{-\ii\chi_p}
  \e^{-\frac12 |S_p|^2}
  \langle2|
  \e^{S_p\widehat a^\dagger}
  (|1\rangle-S_p^\ast|0\rangle)
  \notag\\
  &
  =
  \e^{-\ii\chi_p}
  \e^{-\frac12 |S_p|^2}
  \left[
    \sqrt2\,S_p-\frac{1}{\sqrt2}S_p^\ast S_p^2
  \right]
  \notag\\
  &=
  \frac{S_p}{\sqrt2}
  \left(2-|S_p|^2\right)
  \e^{-\ii\chi_p}\e^{-\frac12 |S_p|^2}.
  \label{if:eq:stimulated-emission-single-mode}
\end{align}
These calculations show the physical features of the thermal
bath: relative to the vacuum, occupied photon modes introduce absorption
and stimulated-emission factors into the same Wilson-line evolution.

\section{Kraus representation approach}
\label{sec:kraus-representation}

In this section, we rederive the Lindblad equation in eq.~\eqref{pion:eq:lindblad-flow} by writing the reduced density matrix in the Kraus representation via the \(S\)-matrix factorization and Wilson-line identities found in Section~\ref{sec:soft-smatrix-wilson-line}, and show that the jump operator and effective Hamiltonian identified in eq.~\eqref{pion:eq:L1-and-H-identification} originate from Wilson-line matrix elements.

\subsection{Zero-temperature Kraus representation}
\label{wl:sec:wilson-line-kraus-representation}

We decompose the initial density matrix $\widehat\rho_i$ of the total system, before the hard scattering, as
\begin{equation}
  \widehat\rho_i
  =
  \widehat\rho_{\rm sys}\otimes\widehat\rho_{\rm env}\, ,
  \label{wl:eq:initial-density-matrix}
\end{equation}
where \(\widehat\rho_{\rm sys}\) describes the matter sector, while
\(\widehat\rho_{\rm env}\) describes the photon sector.
Applying \eqref{wl:eq:general-S-factorization} to
\eqref{wl:eq:initial-density-matrix} gives the final-state density matrix
\begin{align}
  \widehat\rho_f
  &=
  \widehat S\widehat\rho_i\widehat S^\dagger
  \notag\\
  &\simeq
  \left(
    \sum_p |p\rangle\langle p|\otimes\widehat W_p
  \right)
  (\widehat S_0\otimes\widehat I)
  (\widehat\rho_{\rm sys}\otimes\widehat\rho_{\rm env})
  (\widehat S_0^\dagger\otimes\widehat I)
  \left(
    \sum_{p'} |p'\rangle\langle p'|\otimes\widehat W_{p'}^\dagger
  \right)
  \notag\\
  &=
  \left(
    \sum_p |p\rangle\langle p|\otimes\widehat W_p
  \right)
  (\widehat S_0\widehat\rho_{\rm sys}\widehat S_0^\dagger\otimes\widehat\rho_{\rm env})
  \left(
    \sum_{p'} |p'\rangle\langle p'|\otimes\widehat W_{p'}^\dagger
  \right)
  \notag\\
  &=
  \sum_{p,p'}
  \langle p|\widehat S_0\widehat\rho_{\rm sys}\widehat S_0^\dagger|p'\rangle
  |p\rangle\langle p'|
  \otimes
  \widehat W_p\widehat\rho_{\rm env}\widehat W_{p'}^\dagger\, .
  \label{wl:eq:total-state}
\end{align}
To obtain the zero-temperature result, we can simply set
\(\widehat\rho_{\rm env}=|0\rangle\langle0|\).
The reduced density matrix of the hard particles is obtained by tracing over
photon modes in the unresolved window
\(0<\ln(\omega_q/\lambda_{\IR})<\xi\).  We then find,
\begin{align}
  \widehat\rho(\xi)
  &=
  \Tr_{X,\xi}\widehat\rho_f
  \notag\\
  &=
  \sum_X
  \langle X|\widehat\rho_f|X\rangle_{(\xi,0)}
  \notag\\
  &=
  \sum_X\sum_{p,p'}
  \langle p|\widehat S_0\widehat\rho_{\rm sys}\widehat S_0^\dagger|p'\rangle
  |p\rangle\langle p'|
  \langle X|\widehat W_p(\xi,0)|0\rangle
  \langle0|\widehat W_{p'}^\dagger(\xi,0)|X\rangle
  \notag\\
  &=
  \sum_X\sum_{p,p'}
  |p\rangle
  \langle X|\widehat W_p(\xi,0)|0\rangle
  \langle p|\widehat S_0\widehat\rho_{\rm sys}\widehat S_0^\dagger|p'\rangle
  \langle0|\widehat W_{p'}^\dagger(\xi,0)|X\rangle
  \langle p'|
  \notag\\
  &=
  \sum_X
  \left(
    \sum_p
    |p\rangle
    \langle X|\widehat W_p(\xi,0)|0\rangle
    \langle p|
  \right)
  \widehat S_0\widehat\rho_{\rm sys}\widehat S_0^\dagger
  \left(
    \sum_{p'}
    |p'\rangle
    \langle0|\widehat W_{p'}^\dagger(\xi,0)|X\rangle
    \langle p'|
  \right)
  \notag\\
  &\equiv
  \sum_X
  \widehat K_X(\xi,0)\widehat\rho(0)\widehat K_X^\dagger(\xi,0)\, .
  \label{wl:eq:finite-kraus}
\end{align}
The subscript \((\xi,0)\) on
\(\langle X|\widehat\rho_f|X\rangle_{(\xi,0)}\) indicates that the photon trace is
restricted to \(0<\ln(\omega_q/\lambda_{\IR})<\xi\). This is achieved by confining the photon modes in the Wilson-line operator, which we denote as
\(\widehat W_p(\xi,0)\).
Here \(\widehat\rho(0)\equiv
\widehat S_0\widehat\rho_{\rm sys}\widehat S_0^\dagger\) is the hard density matrix after the short-distance interaction, while the Kraus operator $\widehat K_X(\xi,0)$, which implements the trace over unresolved photons, reads
\begin{equation}
  \widehat K_X(\xi,0)
  \equiv
  \sum_p
  |p\rangle
  \langle X|\widehat W_p(\xi,0)|0\rangle
  \langle p|\, .
  \label{wl:eq:finite-K}
\end{equation}
The explicit expression for the Wilson line composed of unresolved photons can be easily motivated from eq.~\eqref{wl:eq:Ssoft-wilson-factorization} as
\begin{equation}
  \widehat W_p(\xi,0)
  \equiv
  \mathcal P
  \exp\!\left[
  -\ii e\int_{\mathcal C_p}\!\dd z^\mu\,
  \widehat A_\mu^{(\xi,0)}(z)
  \right],
  \label{wl:eq:unresolved-wilson-line}
\end{equation}
where the path will be fixed to the straight hard particle trajectories. The photon field restricted to the unresolved window reads,
\begin{equation}
  \widehat A_\mu^{(\xi,0)}(x)
  \equiv
  \sum_\lambda\int\dd\Phi_\xi(q)
  \left[
  \epsilon_\mu(q,\lambda)\widehat a_{q,\lambda}
  \e^{-\ii q\cdot x}
  +
  \epsilon_\mu^\ast(q,\lambda)\widehat a_{q,\lambda}^\dagger
  \e^{+\ii q\cdot x}
  \right].
  \label{wl:eq:unresolved-photon-field}
\end{equation}
Appendix~\ref{pion:app:gauge-invariance} shows that this unresolved Wilson line, and with it the reduced density matrix, is invariant under both ordinary and large gauge transformations.

\subsection{Relation between the Lindblad jump operator and the Wilson line}
\label{wl:sec:from-wilson-line-to-lindblad-equation}

To make the connection between the Wilson line and the Lindblad jump operator, we consider
again the infinitesimal change in \(\widehat\rho(\xi)\) as
\(\xi\) is increased by \(\delta\xi\) shown in Fig.~\ref{pion:fig:resolution-flow}.  This means that we trace out an additional shell of
photons with \(\xi<\ln(\omega_q/\lambda_{\IR})<\xi+\delta\xi\), and ask how
\(\widehat\rho(\xi)\) changes accordingly.  Following the Kraus representation, we have
\begin{equation}
  \widehat\rho(\xi+\delta\xi)
  =
  \sum_X
  \widehat K_X(\xi+\delta\xi,\xi)\widehat\rho(\xi)\widehat K_X^\dagger(\xi+\delta\xi,\xi)\, .
  \label{wl:eq:shell-map-full}
\end{equation}
To proceed, we expand the right-hand side and keep only terms up to
\(O(\delta\xi)\).  Each photon traced over in the infinitesimal shell of eq.~\eqref{wl:eq:unresolved-photon-field} supplies one factor
of the shell measure \(\int_\xi^{\xi+\delta\xi}\dd\Phi=O(\delta\xi)\) because no $\delta$-type function appears in the integrand, so the
integrated \(n\)-photon contribution to the reduced density matrix is
\(O(\delta\xi^n)\).  Thus only the terms built from the vacuum and one-photon
Wilson-line matrix elements contribute through \(O(\delta\xi)\),
\begin{align}
  \widehat\rho(\xi+\delta\xi)
  &=
  \widehat K_0(\xi+\delta\xi,\xi)\widehat\rho(\xi)\widehat K_0^\dagger(\xi+\delta\xi,\xi)
  \notag\\
  &+
  \int_\xi^{\xi+\delta\xi}\!\dd\Phi\,
  \widehat K_1(\xi+\delta\xi,\xi)\widehat\rho(\xi)\widehat K_1^\dagger(\xi+\delta\xi,\xi)
  +
  O(\delta\xi^2)\, .
  \label{wl:eq:shell-map-one-photon}
\end{align}
Here $\widehat K_0$ is obtained by projecting the shell photons onto the vacuum,
\begin{align}
  \widehat K_0(\xi+\delta\xi,\xi)
  &=
  \sum_p
  |p\rangle
  \langle0|\widehat W_p(\xi+\delta\xi,\xi)|0\rangle
  \langle p|
\equiv
  \widehat I+\delta\xi \hspace*{-1mm}\widehat{\,\,\mathbb L}_0(\xi)+O(\delta\xi^2)\, ,
  \label{wl:eq:K0-expansion}
\end{align}
where the $O(\delta\xi)$ term extracted from the zero-photon Kraus operator reads,
\begin{equation}
  \hspace*{-1mm}\widehat{\,\,\mathbb L}_0(\xi)
  =
  \sum_p
  |p\rangle
  \left.
  \frac{\partial}{\partial\delta\xi}
  \langle0|\widehat W_p(\xi+\delta\xi,\xi)|0\rangle
  \right|_{\delta\xi=0}
  \langle p|\, .
  \label{wl:eq:L0-definition}
\end{equation}
The explicit expression for $\displaystyle\frac{\partial}{\partial\delta\xi}
  \langle0|\widehat W_p(\xi+\delta\xi,\xi)|0\rangle
  \big|_{\delta\xi=0}$ in terms of $S_p$ and $\chi_p$
will be presented later in eq.~\eqref{wl:eq:vacuum-matrix-derivative}.  For now, it is sufficient to simply regard it as a regular operator depending on the resolution parameter $\xi$.
Similarly, $\widehat K_1$ corresponds to the one-photon-emission part of the
evolution,
\begin{align}
  \widehat K_1(\xi+\delta\xi,\xi)
  &=
  \sum_p
  |p\rangle
  \langle1|\widehat W_p(\xi+\delta\xi,\xi)|0\rangle
  \langle p|
\equiv
  \hspace*{-1mm}\widehat{\,\,\mathbb L}_1(\xi)+O(\delta\xi)\, ,
  \label{wl:eq:K1-expansion}
\end{align}
where we have defined the one-photon-emission operator at
\(O(\delta\xi^0)\) order,
\begin{equation}
  \hspace*{-1mm}\widehat{\,\,\mathbb L}_1(\xi)
  =
  \sum_p
  |p\rangle
  \left.
  \langle1|\widehat W_p(\xi+\delta\xi,\xi)|0\rangle
  \right|_{\delta\xi^0}
  \langle p|.
  \label{wl:eq:L1-definition}
\end{equation}
Here $\big|_{\delta\xi^0}$ extracts the \(O(1)\) contribution
from the Wilson-line matrix element.  Since the second line in eq.~\eqref{wl:eq:shell-map-one-photon} already carries
one power of \(\delta\xi\) from the shell measure, we only keep this \(O(1)\)
part, namely $\hspace*{-1mm}\widehat{\,\,\mathbb L}_1(\xi)$, which is, in fact, identical to eq.~\eqref{pion:eq:L1-and-H-identification} as we will verify later in this subsection.
Using \(\int_\xi^{\xi+\delta\xi}\dd\Phi=\delta\xi\int\!\dd\nu_\xi+O(\delta\xi^2)\)
with photon polarization sum implied,
and the fact that both $\widehat K_0(\xi+\delta\xi,\xi)$ and $\widehat K_1(\xi+\delta\xi,\xi)$ admit regular expansion at $\delta\xi=0$ as written in eqs.~\eqref{wl:eq:K0-expansion} and~\eqref{wl:eq:K1-expansion},
the evolution of \(\widehat\rho(\xi)\) at \(O(\delta\xi)\) therefore becomes,
\begin{align}
  \widehat\rho(\xi+\delta\xi)
  &=
  \left(\widehat I+\delta\xi\hspace*{-1mm}\widehat{\,\,\mathbb L}_0(\xi)\right)
  \widehat\rho(\xi)
  \left(\widehat I+\delta\xi\hspace*{-1mm}\widehat{\,\,\mathbb L}_0^\dagger(\xi)\right)
  +
  \delta\xi
  \int\!\dd\nu_\xi\,
  \hspace*{-1mm}\widehat{\,\,\mathbb L}_1(\xi)\widehat\rho(\xi)\hspace*{-1mm}\widehat{\,\,\mathbb L}_1^\dagger(\xi)
  +
  O(\delta\xi^2)
  \notag\\
  &=
  \widehat\rho(\xi)
  +
  \delta\xi
  \left[
    \hspace*{-1mm}\widehat{\,\,\mathbb L}_0(\xi)\widehat\rho(\xi)
    +
    \widehat\rho(\xi)\hspace*{-1mm}\widehat{\,\,\mathbb L}_0^\dagger(\xi)
    +
    \int\!\dd\nu_\xi\,
    \hspace*{-1mm}\widehat{\,\,\mathbb L}_1(\xi)\widehat\rho(\xi)\hspace*{-1mm}\widehat{\,\,\mathbb L}_1^\dagger(\xi)
  \right]
  +
  O(\delta\xi^2)\, ,
  \label{wl:eq:shell-expanded}
\end{align}
which implies,
\begin{align}
  \frac{\widehat\rho(\xi+\delta\xi)-\widehat\rho(\xi)}{\delta\xi}
  &=
  \hspace*{-1mm}\widehat{\,\,\mathbb L}_0(\xi)\widehat\rho(\xi)
  +
  \widehat\rho(\xi)\hspace*{-1mm}\widehat{\,\,\mathbb L}_0^\dagger(\xi)
  +
  \int\!\dd\nu_\xi\,
  \hspace*{-1mm}\widehat{\,\,\mathbb L}_1(\xi)\widehat\rho(\xi)\hspace*{-1mm}\widehat{\,\,\mathbb L}_1^\dagger(\xi)
  +
  O(\delta\xi)\, .
  \label{wl:eq:pre-lindblad}
\end{align}
To proceed further, we exploit
the completeness relation of the Kraus operator, which is a direct consequence of the $S$-matrix unitarity and can also be directly verified from its definition~\eqref{wl:eq:finite-K}, allowing us to obtain,
\begin{align}
  \widehat I
  &=
  \sum_X
  \widehat K_X^\dagger(\xi+\delta\xi,\xi)\widehat K_X(\xi+\delta\xi,\xi)
  \notag\\
  &=
  \widehat K_0^\dagger(\xi+\delta\xi,\xi)\widehat K_0(\xi+\delta\xi,\xi)
  +
  \int_\xi^{\xi+\delta\xi}\!\dd\Phi\,
  \widehat K_1^\dagger(\xi+\delta\xi,\xi)\widehat K_1(\xi+\delta\xi,\xi)
  +
  O(\delta\xi^2)
  \notag\\
  &=
  \left[\widehat I+\delta\xi\hspace*{-1mm}\widehat{\,\,\mathbb L}_0^\dagger(\xi)\right]
  \left[\widehat I+\delta\xi\hspace*{-1mm}\widehat{\,\,\mathbb L}_0(\xi)\right]
  +
  \delta\xi
  \int\!\dd\nu_\xi\,
  \hspace*{-1mm}\widehat{\,\,\mathbb L}_1^\dagger(\xi)\hspace*{-1mm}\widehat{\,\,\mathbb L}_1(\xi)
  +
  O(\delta\xi^2)
  \notag\\
  &=
  \widehat I
  +
  \delta\xi
  \left(
    \hspace*{-1mm}\widehat{\,\,\mathbb L}_0(\xi)
    +
    \hspace*{-1mm}\widehat{\,\,\mathbb L}_0^\dagger(\xi)
    +
    \int\!\dd\nu_\xi\,
    \hspace*{-1mm}\widehat{\,\,\mathbb L}_1^\dagger(\xi)\hspace*{-1mm}\widehat{\,\,\mathbb L}_1(\xi)
  \right)
  +
  O(\delta\xi^2)\, ,
  \label{wl:eq:trace-expansion}
\end{align}
Since the above equation holds for arbitrary $\delta\xi$, the coefficient of \(\delta\xi\) must
vanish, thus,
\begin{equation}
  \hspace*{-1mm}\widehat{\,\,\mathbb L}_0(\xi)+\hspace*{-1mm}\widehat{\,\,\mathbb L}_0^\dagger(\xi)
  =
  -
  \int\!\dd\nu_\xi\,
  \hspace*{-1mm}\widehat{\,\,\mathbb L}_1^\dagger(\xi)\hspace*{-1mm}\widehat{\,\,\mathbb L}_1(\xi)\, .
  \label{wl:eq:L0-real-part}
\end{equation}
Physically, eq.~\eqref{wl:eq:L0-real-part} reflects the cancellation of IR divergences between real and virtual corrections, but it determines only the Hermitian part of $\hspace*{-1mm}\widehat{\,\,\mathbb L}_0(\xi)$.  Its
anti-Hermitian part generates the closed-system evolution associated with the
phase of the virtual correction as can be easily seen from its definition in eq.~\eqref{wl:eq:L0-definition} and discussions in Section~\ref{wl:sec:wilson-line-physical-content}. We therefore define, in correspondence with eq.~\eqref{pion:eq:H-matrix-element}, the Hermitian operator,
\begin{equation}
  \widehat {\mathbb H}_{\rm eff}(\xi)
  \equiv
  \frac{\ii}{2}
  \left(
    \hspace*{-1mm}\widehat{\,\,\mathbb L}_0(\xi)-\hspace*{-1mm}\widehat{\,\,\mathbb L}_0^\dagger(\xi)
  \right)\, .
  \label{wl:eq:H-definition}
\end{equation}
Hence $\hspace*{-1mm}\widehat{\,\,\mathbb L}_0(\xi)$ can be written as
\begin{align}
  \hspace*{-1mm}\widehat{\,\,\mathbb L}_0(\xi)
  &=
  \frac12
  \left(
    \hspace*{-1mm}\widehat{\,\,\mathbb L}_0(\xi)+\hspace*{-1mm}\widehat{\,\,\mathbb L}_0^\dagger(\xi)
  \right)
  +
  \frac12
  \left(
    \hspace*{-1mm}\widehat{\,\,\mathbb L}_0(\xi)-\hspace*{-1mm}\widehat{\,\,\mathbb L}_0^\dagger(\xi)
  \right)
  \notag\\
  &=
  -
  \frac12
  \int\!\dd\nu_\xi\,
  \hspace*{-1mm}\widehat{\,\,\mathbb L}_1^\dagger(\xi)\hspace*{-1mm}\widehat{\,\,\mathbb L}_1(\xi)
  -
  \ii\widehat {\mathbb H}_{\rm eff}(\xi).
  \label{wl:eq:L0-decomposition}
\end{align}
Substituting \eqref{wl:eq:L0-decomposition} into \eqref{wl:eq:pre-lindblad} and
taking the limit \(\delta\xi\to0\) leads to the Lindblad equation
\begin{align}
  \frac{\dd\widehat\rho}{\dd\xi}
  &=
  \hspace*{-1mm}\widehat{\,\,\mathbb L}_0(\xi)\widehat\rho(\xi)
  +
  \widehat\rho(\xi)\hspace*{-1mm}\widehat{\,\,\mathbb L}_0^\dagger(\xi)
  +
  \int\!\dd\nu_\xi\,
  \hspace*{-1mm}\widehat{\,\,\mathbb L}_1(\xi)\widehat\rho(\xi)\hspace*{-1mm}\widehat{\,\,\mathbb L}_1^\dagger(\xi)
  \notag\\
  &=
  -\ii[\widehat {\mathbb H}_{\rm eff}(\xi),\widehat\rho(\xi)]
  +
  \int\!\dd\nu_\xi\,
  \left[
    \hspace*{-1mm}\widehat{\,\,\mathbb L}_1(\xi)\widehat\rho(\xi)\hspace*{-1mm}\widehat{\,\,\mathbb L}_1^\dagger(\xi)
    -
    \frac12
    \left\{
     \hspace*{-1mm}\widehat{\,\,\mathbb L}_1^\dagger(\xi)\hspace*{-1mm}\widehat{\,\,\mathbb L}_1(\xi),
      \widehat\rho(\xi)
    \right\}
  \right].
  \label{wl:eq:lindblad}
\end{align}

To identify $\hspace*{-1mm}\widehat{\,\,\mathbb L}_1$ as the Lindblad jump operator and fully reveal the relation of $\hspace*{-1mm}\widehat{\,\,\mathbb L}_0$ and $\hspace*{-1mm}\widehat{\,\,\mathbb L}_1$ to the soft factor~\eqref{eq:Sp} and the imaginary part of the virtual correction, i.e., $\chi_p$ in eq.~\eqref{pion:eq:virtual-phase-decomposition}, let us compute
the Wilson-line matrix element in eqs.~\eqref{wl:eq:L0-definition} and~\eqref{wl:eq:L1-definition} that define them.
The Wilson line restricted to the finite unresolved interval was already
evaluated in \eqref{wl:eq:Wilson-line-soft-factor-form}.  Restricting the same
Wilson line to the infinitesimal interval  $(\xi+\delta\xi,\xi)$ replaces \(\Theta_\xi(q)\) by the corresponding shell window and changes the
phase \(\chi_p(\xi)\) to \(\chi_p(\xi+\delta\xi,\xi)\), yielding
\begin{equation}
  \widehat W_p(\xi+\delta\xi,\xi)
  =
  \exp\!\left[
    \int_\xi^{\xi+\delta\xi}\!\dd\Phi\,
    \Big(
      S_p(q,\lambda)\widehat a^\dagger_{q,\lambda}
      -
      S_p^\ast(q,\lambda)\widehat a_{q,\lambda}
    \Big)
    -
    \ii\chi_p(\xi+\delta\xi,\xi)
  \right]\, ,
  \label{wl:eq:shell-wilson-line-soft-factor-form}
\end{equation}
where the notation $\displaystyle\int_{\xi}^{\xi+\delta\xi}\dd\Phi(q)$ is understood to constrain the photon energy $\omega_q$ in the interval $\e^{\xi}\leq\omega_q/\lambda_{\rm IR}\leq \e^{\xi+\delta\xi}$ as indicated in eq.~\eqref{pion:eq:unresolved-one-photon-measure}.
Since $[\widehat a^\dagger_{q,\lambda},\widehat a_{q',\lambda'}]$ is a $c$-number, eq.~\eqref{wl:eq:shell-wilson-line-soft-factor-form} can be normal-ordered exactly following eq.~\eqref{wl:eq:bch-cnumber-truncation},
\begin{align}
  \widehat W_p(\xi+\delta\xi,\xi)
  &=
  \exp\!\left[
    -\frac12
    \int_\xi^{\xi+\delta\xi}\!\dd\Phi\,
    |S_p(q,\lambda)|^2
    -
    \ii\chi_p(\xi+\delta\xi,\xi)
  \right]
  \notag\\
  &\quad\times
  \exp\!\left[
    \int_\xi^{\xi+\delta\xi}\!\dd\Phi\,
    S_p(q,\lambda)\widehat a^\dagger_{q,\lambda}
  \right]
  \exp\!\left[
    -
    \int_\xi^{\xi+\delta\xi}\!\dd\Phi\,
    S_p^\ast(q,\lambda)\widehat a_{q,\lambda}
  \right]
  \label{wl:eq:normal-ordered-W}
\end{align}
with the first line induced by the commutator.
Then the Wilson-line matrix element in $\hspace*{-1mm}\widehat{\,\,\mathbb L}_0$ can be worked out as,
\begin{align}
  \frac{\partial}{\partial\delta\xi}&
  \langle0|\widehat W_p (\xi+\delta\xi,\xi )|0\rangle
  \Big|_{\delta\xi=0}\notag\\
 &=
  \left.
  \frac{\partial}{\partial\delta\xi}
  \left[
    1
    -
    \frac12
    \int_\xi^{\xi+\delta\xi}\!\dd\Phi\,
    |S_p(q,\lambda)|^2
    -
    \ii\chi_p(\xi+\delta\xi,\xi)
    +
    O(\delta\xi^2)
  \right]
  \right|_{\delta\xi=0}
  \notag\\
 &=
  -
  \frac12
  \int\!\dd\nu_\xi\,
  |S_p(q,\lambda)|^2
  -
  \ii\dot\chi_p(\xi).
  \label{wl:eq:vacuum-matrix-derivative}
\end{align}
Substituting this result into the definition of $\hspace*{-1mm}\widehat{\,\,\mathbb L}_0(\xi)$ in
\eqref{wl:eq:L0-definition} then allows us to obtain the explicit expression for the effective Hamiltonian $\widehat{\mathbb H}_{\rm eff}$, which governs the phase of the reduced density matrix, from its definition in eq.~\eqref{wl:eq:H-definition},
\begin{equation}
  \widehat {\mathbb H}_{\rm eff}(\xi)
  =
  \sum_p
  |p\rangle
  \dot\chi_p(\xi)
  \langle p|\, ,
  \qquad \implies \qquad
   \widehat{\mathbb H}_{\rm eff} (\xi) = \widehat H_{\rm eff}(\xi)\, ,
   \label{wl:eq:no-emission-Hamiltonian}
\end{equation}
 with $\widehat H_{\rm eff}$ defined in eq.~\eqref{pion:eq:L1-and-H-identification}.

Similarly, the one-photon-emission operator $\hspace*{-1mm}\widehat{\,\,\mathbb L}_1(\xi)$ is obtained by taking the matrix element of the same Wilson-line operator between the initial vacuum and the one-photon final state.
Following the definition in eq.~\eqref{wl:eq:K1-expansion} and letting the unresolved photon state $\langle 1|$ carry momentum $q$ with $\omega_q/\lambda_{\rm IR} = \e^{\xi}$ as enforced by the integral measure in eq.~\eqref{wl:eq:lindblad}, we find,
\begin{align}
  &\langle 1|\widehat W_p(\xi+\delta\xi,\xi)|0\rangle
  \Big|_{\delta\xi^0}\notag\\
  &=
  \exp\!\left[
    -\frac12
    \int_\xi^{\xi+\delta\xi}\!\dd\Phi(k)\,
    |S_p(k,\lambda')|^2
    -\ii\chi_p(\xi+\delta\xi,\xi)
  \right]
  \langle1|
  \exp\!\left[
    \int_\xi^{\xi+\delta\xi}\!\dd\Phi(k)\,
    S_p(k,\lambda')\widehat a^\dagger_{k,\lambda'}
  \right]
  |0\rangle
  \left.
  \vphantom{
    \int_\xi^{\xi+\delta\xi}\!\dd\Phi\,
    S_p(q,\lambda)\widehat a^\dagger_{q,\lambda}
  }
  \right|_{\delta\xi^0}
  \notag\\
  &=
  \left.
  \left(1+O(\delta\xi)\right)
  \langle1|
  \left[
    1
    +
    \int\dd\Phi(k)\,
    S_p(k,\lambda')\widehat a^\dagger_{k,\lambda'}\Theta\left(\e^{\xi}\lambda_{\rm IR}\leq\omega_k\leq\e^{\xi+\delta\xi}\lambda_{\rm IR}\right)
    +
    O(\delta\xi)
  \right]
  |0\rangle
  \right|_{\delta\xi^0}
  \notag\\
  &=
  \left.
  \left[
    \langle1|0\rangle
    +
    \langle1|
    \int \dd\Phi(k)\,
    S_p(k,\lambda')\widehat a^\dagger_{k,\lambda'}\Theta\left(\e^{\xi}\lambda_{\rm IR}\leq\omega_k\leq\e^{\xi+\delta\xi}\lambda_{\rm IR}\right)
    |0\rangle
    +
    O(\delta\xi)
  \right]
  \right|_{\delta\xi^0}
  \notag\\
  &=
  \sum_{\lambda'}
  \int
  \frac{\dd^3k}{(2\pi)^3\,2\omega_k}\,
  S_p(k,\lambda')
  (2\pi)^3\,2\omega_q\,
  \delta^{(3)}(\mathbf q-\mathbf k)\delta_{\lambda\lambda'}\Theta\left(\e^{\xi}\lambda_{\rm IR}\leq\omega_k\leq\e^{\xi+\delta\xi}\lambda_{\rm IR}\right)
  \notag\\
  &=
  S_p(q,\lambda)\, ,
  \label{wl:eq:one-photon-matrix-leading}
\end{align}
where in the last line, we have used the fact that the integral remains a constant as $\delta\xi\to 0$ since $\omega_q/\lambda_{\rm IR} =\e^\xi$.
With this matrix element determined, the one-photon-emission operator becomes
\begin{align}
  \hspace*{-1mm}\widehat{\,\,\mathbb L}_1(\xi)
  &=
  \sum_p
  |p\rangle
  \left.
  \langle1|\widehat W_p(\xi+\delta\xi,\xi)|0\rangle
  \right|_{\delta\xi^0}
  \langle p|
 =
  \sum_p
  |p\rangle
  S_p(q,\lambda)
  \langle p|\, ,\notag\\
  \implies& \quad
   \hspace*{-1mm}\widehat{\,\,\mathbb L}_1(\xi) = \widehat L_1(\xi)\, ,
  \label{wl:eq:L1-soft-factor-form}
\end{align}
where $\widehat L_1(\xi)$ is defined in eq.~\eqref{pion:eq:L1-and-H-identification}.
Combining with eq.~\eqref{wl:eq:no-emission-Hamiltonian} then leads to the conclusion that both the amplitude-level and the Wilson-line approaches yield the same Lindblad equation describing the resolution $\xi$ dependence of the reduced density matrix $\widehat\rho$.

The physical meaning of the Lindblad equation~\eqref{wl:eq:lindblad} now becomes more transparent. The infinitesimal
resolution-parameter evolution acts as a two-outcome measurement on the hard
momentum state. To \(O(\delta\xi)\), the hard particles either emit no photon or emit one
photon.  The one-photon event, described by $\hspace*{-1mm}\widehat{\,\,\mathbb L}_1$, measures the hard
momentum through the soft-factor readout \(S_p(q,\lambda)\), so two momentum
states \(p\) and \(p'\) become distinguishable when their soft factors differ.
Less intuitively, the no-photon event also carries hard-momentum information through eq.~\eqref{wl:eq:L0-real-part}.  From
\(S\)-matrix unitarity, the probabilities for no emission and one emission must sum
to one.  Since the one-emission probability is momentum dependent, the
no-emission probability is momentum dependent as well.  In other words, real
emission and the virtual/no-emission correction are complementary
contributions to the decoherence mechanism, and together they determine the
density-matrix evolution.

This QED final-state decoherence has the same physical origin as the loss of interference in the
electron double-slit experiment, where
recording the information about exactly which slit the electron passes through
completely destroys the interference pattern, whereas
leaving the electron's path unrecorded restores the coherent
interference pattern.  In this idealized example, the path information is essentially
all-or-nothing, because the electron path is either recorded or not by the measurement device and the path information can be regarded as a ``hard destroyer'' of the electron's quantum state causing the total collapse of the wavefunction and the elimination of quantum properties.  In comparison, the unresolved photons in our current consideration effectively measure the quantum information of the hard particles in a ``softer way" where the decoherence between the quantum states of the system, as removed from the environment, varies continuously as a function of the resolution parameter $\xi$.
In other words, $\xi$ classifies each photon mode as ``system'' (resolved by the detector) or ``environment'' (unresolved and traced over), and its variation consequently allows one to detect the change in quantum coherence of the hard particles as photons close to the resolution boundary, i.e., $\ln(\omega_q/\lambda_{\rm IR})\sim\xi$, acquire new identities:  system photon~$\leftrightarrow$~environment photon.
As we have demonstrated, the Wilson-line operator serves as the generator of the Lindblad jump operator, as a consequence of the relation between Wilson line and soft photon emissions.

\subsection{Thermal Kraus representation}
\label{wl:sec:finite-temperature-thermal-state}

The previous sections used a pure photon vacuum as the initial environmental
state, which is essentially equivalent to the zero-temperature condition. We now generalize our discussions to the finite-temperature case by replacing the pure photon vacuum  with a thermal mixed state at temperature $T$:
\begin{equation}
  \widehat\rho_{\rm env}
  =
  \sum_N P_N |N\rangle\langle N|,
  \qquad
  P_N=\frac{\e^{-\omega_N/(k_BT)}}{\sum_M \e^{-\omega_M/(k_BT)}}\, .
  \label{wl:eq:thermal-photon-density-number-basis}
\end{equation}
Here $N$ denotes an $N$-photon state and \(\omega_N\) is
the energy of the corresponding photon state.
With the hard scattering already included in \(\widehat\rho(0)\), the full final
density matrix in a thermal photon bath reads,
\begin{align}
  \widehat\rho_{f,\rm th}
  &\equiv
  \left(
    \sum_p |p\rangle\langle p|\otimes\widehat W_p
  \right)
  (\widehat\rho(0)\otimes\widehat\rho_{\rm env})
  \left(
    \sum_{p'} |p'\rangle\langle p'|\otimes\widehat W_{p'}^\dagger
  \right)
  \notag\\
  &=
  \sum_N P_N
  \left(
    \sum_p |p\rangle\langle p|\otimes\widehat W_p
  \right)
  (\widehat\rho(0)\otimes |N\rangle\langle N|)
  \left(
    \sum_{p'} |p'\rangle\langle p'|\otimes\widehat W_{p'}^\dagger
  \right)\, .
  \label{wl:eq:thermal-final-density-matrix}
\end{align}
As before, the reduced hard density matrix is obtained by tracing over the unresolved
photons in the photon-number basis:
\begin{align}
  \widehat\rho_{\rm th}(\xi)
  &\equiv
  \Tr_{X,\xi}\widehat\rho_{f,\rm th}
  =
  \sum_X
  \langle X|\widehat\rho_{f,\rm th}|X\rangle_{(\xi,0)}
\equiv
  \sum_{X,N}
  \widehat K_{XN}(\xi,0)
  \widehat\rho(0)
  \widehat K_{XN}^\dagger(\xi,0)\, ,
  \label{wl:eq:thermal-kraus-map}
\end{align}
with the finite-temperature Kraus operators defined as,
\begin{equation}
  \widehat K_{XN}(\xi,0)
  =
  \sqrt{P_N}
  \sum_p
  |p\rangle
  \langle X|\widehat W_p(\xi,0)|N\rangle
  \langle p| \, .
  \label{wl:eq:thermal-kraus-operator-full}
\end{equation}
This is the direct finite-temperature analog of \eqref{wl:eq:finite-K}.  The new
feature is that the Kraus operator depends on both the final unresolved photon
mode $\langle X|$ and the initial thermal state $|N\rangle$.  The prefactor $\sqrt{P_N}$ provides the weight of the initial finite-temperature state.

We next repeat the analysis on the infinitesimal shell of the resolution parameter $(\xi+\delta\xi,\xi)$.
Let $\langle M|$ and $|N\rangle$ be respectively the final and initial
photon states that give rise to the difference $\widehat\rho_{\rm th}(\xi+\delta\xi)-\widehat\rho_{\rm th}(\xi)$.  The shell Kraus operator then reads,
\begin{equation}
  \widehat K_{MN}(\xi+\delta\xi,\xi)
  =
  \sqrt{P_N}
  \sum_p
  |p\rangle
  \langle M|\widehat W_p(\xi+\delta\xi,\xi)|N\rangle
  \langle p|\, .
  \label{wl:eq:thermal-shell-kraus-MN}
\end{equation}
Since the emission of \(n\) soft photons in the interval $(\xi+\delta\xi,\xi)$ is suppressed by
 $O(\delta\xi^n)$, we only need to consider processes in which the initial and final photon states differ by zero or one photon number, namely,
\begin{equation}
  M=N,
  \qquad
  M=N+1,
  \qquad
  M=N-1\, .
  \label{wl:eq:thermal-allowed-shell-branches}
\end{equation}
The \(M=N+1\) outcome
corresponds to a single real-photon emission,
and the \(M=N-1\) outcome implies absorption of one photon initially
present in the thermal state. Hence, we restrict the
Fock basis in \eqref{wl:eq:thermal-photon-density-number-basis} to
one shell mode $(q,\lambda)$ that is responsible for the changes in $\widehat\rho_{\rm th}(\xi)$ as the resolution parameter varies from $\xi$ to $\xi+\delta\xi$.  We therefore find,
\begin{align}
  \widehat\rho_{\rm th}(\xi+\delta\xi)
  &=
  \sum_{N=0}^{\infty}
  \widehat K_{NN}(\xi+\delta\xi,\xi)\widehat\rho_{\rm th}(\xi)
  \widehat K_{NN}^\dagger(\xi+\delta\xi,\xi)
  \notag\\
  &+
  \int_\xi^{\xi+\delta\xi}\!\dd\Phi
  \sum_{N=0}^{\infty}
  \widehat K_{N+1,N}(\xi+\delta\xi,\xi)\widehat\rho_{\rm th}(\xi)
  \widehat K_{N+1,N}^\dagger(\xi+\delta\xi,\xi)
  \notag\\
  &+
  \int_\xi^{\xi+\delta\xi}\!\dd\Phi
  \sum_{N=1}^{\infty}
  \widehat K_{N-1,N} (\xi+\delta\xi,\xi)\widehat\rho_{\rm th}(\xi)
  \widehat K_{N-1,N}^\dagger(\xi+\delta\xi,\xi)
  +O(\delta\xi^2)\, .
  \label{wl:eq:thermal-shell-map-three-branches}
\end{align}
As in Section~\ref{wl:sec:from-wilson-line-to-lindblad-equation}, let us define the finite-temperature jump operators.
The Kraus operator of equal photon number $ \widehat K_{NN}$ in the infinitesimal resolution shell reads,
\begin{align}
  \widehat K_{NN}(\xi+\delta\xi,\xi)
  &=
  \sqrt{P_N}
  \sum_p
  |p\rangle
  \langle N|\widehat W_p(\xi+\delta\xi,\xi)|N\rangle
  \langle p|
  \notag\\
  &=
  \sqrt{P_N}
  \left[
    \widehat I+\delta\xi\,\hspace*{-1mm}\widehat{\,\,\mathbb L}_{0,N}(\xi)+O(\delta\xi^2)
  \right],
  \label{wl:eq:thermal-Knn-abstract}
\end{align}
from which the definition of $\hspace*{-1mm}\widehat{\,\,\mathbb L}_{0,N}(\xi)$ is obvious,
\begin{equation}
  \hspace*{-1mm}\widehat{\,\,\mathbb L}_{0,N}(\xi)
  \equiv
  \sum_p
  |p\rangle
  \left.
  \frac{\partial}{\partial\delta\xi}
  \langle N|\widehat W_p(\xi+\delta\xi,\xi)|N\rangle
  \right|_{\delta\xi=0}
  \langle p|\, .
  \label{wl:eq:thermal-L0n-W-matrix-element}
\end{equation}
It is clear that $\hspace*{-1mm}\widehat{\,\,\mathbb L}_{0,N}(\xi)$ is the generator of $\widehat K_{NN}(\xi+\delta\xi,\xi)$.
Summing the $M=N$ contribution over the initial thermal states in the infinitesimal shell we derive the contribution from the first line of eq.~\eqref{wl:eq:thermal-shell-map-three-branches},
\begin{align}
  \sum_{N=0}^{\infty}
  \widehat K_{NN}&(\xi+\delta\xi,\xi)\widehat\rho_{\rm th}(\xi)
  \widehat K_{NN}^\dagger(\xi+\delta\xi,\xi)\notag\\
  &=
  \sum_{N=0}^{\infty}
  P_N
  \left[
    \widehat I+\delta\xi\,\hspace*{-1mm}\widehat{\,\,\mathbb L}_{0,N}(\xi)
  \right]
  \widehat\rho_{\rm th}(\xi)
  \left[
    \widehat I+\delta\xi\,\hspace*{-1mm}\widehat{\,\,\mathbb L}_{0,N}^\dagger(\xi)
  \right]
  +O(\delta\xi^2)
  \notag\\
  &=
  \left(\sum_{N=0}^{\infty}P_N\right)\widehat\rho_{\rm th}(\xi)
  +
  \delta\xi
  \sum_{N=0}^{\infty}
  P_N
  \left[
    \hspace*{-1mm}\widehat{\,\,\mathbb L}_{0,N}(\xi)\widehat\rho_{\rm th}(\xi)
    +
    \widehat\rho_{\rm th}(\xi)\hspace*{-1mm}\widehat{\,\,\mathbb L}_{0,N}^\dagger(\xi)
  \right]
  +O(\delta\xi^2)
  \notag\\
  &=
  \widehat\rho_{\rm th}(\xi)
  +
  \delta\xi
  \sum_{N=0}^{\infty}
  P_N
  \left[
    \hspace*{-1mm}\widehat{\,\,\mathbb L}_{0,N}(\xi)\widehat\rho_{\rm th}(\xi)
    +
    \widehat\rho_{\rm th}(\xi)\hspace*{-1mm}\widehat{\,\,\mathbb L}_{0,N}^\dagger(\xi)
  \right]
  +O(\delta\xi^2)
  \notag\\
  &=
  \widehat\rho_{\rm th}(\xi)
  +
  \delta\xi
  \left[
    \hspace*{-1mm}\widehat{\,\,\mathbb L}_{0,\rm th}(\xi)\widehat\rho_{\rm th}(\xi)
    +
    \widehat\rho_{\rm th}(\xi)\hspace*{-1mm}\widehat{\,\,\mathbb L}_{0,\rm th}^\dagger(\xi)
  \right]
  +O(\delta\xi^2),
  \label{wl:eq:thermal-no-change-abstract}
\end{align}
where in the last line we have defined the thermal average
\begin{equation}
  \hspace*{-1mm}\widehat{\,\,\mathbb L}_{0,\rm th}(\xi)
  \equiv
  \sum_{N=0}^{\infty}P_N\,\hspace*{-1mm}\widehat{\,\,\mathbb L}_{0,N}(\xi).
  \label{wl:eq:thermal-L0-thermal-average-definition}
\end{equation}
For the one-photon emission and absorption processes, the corresponding
operators are expanded as
\begin{align}
  \widehat K_{N\pm1,N}(\xi+\delta\xi,\xi)
  &=
  \sqrt{P_N}
  \sum_p
  |p\rangle
  \langle N\pm1|\widehat W_p(\xi+\delta\xi,\xi)|N\rangle
  \langle p|
  \notag\\
  &=
  \sqrt{P_N}
  \sum_p
  |p\rangle
  \left.
  \langle N\pm1|\widehat W_p(\xi+\delta\xi,\xi)|N\rangle
  \right|_{\delta\xi^0}
  \langle p|
  +O(\delta\xi)
  \notag\\
  &\equiv
  \hspace*{-1mm}\widehat{\,\,\mathbb L}_{\pm,N}(\xi)
  +O(\delta\xi).
  \label{wl:eq:thermal-LpmN-W-matrix-element}
\end{align}
Here the notation \(\left.\cdots\right|_{\delta\xi^0}\) has the same meaning as
in \eqref{wl:eq:L1-definition}.
As before, the phase-space measure over the shell $(\xi+\delta\xi,\xi)$ in eq.~\eqref{wl:eq:thermal-shell-map-three-branches}
already supplies an overall \(O(\delta\xi)\) of a one-quantum channel, so the
matrix element itself is kept only at \(O(1)\).  With the corresponding
ranges of $N$ understood as $N\ge0$ and $N\ge1$ respectively for the $\widehat K_{N\pm1,N}$ Kraus operators, the contributions from the two real-radiation channels read,
\begin{align}
  \int_\xi^{\xi+\delta\xi}\!\dd\Phi
  &\sum_N
  \widehat K_{N\pm1,N}(\xi+\delta\xi,\xi)\widehat\rho_{\rm th}(\xi)
  \widehat K_{N\pm1,N}^\dagger(\xi+\delta\xi,\xi)
  \notag\\
  &=
  \int_\xi^{\xi+\delta\xi}\!\dd\Phi
  \sum_N
  \left[
    \hspace*{-1mm}\widehat{\,\,\mathbb L}_{\pm,N}(\xi)+O(\delta\xi)
  \right]
  \widehat\rho_{\rm th}(\xi)
  \left[
    \hspace*{-1mm}\widehat{\,\,\mathbb L}_{\pm,N}^\dagger(\xi)+O(\delta\xi)
  \right]
  \notag\\
  &=
  \delta\xi
  \int\!\dd\nu_\xi
  \sum_N
  \hspace*{-1mm}\widehat{\,\,\mathbb L}_{\pm,N}(\xi)\widehat\rho_{\rm th}(\xi)
  \hspace*{-1mm}\widehat{\,\,\mathbb L}_{\pm,N}^\dagger(\xi)
  +O(\delta\xi^2).
  \label{wl:eq:thermal-real-branch-abstract}
\end{align}
Substituting
\eqref{wl:eq:thermal-no-change-abstract},
\eqref{wl:eq:thermal-real-branch-abstract} into
\eqref{wl:eq:thermal-shell-map-three-branches} gives the thermal analog of
\eqref{wl:eq:pre-lindblad},
\begin{align}
  \frac{\widehat\rho_{\rm th}(\xi+\delta\xi)-\widehat\rho_{\rm th}(\xi)}{\delta\xi}
  &=
  \hspace*{-1mm}\widehat{\,\,\mathbb L}_{0,\rm th}(\xi)\widehat\rho_{\rm th}(\xi)
  +
  \widehat\rho_{\rm th}(\xi)\hspace*{-1mm}\widehat{\,\,\mathbb L}_{0,\rm th}^\dagger(\xi)
  \notag\\
  &
 +
  \sum_{i=\pm}\int\!\dd\nu_\xi
  \sum_N
  \hspace*{-1mm}\widehat{\,\,\mathbb L}_{i,N}(\xi)\widehat\rho_{\rm th}(\xi)
  \hspace*{-1mm}\widehat{\,\,\mathbb L}_{i,N}^\dagger(\xi)
  +O(\delta\xi).
  \label{wl:eq:thermal-pre-lindblad-W-matrix-elements}
\end{align}
We again impose the trace-preservation condition guaranteed by the $S$-matrix unitarity,
\begin{align}
  \hspace*{-1mm}\widehat{\,\,\mathbb L}_{0,\rm th}(\xi)+\hspace*{-1mm}\widehat{\,\,\mathbb L}_{0,\rm th}^\dagger(\xi)
  &=
  -
  \int\!\dd\nu_\xi
  \sum_N
  \left[
    \hspace*{-1mm}\widehat{\,\,\mathbb L}_{+,N}^\dagger(\xi)
    \hspace*{-1mm}\widehat{\,\,\mathbb L}_{+,N}(\xi)
    +
    \hspace*{-1mm}\widehat{\,\,\mathbb L}_{-,N}^\dagger(\xi)
    \hspace*{-1mm}\widehat{\,\,\mathbb L}_{-,N}(\xi)
  \right]\, ,
  \label{wl:eq:thermal-trace-preservation-abstract}
\end{align}
and consequently define the thermal effective Hamiltonian operator,
\begin{equation}
  \widehat H_{\rm eff,th}(\xi)
  \equiv
  \frac{\ii}{2}
  \left(
    \hspace*{-1mm}\widehat{\,\,\mathbb L}_{0,\rm th}(\xi)-\hspace*{-1mm}\widehat{\,\,\mathbb L}_{0,\rm th}^\dagger(\xi)
  \right)\, .
  \label{wl:eq:thermal-H-definition}
\end{equation}
Since the anti-Hermitian part of $\hspace*{-1mm}\widehat{\,\,\mathbb L}_{0,N}(\xi)$ equals $-\ii\sum_p|p\rangle\dot\chi_p(\xi)\langle p|$ for every $N$, as can be easily confirmed using eq.~\eqref{wl:eq:normal-ordered-W}, the thermal average leaves it unchanged, and $\widehat H_{\rm eff,th}(\xi)=\widehat H_{\rm eff}(\xi)$ of eq.~\eqref{wl:eq:no-emission-Hamiltonian}. The finite-temperature Lindblad equation then takes the form,
\begin{align}
  \frac{\dd\widehat\rho_{\rm th}}{\dd\xi}
  &=
  -\ii[\widehat H_{\rm eff}(\xi),\widehat\rho_{\rm th}(\xi)]
+
  \int\!\dd\nu_\xi
  \sum_{N,i=\pm}
  \left[
    \hspace*{-1mm}\widehat{\,\,\mathbb L}_{i,N}(\xi)\widehat\rho_{\rm th}(\xi)
    \hspace*{-1mm}\widehat{\,\,\mathbb L}_{i,N}^\dagger(\xi)
    -
    \frac12
    \left\{
      \hspace*{-1mm}\widehat{\,\,\mathbb L}_{i,N}^\dagger(\xi)\hspace*{-1mm}\widehat{\,\,\mathbb L}_{i,N}(\xi),
      \widehat\rho_{\rm th}(\xi)
    \right\}
  \right]\, .
  \label{wl:eq:thermal-lindblad-W-matrix-elements}
\end{align}

\noindent
Let us now recast the Lindblad equation~\eqref{wl:eq:thermal-lindblad-W-matrix-elements} into a more explicit and familiar form describing the effects of unresolved photons in a finite-temperature bath by evaluating the Wilson-line matrix elements central to the thermal jump operators.  Using the normal-ordered Wilson line in
\eqref{wl:eq:normal-ordered-W}, the one-quantum matrix elements are
\begin{align}
  \left.
  \langle N+1|\widehat W_p(\xi+\delta\xi,\xi)|N\rangle
  \right|_{\delta\xi^0}
  &=
  \left.
  \left(1+O(\delta\xi)\right)
  \langle N+1|
  \left[
    1
    +
    \int_\xi^{\xi+\delta\xi}\!\dd\Phi\,
    S_p(q,\lambda)\widehat a_{q,\lambda}^\dagger
    +
    O(\delta\xi)
  \right]
  |N\rangle
  \right|_{\delta\xi^0}
  \notag\\
  &=
  \sqrt{N+1}\,S_p(q,\lambda),
  \notag\\
  \left.
  \langle N-1|\widehat W_p(\xi+\delta\xi,\xi)|N\rangle
  \right|_{\delta\xi^0}
  &=
  \left.
  \left(1+O(\delta\xi)\right)
  \langle N-1|
  \left[
    1
    -
    \int_\xi^{\xi+\delta\xi}\!\dd\Phi\,
    S_p^\ast(q,\lambda)\widehat a_{q,\lambda}
    +
    O(\delta\xi)
  \right]
  |N\rangle
  \right|_{\delta\xi^0}
  \notag\\
  &=
  -\sqrt{N}\,S_p^\ast(q,\lambda).
  \label{wl:eq:thermal-W-matrix-elements-soft-factor}
\end{align}
Substituting \eqref{wl:eq:thermal-W-matrix-elements-soft-factor} into
\eqref{wl:eq:thermal-LpmN-W-matrix-element} gives the relation between $\hspace*{-1mm}\widehat{\,\,\mathbb L}_{+,N}(\xi)$, which is induced by the $N$-photon thermal state, and the zero-temperature emission jump operator $\hspace*{-1mm}\widehat{\,\,\mathbb L}_1(\xi)=\widehat L_1(\xi)$ in
\eqref{wl:eq:L1-soft-factor-form},
\begin{align}
  \hspace*{-1mm}\widehat{\,\,\mathbb L}_{+,N}(\xi)
  &=
  \sqrt{P_N(N+1)}\,
  \widehat L_1(\xi)\, ,&
  \hspace*{-1mm}\widehat{\,\,\mathbb L}_{-,N}(\xi)
  &=
  -\sqrt{P_NN}\,
  \widehat L_1^\dagger(\xi)\, .
  \label{wl:eq:thermal-LpmN-soft-factor}
\end{align}
The photons selected by the $(\xi+\delta\xi,\xi)$ shell necessarily carry the unique energy $\omega_q$ which implies the probability distribution $P_N=p_N(\omega_q) = (1-\e^{-\omega_q/(k_BT)})\e^{-N\omega_q/(k_BT)}$. We can therefore obtain the average photon number $\bar n(\omega_q)$ as,
\begin{align}
    \bar n(\omega_q)\equiv\sum_N N p_N(\omega_q) = \sum_N N\,(1-\e^{-\omega_q/(k_BT)})\e^{-N\omega_q/(k_BT)}=1/(\e^{\omega_q/(k_BT)}-1)\, ,
\end{align}
and then
\(\sum_N p_N(\omega_q)(N+1)=1+\bar n(\omega_q)\) where the contribution $1$ and $\bar n(\omega_q)$ can be regarded as spontaneous and stimulated photon emissions, respectively.
The effective jump operators for the two photon-number changing channels can therefore be derived as,
\begin{align}
  \sum_N
  \hspace*{-1mm}\widehat{\,\,\mathbb L}_{+,N}(\xi)\widehat\rho_{\rm th}(\xi)\hspace*{-1mm}\widehat{\,\,\mathbb L}_{+,N}^\dagger(\xi)
  &=
  \left[
    \sum_N (N+1) p_N(\omega_q)
  \right]
  \widehat L_1(\xi)\widehat\rho_{\rm th}(\xi)\widehat L_1^\dagger(\xi)
   =
  \widehat L_+(\xi)\widehat\rho_{\rm th}(\xi)\widehat L_+^\dagger(\xi),
  \notag\\
  \widehat L_+(\xi)
  &=
  \sqrt{1+\bar n(\omega_q)}\,
  \widehat L_1(\xi)\, ,
  \notag\\[2mm]
  \sum_N
  \hspace*{-1mm}\widehat{\,\,\mathbb L}_{-,N}(\xi)\widehat\rho_{\rm th}(\xi)\hspace*{-1mm}\widehat{\,\,\mathbb L}_{-,N}^\dagger(\xi)
  &=
  \left[
    \sum_N Np_N(\omega_q)
  \right]
  \widehat L_1^\dagger(\xi)\widehat\rho_{\rm th}(\xi)\widehat L_1(\xi)
=
  \widehat L_-(\xi)\widehat\rho_{\rm th}(\xi)\widehat L_-^\dagger(\xi),
  \notag\\
  \widehat L_-(\xi)
  &=
  -\sqrt{\bar n(\omega_q)}\,
  \widehat L_1^\dagger(\xi)\, .
  \label{wl:eq:thermal-Lpm-effective-soft-factor}
\end{align}
With these effective jumps, the Lindblad equation in
\eqref{wl:eq:thermal-lindblad-W-matrix-elements},
which includes interactions between the hard particles and the photons in the thermal bath, can be written in the compact form,
\begin{align}
  \frac{\dd\widehat\rho_{\rm th}}{\dd\xi}
  =
  -\ii[\widehat H_{\rm eff}(\xi),\widehat\rho_{\rm th}(\xi)]
  +\sum_{i=\pm}
  \int\!\dd\nu_\xi
  \left[
    \widehat L_i(\xi)\widehat\rho_{\rm th}(\xi)
    \widehat L_i^\dagger(\xi)
    -
    \frac12
    \left\{
      \widehat L_i^\dagger(\xi)\widehat L_i(\xi),
      \widehat\rho_{\rm th}(\xi)
    \right\}
  \right]\, .
  \label{wl:eq:thermal-lindblad-effective-jumps}
\end{align}
The $+$ jump operator induces emission into the unresolved shell.
The \(-\) jump comes from
absorbing photons already present in the thermal environment, with a weight of
\(\bar n(\omega_q)\).

It is useful to explicitly display the resulting thermal decoherence kernel
by taking the hard momentum-space matrix element of
\eqref{wl:eq:thermal-lindblad-effective-jumps}.  Then, substituting
\eqref{wl:eq:thermal-Lpm-effective-soft-factor} and keeping only the real part of the second term on the right-hand side of eq.~\eqref{wl:eq:thermal-lindblad-effective-jumps}, which induces the dissipative/decoherence effects,
gives,
\begin{align}
  \left.
  \frac{\dd\rho_{pp'}(\xi)}{\dd\xi}
  \right|_{\rm decoh}
  &=
  \int\!\dd\nu_\xi\,
  \Bigg[
    \big(1+\bar n(\omega_q)\big)
    S_p(q,\lambda)S_{p'}^\ast(q,\lambda)
    +
    \bar n(\omega_q)
    S_p^\ast(q,\lambda)S_{p'}(q,\lambda)
  \notag\\
  &-
    \frac12
    \big(1+2\bar n(\omega_q)\big)
    \left(
      |S_p(q,\lambda)|^2
      +
      |S_{p'}(q,\lambda)|^2
    \right)
  \Bigg]\rho_{pp'}(\xi)
  \notag\\
  &=
  \int\!\dd\nu_\xi\,
  \Bigg[
    \big(1+2\bar n(\omega_q)\big)
    \Ree\!\left[
      S_p(q,\lambda)S_{p'}^\ast(q,\lambda)
    \right]
  \notag\\
  &-
    \frac12
    \big(1+2\bar n(\omega_q)\big)
    \left(
      |S_p(q,\lambda)|^2
      +
      |S_{p'}(q,\lambda)|^2
    \right)
  \Bigg]\rho_{pp'}(\xi)
  \notag\\
  &=
  -
  \frac12
  \int\!\dd\nu_\xi\,
  \big(1+2\bar n(\omega_q)\big)
  |S_p(q,\lambda)-S_{p'}(q,\lambda)|^2
  \rho_{pp'}(\xi)
  \notag\\
  &=
  -
  \frac12
  \int\!\dd\nu_\xi\,
  \coth\!\left(\frac{\omega_q}{2k_BT}\right)
  |S_p(q,\lambda)-S_{p'}(q,\lambda)|^2
  \rho_{pp'}(\xi)\, ,
  \label{wl:eq:thermal-coth-decoherence-kernel}
\end{align}
where in the second line we have relegated the temperature-independent imaginary part of $S_pS_{p'}^\ast$ into the Hamiltonian contribution that generates the phase $\chi_{pp'}$ for the density matrix.  The last line uses
\(1+2\bar n(\omega_q)=\coth[\omega_q/(2k_BT)]\).  By comparing to eq.~\eqref{pion:eq:matrix-flow-with-H}, we therefore conclude that
the decoherence kernel for a finite-temperature photon environment differs from its vacuum counterpart by an overall Bose thermal factor $\coth[\omega_q/(2k_BT)]$.

\section{Influence functional approach}
\label{sec:influence-functional}

The leading-power evolution from \(\widehat\rho(0)\) to \(\widehat\rho(\xi)\) in a general photon environmental state \(\widehat\rho_{\rm env}\) can equivalently be
written in the Feynman--Vernon influence-functional form by reorganizing the
photon trace in eq.~\eqref{wl:eq:total-state} as follows:
\begin{align}
  \widehat\rho(\xi)
  &=
  \sum_X\sum_{p,p'}
  \langle p|\widehat\rho(0)|p'\rangle
  |p\rangle\langle p'|
  \langle X|\widehat W_p(\xi,0)\,\widehat\rho_{\rm env}\,\widehat W_{p'}^\dagger(\xi,0)|X\rangle
  \notag\\
  &=
  \sum_{p,p'}
  \langle p|\widehat\rho(0)|p'\rangle
  |p\rangle\langle p'|
  \Tr_{X,\xi}\big[\widehat W_p(\xi,0)\,\widehat\rho_{\rm env}\,\widehat W_{p'}^\dagger(\xi,0)\big]
  \notag\\
  &=
  \sum_{p,p'}
  \langle p|\widehat\rho(0)|p'\rangle
  |p\rangle\langle p'|
  \Tr_{X,\xi}\big[\widehat\rho_{\rm env}\,\widehat W_{p'}^\dagger(\xi,0)\widehat W_p(\xi,0)\big]
  \notag\\
  &=
  \sum_{p,p'}
  \langle p|\widehat\rho(0)|p'\rangle\,
  \mathcal F_{pp'}(\xi,0)\,
  |p\rangle\langle p'|,
  \label{if:eq:influence-functional-form}
\end{align}
where
\(\mathcal F_{pp'}(\xi,0)\equiv
\Tr_{X,\xi}\big[\widehat\rho_{\rm env}\widehat W_{p'}^\dagger(\xi,0)\widehat W_p(\xi,0)\big]\) is the
Feynman--Vernon influence functional which captures the non-unitary evolution
of the reduced hard density-matrix element
\(\langle p|\widehat\rho(0)|p'\rangle\) induced by the
environment~\cite{Hu:1993vs,Hu:1993qa,Hanggi:2004toz,Halliwell:1995mh,Fleming:2010bow,Romero:1996bm,Blume-Kohout:2008cdr,Jana:2020vyx,Calzetta:1986cq}.

\subsection{Zero-temperature influence functional}
\label{if:subsec:vacuum-soft-correlator}

Starting from the influence-functional form in
\eqref{if:eq:influence-functional-form} with $\widehat\rho_{\rm env}=|0\rangle\langle0|$, we now evaluate the zero-temperature influence functional
\(\mathcal F^{(0)}_{pp'}(\xi,0)\).  Inserting the normal-ordered Wilson-line operators given in eq.~\eqref{if:eq:normal-ordered-soft-evolution}
yields,
\begin{align}
  \mathcal F^{(0)}_{pp'}(\xi,0)
  &=
  \e^{-\ii(\chi_p-\chi_{p'})}
  \e^{-\frac12 |S_p|^2-\frac12 |S_{p'}|^2}
  \langle0|
  \e^{-S_{p'}\widehat a^\dagger}
  \e^{S_{p'}^\ast\widehat a}
  \e^{S_p\widehat a^\dagger}
  \e^{-S_p^\ast\widehat a}
  |0\rangle
  \notag\\
  &=
  \e^{-\ii(\chi_p-\chi_{p'})}
  \e^{-\frac12 |S_p|^2-\frac12 |S_{p'}|^2}
  \langle0|
  \e^{S_{p'}^\ast\widehat a}
  \e^{S_p\widehat a^\dagger}
  |0\rangle
  \notag\\
  &=
  \e^{-\ii(\chi_p-\chi_{p'})}
  \e^{-\frac12 |S_p|^2-\frac12 |S_{p'}|^2}
  \sum_{m,n=0}^{\infty}
  \frac{(S_{p'}^\ast)^m(S_p)^n}{m!\,n!}
  \langle0|\widehat a^m(\widehat a^\dagger)^n|0\rangle
  \notag\\
  &=
  \e^{-\ii(\chi_p-\chi_{p'})}
  \e^{-\frac12 |S_p|^2-\frac12 |S_{p'}|^2}
  \sum_{n=0}^{\infty}
  \frac{(S_{p'}^\ast S_p)^n}{n!}
  \notag\\
  &=
  \e^{-\ii(\chi_p-\chi_{p'})}
  \e^{S_pS_{p'}^\ast-\frac12 |S_p|^2-\frac12 |S_{p'}|^2}
  \notag\\
  &=
  \e^{-\frac12 |S_p-S_{p'}|^2}
  \e^{-\ii[\chi_p-\chi_{p'}-\Imm(S_p S_{p'}^\ast )]} .
  \label{if:eq:vacuum-soft-correlator}
\end{align}
Thus, the influence functional method reproduces the same resolution-space
evolution for the zero-temperature case as the amplitude-level matrix-element flow in
\eqref{pion:eq:matrix-flow-full-exponent} and, equivalently, the Lindblad equation in \eqref{wl:eq:lindblad}.

\subsection{Thermal influence functional}
\label{if:subsec:thermal-soft-correlator}

At finite temperature, we simply repeat the calculation after replacing the
photon vacuum projector \(|0\rangle\langle0|\) by the same thermal photon bath given in eq.~\eqref{wl:eq:thermal-photon-density-number-basis}.
The reduced density matrix for a finite-temperature photon background then reads,
\begin{align}
  \rho_{pp'}(\xi)
  &=
  \langle p|
  \Tr_{X,\xi}\!\left[
    \widehat S\widehat\rho_i\widehat S^\dagger
  \right]
  |p'\rangle
  \notag\\
  &\simeq
  \langle p|\widehat\rho(0)|p'\rangle\,
  \Tr_{X,\xi}\!\left[
    \widehat W_p(\xi,0)\widehat\rho_{\rm env}\widehat W_{p'}^\dagger(\xi,0)
  \right]
  \notag\\
  &=
  \langle p|\widehat\rho(0)|p'\rangle
  \sum_N P_N
  \langle N|\widehat W_{p'}^\dagger(\xi,0)\widehat W_p(\xi,0)|N\rangle \notag\\
  &\equiv\langle p|\widehat\rho(0)|p'\rangle \mathcal F^{\rm (th)}_{pp'}(\xi,0)\, ,
  \label{if:eq:thermal-soft-correlator}
\end{align}
where we have used the leading-power approximation in eq.~\eqref{wl:eq:general-S-factorization} to get to the second line.
Since the thermal state and the Wilson line both
factorize over the unresolved photon modes whose energies satisfy the condition $0\leq \ln(\omega_q/\lambda_{\rm IR})\leq \xi$, we can
first consider states composed of $N$ photons each carrying
the same energy $\omega_q$, denoted simply as $|N\rangle$ below, and subsequently construct the complete unresolved-photon Fock space as direct products of states composed of photons with different energies within the unresolved window. The probability distribution for the subsequent thermal state then reads,
\(p_N(\omega_q)=(1-r)r^N\) with
\(r=\e^{-\omega_q/(k_BT)}\).  Then, using eq.~\eqref{if:eq:normal-ordered-soft-evolution},  we find that the total contribution of thermal photons with energy $\omega_q$ to ${\cal F}_{pp'}^{\rm (th)}$ reads,
\begin{align}
  \mathcal F^{\rm (th)}_{pp',\omega_q}(\xi,0)
  &=
  (1-r)
  \e^{-\ii(\chi_p-\chi_{p'})}
  \e^{-\frac12|S_{p'}|^2-\frac12|S_p|^2+S_{p'}^\ast S_p}
  \sum_{N=0}^{\infty}
  r^N
  \langle N|
  \e^{-S_{p'}\widehat a^\dagger}
  \e^{S_p\widehat a^\dagger}
  \e^{S_{p'}^\ast\widehat a}
  \e^{-S_p^\ast\widehat a}
  |N\rangle
  \notag\\
  &=
  (1-r)
  \e^{-\ii(\chi_p-\chi_{p'})}
  \e^{-\frac12|S_{p'}|^2-\frac12|S_p|^2+S_{p'}^\ast S_p}
  \sum_{N=0}^{\infty}
  r^N
  \langle N|
  \e^{(S_p-S_{p'})\widehat a^\dagger}
  \e^{-(S_p-S_{p'})^\ast\widehat a}
  |N\rangle
  \notag\\
  &=
  (1-r)
  \e^{-\ii(\chi_p-\chi_{p'})}
  \e^{\ii\Imm(S_pS_{p'}^\ast)-\frac12|S_p-S_{p'}|^2}
  \sum_{N=0}^{\infty}r^N
  \sum_{\ell=0}^{N}
  \frac{(-|S_p-S_{p'}|^2)^\ell}{(\ell!)^2}
  \frac{N!}{(N-\ell)!}
  \notag\\
  &=
  (1-r)
  \e^{-\ii(\chi_p-\chi_{p'})}
  \e^{\ii\Imm(S_pS_{p'}^\ast)-\frac12|S_p-S_{p'}|^2}
  \sum_{N=0}^{\infty}r^N \mathrm L_N(|S_p-S_{p'}|^2)
  \notag\\
  &=
  \e^{-\ii(\chi_p-\chi_{p'})}
  \e^{\ii\Imm(S_pS_{p'}^\ast)-\frac12|S_p-S_{p'}|^2}
  \e^{-\frac{r}{1-r}|S_p-S_{p'}|^2}
  \notag\\
  &=
  \e^{-\frac12\coth\left(\frac{\omega_q}{2k_BT}\right)|S_p-S_{p'}|^2}
  \e^{-\ii[\chi_p-\chi_{p'}-\Imm(S_pS_{p'}^\ast)]}\, ,
  \label{if:eq:single-mode-thermal-correlator}
\end{align}
where we have used
\(\langle N|(\widehat a^\dagger)^\ell\widehat a^\ell|N\rangle=N!/(N-\ell)!\) with $\ell\leq N$ for nonzero result, and the Laguerre polynomial
$\mathrm L_N(x)=\sum_{\ell=0}^{N}\binom{N}{\ell}(-x)^\ell/\ell!$ satisfies
$\sum_{N=0}^{\infty}r^N\mathrm L_N(x)=(1-r)^{-1}\e^{-rx/(1-r)}$.
The last line uses \((1+r)/(1-r)=\coth[\omega_q/(2k_BT)]\).
The matrix element in the total trace over thermal states can be written schematically as
\begin{align}
    \langle N|\widehat W_{p'}^\dagger(\xi,0)\widehat W_p(\xi,0)| N\rangle = \lim_{M\to\infty} \prod_{i=0}^{M-1}\langle N(\omega_{q_i})|\widehat W_{p'}^\dagger(\xi_{i+1},\xi_i)\widehat W_p(\xi_{i+1},\xi_i)| N(\omega_{q_i})\rangle \label{eq:thermal_mode_product}
\end{align}
with boundary conditions $\xi_0=0, \xi_M = \xi$, and the energy $\omega_{q_i}$ lies within the unresolved window of $(\xi_{i+1},\xi_i)$. So, the phase-space integration measure of the Wilson line $\widehat W_p(\xi_{i+1},\xi_i)$ becomes
\begin{align}
    \int^{\e^{\xi_{i+1}}\lambda_{\rm IR}}_{\e^{\xi_{i}}\lambda_{\rm IR}} \dd\Phi(q)\, ,
    \label{eq:phase_xi_window}
\end{align}
with the integrand containing exactly the same $q$-dependent functions and operators as in eq.~\eqref{wl:eq:Wilson-line-soft-factor-form}. The thermal weight $P_N$ of eq.~\eqref{if:eq:thermal-soft-correlator} is likewise the product of the $p_N(\omega_{q_i})$, so that summing eq.~\eqref{eq:thermal_mode_product} over $N$ with this weight gives
\begin{align}
    \mathcal F^{\rm (th)}_{pp'}(\xi,0) = \lim_{M\to\infty} \prod_{i=0}^{M-1}\mathcal F^{\rm (th)}_{pp',\omega_{q_i}}(\xi_{i+1},\xi_i)\, ,
    \label{eq:thermal_F_product}
\end{align}
where each factor is eq.~\eqref{if:eq:single-mode-thermal-correlator} on the window $(\xi_{i+1},\xi_i)$. Since the contribution from a single partition of thermal modes with energy $\omega_q$ takes the exponential form, eq.~\eqref{if:eq:single-mode-thermal-correlator}, the product of different energy modes in eq.~\eqref{eq:thermal_F_product} becomes a summation in the exponents. The lower and upper limits of consecutive windows $(\xi_{i+1},\xi_i)$ and $(\xi_{i+2},\xi_{i+1})$ therefore combine into a single phase-space integral in $(\xi_{i+2},\xi_i)$, and so do the phases $\chi_p(\xi_{i+1},\xi_i)$, which are additive as can be easily seen from eq.~\eqref{pion:eq:virtual-eikonal-current}, with the unresolved window again modified to eq.~\eqref{eq:phase_xi_window}, and its definition in eq.~\eqref{pion:eq:virtual-phase-decomposition} meaning that $\chi_p(\xi_{i+2},\xi_{i+1}) + \chi_p(\xi_{i+1},\xi_i) = \chi_p(\xi_{i+2},\xi_i)$.
It is therefore straightforward to write the complete expression of ${\cal F}^{\rm (th)}_{pp'}$. In particular, the full thermal decoherence exponent in the convention of eq.~\eqref{pion:eq:combined-coherence-factor} reads,
\begin{equation}
  \Gamma^{(\rm th)}_{pp'}(\xi)
  =
  \frac12
  \sum_\lambda\int\dd\Phi_\xi(q)\,
  \coth\!\left(\frac{\omega_q}{2k_BT}\right)
  |S_p(q,\lambda)-S_{p'}(q,\lambda)|^2 .
  \label{if:eq:thermal-decoherence-exponent}
\end{equation}

\noindent
Taking the derivative ${\rm d}/{\rm d}\xi$ of eq.~\eqref{if:eq:thermal-soft-correlator}, we recover the thermal Lindblad equation of Section~\ref{wl:sec:finite-temperature-thermal-state}, whose decoherence kernel is eq.~\eqref{wl:eq:thermal-coth-decoherence-kernel}.

\subsection{Decoherence rate}
\label{if:subsec:decoherence-rate}
\label{if:subsec:thermal-resolution-flow}

In the same kinematic regime as in
Section~\ref{pion:sec:angular-decoherence}, direct substitution into the
thermal decoherence rate gives,
\begin{align}
  \dot\Gamma^{(\rm th)}_{pp'}(\xi)
  &=
  \frac12
  \int\dd\nu_\xi\,
  \coth\!\left(\frac{\omega_q}{2k_BT}\right)
  |S_p(q,\lambda)-S_{p'}(q,\lambda)|^2
  \notag\\
  &=
  \frac12
  \sum_\lambda
  \int_0^\infty
  \frac{\omega_q\,\dd\omega_q\,\dd\Omega}
       {2(2\pi)^3}
  \delta\!\left(\xi-\ln\frac{\omega_q}{\lambda_{\IR}}\right)
  \coth\!\left(\frac{\omega_q}{2k_BT}\right)
  \frac{4e^2v^2}{\omega_q^2}
  \left|
    (\mathbf n-\mathbf n')\cdot\epsilon^\ast(q,\lambda)
  \right|^2
  \notag\\
  &=
  \frac{e^2v^2}{(2\pi)^3}
  \sum_\lambda
  \int_0^\infty\frac{\dd\omega_q}{\omega_q}\,
  \delta\!\left(\xi-\ln\frac{\omega_q}{\lambda_{\IR}}\right)
  \coth\!\left(\frac{\omega_q}{2k_BT}\right)
  \int\dd\Omega
  \left|
    (\mathbf n-\mathbf n')\cdot\epsilon^\ast(q,\lambda)
  \right|^2
  \notag\\
  &=
  \frac{e^2v^2}{(2\pi)^3}
  \coth\!\left(\frac{\lambda_{\rm det}}{2k_BT}\right)
  \sum_\lambda
  \int\dd\Omega
  \left|
    (\mathbf n-\mathbf n')\cdot\epsilon^\ast(q,\lambda)
  \right|^2
  \notag\\
  &=
  \frac{e^2v^2}{(2\pi)^3}
  \coth\!\left(\frac{\lambda_{\rm det}}{2k_BT}\right)
  \left[
    4\pi|\mathbf n-\mathbf n'|^2
    -
    \frac{4\pi}{3}|\mathbf n-\mathbf n'|^2
  \right]
  \notag\\
  &\simeq
  \frac{4\alpha}{3\pi}\,
  v^2(\delta\theta)^2
  \coth\!\left(\frac{\lambda_{\rm det}}{2k_BT}\right)\, ,
  \label{if:eq:thermal-angular-rate}
\end{align}
with $\lambda_{\rm det} = \lambda_{\rm IR}\e^\xi$ and we have applied the small angle approximation in the last step. The angular integral can be worked out easily following eq.~\eqref{pion:eq:angular-decoherence-main}.

\begin{figure}[!t]
  \centering
  \includegraphics[width=0.68\linewidth]{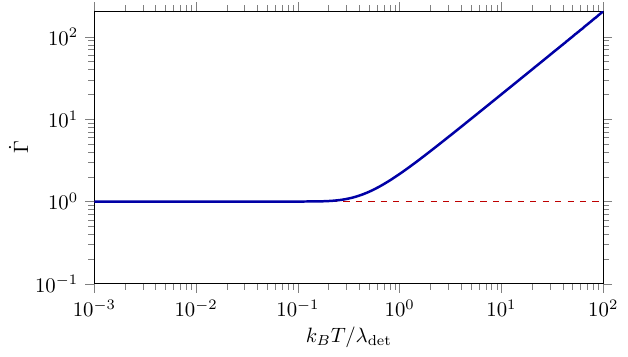}
  \caption{Plot of the normalized thermal decoherence rate
  \(\dot\Gamma^{(\rm th)}_{pp'}/\dot\Gamma_{pp'}\), labeled \(\dot\Gamma\) on the vertical axis, against \(k_BT/\lambda_{\rm det}\). The red dashed line is the constant
zero-temperature decoherence rate for reference, while the blue curve is the finite-temperature decoherence rate given in eq.~\eqref{if:eq:thermal-angular-rate}. }
  \label{if:fig:thermal-shell-rate}
\end{figure}

In Fig.~\ref{if:fig:thermal-shell-rate}, we demonstrate the distinct features of the decoherence rate in different thermal regimes.
When \(k_BT\ll\lambda_{\rm det}\), the blue curve
coincides with the red line, because the unresolved photons being traced over are mostly
final-state radiation from the charged pions. On the other hand, when $k_BT\gtrsim\lambda_{\rm det}$, the blue curve quickly rises above the red reference because the resolution shell $(\xi+\delta\xi,\xi)$ becomes populated by thermal photons where the decoherence rate is significantly enhanced by the interactions between the hard particles and the photon bath, so that the quantum correlations between the hard momentum states are quickly destroyed and dissipated into the environment.

\section{Conclusion}
\label{sec:conclusion}

In this work, we have taken scalar QED as a primary example and treated the unresolved soft-photon radiation
as the environment. We used the detector
resolution parameter \(\xi=\ln(\lambda_{\rm det}/\lambda_{\IR})\) to organize
the reduced density matrix of the observed hard scalar-particle pair.  At the amplitude level, the leading soft theorem gives an exponentiated
form in eq.~\eqref{pion:eq:combined-coherence-factor} for the reduced density matrix as a consequence of the interplay between the virtual and real contributions.
For the diagonal elements of the reduced density matrix,
the real emission amplitude cancels the real part of the virtual correction guaranteeing the infrared finiteness of the inclusive cross section.
In contrast, the sum of the real parts of the real and virtual corrections is always negative, hence suppressing the off-diagonal elements of the density matrix.  For hard charged scalars with momentum configurations $p$ and $p'$ differing by a small angle $\delta\theta$, the vacuum decoherence rate is proportional to $\alpha v^2(\delta\theta)^2$, and in the presence of a thermal photon bath, the rate is increased by an overall factor of $\coth[\lambda_{\rm det}/(2k_BT)]$.

The operator origin of this evolution over the resolution parameter follows from the leading-power factorization of the scalar-QED meaning that the LP $S$-matrix can be factorized into the hard production operator and a Wilson line conditioned on the momenta of the scalars.  Because the Abelian field commutator is a $c$-number, path ordering reduces to a momentum-dependent phase multiplying an exponentiated operator composed of ladder operators weighted by the soft factor, see eq.~\eqref{wl:eq:Wilson-line-soft-factor-form}.
The vacuum and one-photon matrix elements of this Wilson-line operator give,
respectively, the no-emission and single-emission Kraus operators.  Applying the same construction to an infinitesimal resolution shell identifies the one-photon Wilson-line matrix element with the Lindblad jump operator, while the trace-preserving condition fixes the Hermitian part of $\hspace*{-1mm}\widehat{\,\,\mathbb L}_0$ describing zero emissions by its single-emission counterpart $\hspace*{-1mm}\widehat{\,\,\mathbb L}_1$, see, e.g., eq.~\eqref{wl:eq:L0-real-part}.
At finite temperature, emission and absorption provide two jump channels whose bosonic weights combine into a single thermal factor reproducible both in the Kraus and influence-functional approaches, as shown in eqs.~\eqref{wl:eq:thermal-coth-decoherence-kernel} and~\eqref{if:eq:thermal-decoherence-exponent}.

As a consistency check, we demonstrated that the same results can also be derived at the matrix element level using the Feynman--Vernon influence functional approach.
Physically, the detector resolution acts as a flow parameter distinguishing which information is kept in the hard system: enlarging the unresolved photon window, which implies more photons are being classified as part of the environment,
preserves inclusive probabilities but progressively suppresses coherence between hard states with distinguishable momentum configurations, e.g., by a small angle $\delta\theta$, since the quantum information is leaked into the environment. We note that our derivations for the reduced density matrix remain valid if the produced hard particles are $U(1)$-charged massive fermions, since at leading power the spins of the hard fermions are invisible to the soft photons. The same relation between the Wilson-line and jump operators therefore still applies, with the spin dependence only appearing in the hard production amplitude.

\section*{Acknowledgements}
We thank Ding-Yu Shao for illuminating discussions.
We acknowledge the support in part by the National Natural Science Foundation of China with Grant No.12535006, by the Guangdong Basic and Applied Basic Research Foundation under grant No. 2026A1515011306, and by the University Development Fund of the Chinese University of Hong Kong, Shenzhen, under the grant No. UDF0100386.

\appendix

\section{Proof of CPTP and CP divisibility}
\label{pion:app:cptp}

This appendix proves that the dissipative dephasing part $\dot\Gamma_{pp'}(\xi)$ is completely positive trace-preserving (CPTP) and completely positive (CP) divisible as claimed in Section~\ref{pion:sec:lindblad-flow}.  The phase term $-\ii\dot\chi_{pp'}(\xi)$ in
eq.~\eqref{pion:eq:matrix-flow-with-H} is separated from the real dephasing kernel $-\dot\Gamma_{pp'}(\xi)$ as dictated by eq.~\eqref{pion:eq:matrix-flow-with-H}.
For the diagonal matrix elements with $p=p'$, $\chi_{pp'}(\xi)=\Gamma_{pp'}(\xi)=0$ by eqs.~\eqref{pion:eq:phase-difference}, \eqref{pion:eq:gamma-square}, and~\eqref{pion:eq:integrated-soft-overlap} and therefore the trace of the density matrix is $\xi$-independent, i.e., trace-preserving.

Next, we show the complete positivity of the dephasing contribution $-\dot\Gamma_{pp'}(\xi)$.
Let \(\mathcal V_{\rm diss}(\xi+\delta\xi,\xi)\) denote the infinitesimal dephasing step,
\begin{equation}
  \widehat\rho(\xi+\delta\xi)=\mathcal V_{\rm diss}(\xi+\delta\xi,\xi)[\widehat\rho(\xi)] \, ,
\end{equation}
which acts on the hard momentum basis and yields,
\begin{equation}
  \mathcal V_{\rm diss}(\xi+\delta\xi,\xi)[|p\rangle\langle p'|]
  =
  \e^{-\delta\xi\,\dot\Gamma_{pp'}(\xi)}
  |p\rangle\langle p'| \, .
\end{equation}
We further define
\begin{equation}
  \langle A,B\rangle_\xi\equiv
  \int\!\dd\nu_\xi\,A^\ast(q,\lambda)B(q,\lambda),
\end{equation}
and the corresponding norm by $|A|_\xi^2\equiv\langle A,A\rangle_\xi$. Then from eq.~\eqref{pion:eq:gamma-square}, we have,
\begin{equation}
  \dot\Gamma_{pp'}(\xi)=\frac{1}{2}|S_p-S_{p'}|_\xi^2 .
\end{equation}
For any finite family of hard momentum states $\{p_i\}$ and real coefficients
$\{c_i\}$ satisfying $\sum_i c_i=0$,
\begin{align}
  -\sum_{ij}c_i c_j\dot\Gamma_{p_i p_j}(\xi)
  &=
  -\frac{1}{2}\sum_{ij}c_i c_j
  \left(|S_{p_i}|_\xi^2+|S_{p_j}|_\xi^2
  -\langle S_{p_i},S_{p_j}\rangle_\xi
  -\langle S_{p_j},S_{p_i}\rangle_\xi\right)
  \notag\\
  &=
  \frac{1}{2}\bigl(
  \big\langle \sum\nolimits_i c_iS_{p_i},
  \sum\nolimits_j c_jS_{p_j}\big\rangle_\xi
  +
  \big\langle \sum\nolimits_j c_jS_{p_j},
  \sum\nolimits_i c_iS_{p_i}\big\rangle_\xi
  \bigr)
  \notag\\
  &=
  \bigl|\sum\nolimits_i c_iS_{p_i}\bigr|_\xi^2
  \notag\\
  &\ge 0 ,\label{eq:dotGamma}
\end{align}
where the terms proportional to $|S_{p_i}|_\xi^2$ cancel because $\sum_i c_i=0$.
This means $-\dot\Gamma$ is conditionally positive semidefinite. Here ``conditionally" refers to the single constraint $\sum_ic_i=0$ for the real constants $\{c_i\}$. Schoenberg's theorem~\cite{Schoenberg:1938} then implies that for every $\delta\xi\ge 0$,
$\big[\e^{-\delta\xi\,\dot\Gamma_{p_i p_j}(\xi)}\big]_{ij}$ is a positive semidefinite matrix on any
finite momentum block. On such a block, $\mathcal V_{\rm diss}(\xi+\delta\xi,\xi)$ acts by entrywise
multiplication, namely, each matrix element $\rho_{p_i p_j}(\xi)$ is multiplied by
$\e^{-\delta\xi\,\dot\Gamma_{p_i p_j}(\xi)}$. By the Schur product theorem~\cite{HornJohnson:2012}, which states that entrywise
multiplication by a positive semidefinite matrix is completely positive,  the infinitesimal dephasing step,
\begin{align}
    \langle p_i|\mathcal V_{\rm diss}(\xi+\delta\xi,\xi)[\widehat\rho(\xi)]|p_j\rangle=\big[\e^{-\delta\xi\,\dot\Gamma_{p_i p_j}(\xi)}\big]_{ij}\rho_{p_ip_j}(\xi)
    \label{eq:Vdiss_inf}
\end{align}
is therefore completely positive, with $\rho_{p_ip_j}(\xi)$ being a positive semidefinite matrix. The latter statement can be justified iteratively from the semidefiniteness of the full density matrix $\rho_{p_ip_j}(0)$ with no unresolved photon tracing.
Combined with $\dot\Gamma_{pp}(\xi)=0$ guaranteeing that the diagonal entries are unchanged and the trace
is preserved, we conclude that each infinitesimal shell map is CPTP.
Eq.~\eqref{eq:Vdiss_inf} implies a first-order differential equation. Since all quantities are $c$-valued and $\mathcal V_{\rm diss}$ acts on ${\rho}_{pp'}$ entrywise multiplicatively without subtleties regarding path-ordering, with the initial condition $\mathcal V_{\rm diss}(\xi,\xi)=1$, its solution can be trivially found,
\begin{align}
\mathcal V_{\rm diss}(\xi_2,\xi_1)
[|p\rangle\langle p'|]
=
\exp\!\left[
-\int_{\xi_1}^{\xi_2}\dd\eta\,\dot\Gamma_{pp'}(\eta)
\right]|p\rangle\langle p'|\, .\label{eq:Vdiss-xi12}
\end{align}
It is clear that the integral in eq.~\eqref{eq:Vdiss-xi12} remains conditionally positive semidefinite from eq.~\eqref{eq:dotGamma} for arbitrary $\xi$. The same Schoenberg and Schur arguments then again establish complete positivity of $\mathcal V_{\rm diss}(\xi_2,\xi_1)$. With \(\dot\Gamma_{pp}=0\) additionally ensuring trace preservation, we conclude $\mathcal V_{\rm diss}(\xi_2,\xi_1)$ is CPTP. The $c$-valued integral in the dissipative factor also implies,
\begin{align}
\mathcal V_{\rm diss}(\xi_2,0)
=
\mathcal V_{\rm diss}(\xi_2,\xi_1)
\mathcal V_{\rm diss}(\xi_1,0)\, .
\end{align}
Thus $\mathcal V_{\rm diss}(\xi_2,\xi_1)$ can be divided into consecutive CPTP maps establishing CP-divisibility.
Specifically, with $\xi_1=0$, the finite map \(\mathcal V_{\rm diss}(\xi,0)\) is CPTP
and CP-divisible.

\section{Leading-power Hamiltonian in scalar QED}
\label{app:scalarQED}

In this appendix, we provide detailed derivations of the necessary ingredients for constructing and simplifying the QED Wilson lines in the LP soft approximation.
For a complex scalar field with covariant derivative \(D_\mu=\partial_\mu+\ii eA_\mu\), the Lagrangian
density reads,
\begin{equation}
  \mathcal L
  =
  -\frac14 F_{\mu\nu}F^{\mu\nu}
  +(D_\mu\phi)^\dagger(D^\mu\phi)
  -m_0^2\phi^\dagger\phi \, ,
  \label{sqed:eq:sqed-lagrangian}
\end{equation}
where \(m_0\) is the bare mass of the pion, and the interaction part takes the form,
\begin{equation}
  \mathcal L_{\rm int}
  =
  -eA_\mu J^\mu
  +e^2A_\mu A^\mu\phi^\dagger\phi \, ,
  \label{sqed:eq:interaction-lagrangian}
\end{equation}
where the charged scalar current is
\begin{equation}
  J^\mu
  =
  \ii\left[
  \phi^\dagger\partial^\mu\phi
  -(\partial^\mu\phi^\dagger)\phi
  \right]\, .
  \label{sqed:eq:scalar-current}
\end{equation}
The mode expansion of the scalar and photon fields in the second quantization takes the following form,
\begin{align}
  \widehat\phi(x)
  &=
  \int\dd\Pi(p)
  \left[
  \widehat b_{\mathbf p}\e^{-\ii p\cdot x}
  +\widehat d_{\mathbf p}^\dagger \e^{+\ii p\cdot x}
  \right],
  \notag\\
  \widehat A_\mu(x)
  &=
  \sum_\lambda\int\dd\Phi(q)
  \left[
  \epsilon_\mu(q,\lambda)\widehat a_{q,\lambda}\e^{-\ii q\cdot x}
  +\epsilon_\mu^\ast(q,\lambda)\widehat a_{q,\lambda}^\dagger \e^{+\ii q\cdot x}
  \right],
  \label{sqed:eq:mode-expansions}
\end{align}
where $ p^0=E_{\mathbf p},  q^0=\omega_q$, and
\begin{align}
    \bigl[\widehat b_{\mathbf p},\widehat b_{\mathbf k}^\dagger\bigr]
    &=
    \bigl[\widehat d_{\mathbf p},\widehat d_{\mathbf k}^\dagger\bigr]
    =
    (2\pi)^3 2E_{\mathbf p}\,
    \delta^{(3)}(\mathbf p-\mathbf k)\, ,&
    \bigl[\widehat a_{q,\lambda},\widehat a_{k,\lambda'}^\dagger\bigr]
    &=
    (2\pi)^3 2\omega_q\,
    \delta^{(3)}(\mathbf q-\mathbf k)\delta_{\lambda\lambda'}\, .
  \label{sqed:eq:momentum-measure}
\end{align}

\noindent
The interaction Lagrangian~\eqref{sqed:eq:interaction-lagrangian} contains both a linear and a quadratic photon coupling. At leading power we only need the linear coupling to the external eikonal current, whereas the quadratic four-point interaction contributes only at the subleading order.

By inserting the mode expansion~\eqref{sqed:eq:mode-expansions} into eq.~\eqref{sqed:eq:scalar-current}, the current operator reads,
\begin{align}
  \widehat J^\mu(x)
  &=
  \ii\left[
  \widehat\phi^\dagger(x)\partial^\mu\widehat\phi(x)
  -(\partial^\mu\widehat\phi^\dagger(x))\widehat\phi(x)
  \right]
  \notag\\
  &=
  \ii\int\dd\Pi(p)\dd\Pi(p')
  \left[
  \widehat b_{\mathbf p}^\dagger \e^{+\ii p\cdot x}
  +\widehat d_{\mathbf p}\e^{-\ii p\cdot x}
  \right]
  \left[
  -\ii p'^\mu\widehat b_{\mathbf p'}\e^{-\ii p'\cdot x}
  +\ii p'^\mu\widehat d_{\mathbf p'}^\dagger \e^{+\ii p'\cdot x}
  \right]
  \notag\\
  &-\ii\int\dd\Pi(p)\dd\Pi(p')
  \left[
  \ii p^\mu\widehat b_{\mathbf p}^\dagger \e^{+\ii p\cdot x}
  -\ii p^\mu\widehat d_{\mathbf p}\e^{-\ii p\cdot x}
  \right]
  \left[
  \widehat b_{\mathbf p'}\e^{-\ii p'\cdot x}
  +\widehat d_{\mathbf p'}^\dagger \e^{+\ii p'\cdot x}
  \right]
  \notag\\
  &=
  \widehat J_{bb}^\mu(x)+\widehat J_{dd}^\mu(x)+\widehat J_{bd}^\mu(x)+\widehat J_{db}^\mu(x),
  \label{sqed:eq:current-sector-decomposition}
\end{align}
where the definitions of the four operators $\widehat J_{bb}^\mu(x),\,\cdots$ are self-explanatory.
\noindent
The corresponding mode-expanded Hamiltonians can then be written as,
\begin{align}
  \widehat H_{bb}^{(1)}(t)
  &\equiv
  e\int\dd^3x\,\widehat A_\mu(t,\mathbf x)\widehat J_{bb}^\mu(t,\mathbf x)
  \notag\\
  &=
  e\sum_\lambda\int\dd\Phi(q)\int\dd\Pi(p)
  \frac{(p+p_{\mathbf p-\mathbf q})^\mu\epsilon_\mu(q,\lambda)}
  {2E_{\mathbf p-\mathbf q}}
  \widehat b_{\mathbf p}^\dagger\widehat b_{\mathbf p-\mathbf q}\widehat a_{q,\lambda}
  \e^{+\ii(E_{\mathbf p}-E_{\mathbf p-\mathbf q}-\omega_q)t}
  \notag\\
  &+
  e\sum_\lambda\int\dd\Phi(q)\int\dd\Pi(p)
  \frac{(p+p_{\mathbf p+\mathbf q})^\mu\epsilon_\mu^\ast(q,\lambda)}
  {2E_{\mathbf p+\mathbf q}}
  \widehat b_{\mathbf p}^\dagger\widehat b_{\mathbf p+\mathbf q}\widehat a_{q,\lambda}^\dagger
  \e^{+\ii(E_{\mathbf p}-E_{\mathbf p+\mathbf q}+\omega_q)t}\, ,
  \label{sqed:eq:Hbb-exact-two-terms}
\end{align}
where $\omega_q$ is the energy of the photon.
For any three-momentum \(\mathbf r\), we denote the corresponding on-shell four-momentum by
\begin{equation}
  p_{\mathbf r}^\mu\equiv(E_{\mathbf r},\mathbf r),
  \qquad
  E_{\mathbf r}\equiv\sqrt{\mathbf r^2+m^2}\, .
  \label{sqed:eq:on-shell-momentum-r}
\end{equation}

\noindent
Similarly, for the \(dd\), \(bd\), and \(db\) parts, we have
\begin{align}
  \widehat H_{dd}^{(1)}(t)
  &=
  -e\sum_\lambda\int\dd\Phi(q)\int\dd\Pi(p)
  \frac{(p+p_{\mathbf p-\mathbf q})^\mu\epsilon_\mu(q,\lambda)}
  {2E_{\mathbf p-\mathbf q}}
  \widehat d_{\mathbf p}^\dagger
  \widehat d_{\mathbf p-\mathbf q}
  \widehat a_{q,\lambda}
  \e^{+\ii(E_{\mathbf p}-E_{\mathbf p-\mathbf q}-\omega_q)t}
  \notag\\
  &-
  e\sum_\lambda\int\dd\Phi(q)\int\dd\Pi(p)
  \frac{(p+p_{\mathbf p+\mathbf q})^\mu\epsilon_\mu^\ast(q,\lambda)}
  {2E_{\mathbf p+\mathbf q}}
  \widehat d_{\mathbf p}^\dagger
  \widehat d_{\mathbf p+\mathbf q}
  \widehat a_{q,\lambda}^\dagger
  \e^{+\ii(E_{\mathbf p}-E_{\mathbf p+\mathbf q}+\omega_q)t},
  \notag\\
  \widehat H_{bd}^{(1)}(t)
  &=
  e\sum_\lambda\int\dd\Phi(q)\int\dd\Pi(p)
  \frac{(p-p_{\mathbf q-\mathbf p})^\mu\epsilon_\mu(q,\lambda)}
  {2E_{\mathbf q-\mathbf p}}
  \widehat b_{\mathbf p}^\dagger
  \widehat d_{\mathbf q-\mathbf p}^\dagger
  \widehat a_{q,\lambda}
  \e^{+\ii(E_{\mathbf p}+E_{\mathbf q-\mathbf p}-\omega_q)t}
  \notag\\
  &+
  e\sum_\lambda\int\dd\Phi(q)\int\dd\Pi(p)
  \frac{(p-p_{-\mathbf p-\mathbf q})^\mu\epsilon_\mu^\ast(q,\lambda)}
  {2E_{-\mathbf p-\mathbf q}}
  \widehat b_{\mathbf p}^\dagger
  \widehat d_{-\mathbf p-\mathbf q}^\dagger
  \widehat a_{q,\lambda}^\dagger
  \e^{+\ii(E_{\mathbf p}+E_{-\mathbf p-\mathbf q}+\omega_q)t},
  \notag\\
  \widehat H_{db}^{(1)}(t)
  &=
  -e\sum_\lambda\int\dd\Phi(q)\int\dd\Pi(p)
  \frac{(p-p_{-\mathbf p-\mathbf q})^\mu\epsilon_\mu(q,\lambda)}
  {2E_{-\mathbf p-\mathbf q}}
  \widehat d_{\mathbf p}
  \widehat b_{-\mathbf p-\mathbf q}
  \widehat a_{q,\lambda}
  \e^{-\ii(E_{\mathbf p}+E_{-\mathbf p-\mathbf q}+\omega_q)t}
  \notag\\
  &-
  e\sum_\lambda\int\dd\Phi(q)\int\dd\Pi(p)
  \frac{(p-p_{\mathbf q-\mathbf p})^\mu\epsilon_\mu^\ast(q,\lambda)}
  {2E_{\mathbf q-\mathbf p}}
  \widehat d_{\mathbf p}
  \widehat b_{\mathbf q-\mathbf p}
  \widehat a_{q,\lambda}^\dagger
  \e^{-\ii(E_{\mathbf p}+E_{\mathbf q-\mathbf p}-\omega_q)t}.
  \label{sqed:eq:Hdd-Hbd-Hdb-exact}
\end{align}

The soft power expansion in $O(\omega_q/E_{\mathbf p})$ can be constructed systematically by analyzing the Hamiltonian mode expansion. Since we are interested in studying the effects of soft-photon emissions on the reduced density matrix of hard particles, the temporal integral for the scalar-QED Hamiltonian runs from the hard-production point at $t=0$ to the far future at $t=\infty$.
Let us first analyze the Fourier modes of the Hamiltonian operators in eq.~\eqref{sqed:eq:Hbb-exact-two-terms}.
With \(\mathbf v_{\mathbf p}=\mathbf p/E_{\mathbf p}\), the corresponding phases are
\begin{align}
  E_{\mathbf p}-E_{\mathbf p\pm\mathbf q}\pm\omega_q
  &=
  E_{\mathbf p}
  -\left[
  E_{\mathbf p}
  \pm\mathbf v_{\mathbf p}\cdot\mathbf q
  +O\!\left(\frac{|\mathbf q|^2}{E_{\mathbf p}}\right)
  \right]
  \pm\omega_q
  \notag\\
  &=
  \pm\left(
  \omega_q
  -\mathbf v_{\mathbf p}\cdot\mathbf q
  \right)
  +O\!\left(\frac{|\mathbf q|^2}{E_{\mathbf p}}\right)\, ,
  \label{sqed:eq:soft-phase-pm}
\end{align}
which is soft compared to $E_{\mathbf p}$ leading to enhancement of the $\widehat H_{bb}^{(1)}(t)$ due to integration over a slow-varying oscillating function with $t\geq 0$. The same conclusion can also be made for the other particle-number preserving Hamiltonian $\widehat H_{dd}^{(1)}(t)$.
This, however, does not hold for the Hamiltonians that change the particle number. For $\widehat H_{bd}^{(1)}$,
the energy combination in the Fourier exponent is at $O(E_{\mathbf p})$ inducing large oscillations in the Fourier integral,
\begin{align}
  E_{\mathbf p}+E_{-\mathbf p\pm\mathbf q}\mp\omega_q
  &=
  E_{\mathbf p}
  +
  \left[
  E_{\mathbf p}
  \mp\mathbf v_{\mathbf p}\cdot\mathbf q
  +O\!\left(\frac{|\mathbf q|^2}{E_{\mathbf p}}\right)
  \right]
  \mp\omega_q
  \notag\\
  &=
  2E_{\mathbf p}+O(|\mathbf q|).
  \label{sqed:eq:Hbd-phase-not-soft}
\end{align}
The temporal integrals
over the Hamiltonians $\widehat H_{bb}^{(1)}(t)$ and $\widehat H_{dd}^{(1)}(t)$ therefore give the dominant contribution by producing an eikonal denominator, while the other two, $\widehat H_{bd}^{(1)}(t)$ and $\widehat H_{db}^{(1)}(t)$, are instead power suppressed as we show below,
\begin{align}
  \lim_{\varepsilon\to0^+}
  \int_0^\infty\!\dd t\,
  \e^{\pm\ii(\omega_q-\mathbf v_{\mathbf p}\cdot\mathbf q)t-\varepsilon t}
  &=
  \frac{\pm\ii}
  {\omega_q-\mathbf v_{\mathbf p}\cdot\mathbf q\pm\ii0}\, ,
  \notag\\
  \lim_{\varepsilon\to0^+}
  \int_0^\infty\!\dd t\,
  \e^{\pm\ii(2E_{\mathbf p}+O(|\mathbf q|))t-\varepsilon t}
  &=
  \frac{\pm\ii}
  {2E_{\mathbf p}+O(|\mathbf q|)\pm\ii0}\, .
  \label{sqed:eq:eikonal-versus-pair-changing-scaling}
\end{align}
The particle-number-changing terms are therefore power suppressed and hence discarded in the leading-power approximation, while both $\widehat H_{bb}^{(1)}$ and $\widehat H_{dd}^{(1)}$ are kept. Together with the subleading four-point interaction, this also excludes matter vacuum-polarization diagrams from the leading-power expansion.

We now simplify the remaining number-preserving Hamiltonians by Taylor expanding the $c$-functions of their integrands for \(|\mathbf q|/E_{\mathbf p}\ll1\). In the following formulas, we write \(\epsilon_\mu^+(q,\lambda)\equiv\epsilon_\mu^\ast(q,\lambda)\) and \(\epsilon_\mu^-(q,\lambda)\equiv\epsilon_\mu(q,\lambda)\). The $c$-functions then become
\begin{align}
  \frac{(p+p_{\mathbf p\pm\mathbf q})^\mu\epsilon_\mu^\pm(q,\lambda)}
  {2E_{\mathbf p\pm\mathbf q}}
  &=
  \frac{(p+p_{\mathbf p\pm\mathbf q})\cdot\epsilon^\pm(q,\lambda)}
  {2E_{\mathbf p}}
  \left[
  1+O\!\left(\frac{|\mathbf q|}{E_{\mathbf p}}\right)
  \right]
  \notag\\
  &=
  \frac{p\cdot\epsilon^\pm(q,\lambda)}{E_{\mathbf p}}
  +O\!\left(\frac{|\mathbf q|}{E_{\mathbf p}}\right).
  \label{sqed:eq:soft-numerator-pm}
\end{align}
Similarly, the creation-annihilation bilinear is expanded in the same soft limit,
\begin{equation}
  \widehat b_{\mathbf p}^\dagger\widehat b_{\mathbf p\pm\mathbf q}
  =
  \widehat b_{\mathbf p}^\dagger\widehat b_{\mathbf p}
  +O(\mathbf q\cdot\partial_{\mathbf p}).
  \label{sqed:eq:recoil-expansion}
\end{equation}
Substituting eqs.~\eqref{sqed:eq:soft-phase-pm}, \eqref{sqed:eq:soft-numerator-pm}, and \eqref{sqed:eq:recoil-expansion} into eq.~\eqref{sqed:eq:Hbb-exact-two-terms} gives the leading-power Hamiltonian in the \(bb\) sector,
\begin{align}
  \widehat H_{bb}^{(1)}(t)
  =
  e\sum_\lambda\int\dd\Pi(p)
  \int\dd\Phi(q)
  \frac{p^\mu}{E_{\mathbf p}}
  \widehat b_{\mathbf p}^\dagger\widehat b_{\mathbf p}
  \Big[
  &\epsilon_\mu^\ast(q,\lambda)\widehat a_{q,\lambda}^\dagger
  \e^{+\ii(\omega_q-\mathbf v_{\mathbf p}\cdot\mathbf q)t}
  \notag\\
  &+\epsilon_\mu(q,\lambda)\widehat a_{q,\lambda}
  \e^{-\ii(\omega_q-\mathbf v_{\mathbf p}\cdot\mathbf q)t}
  \Big]
  +O(q^0).
  \label{sqed:eq:Hbb-leading-soft-mode}
\end{align}
Applying the same expansion to eq.~\eqref{sqed:eq:Hdd-Hbd-Hdb-exact} gives the leading-power antiparticle number-preserving Hamiltonian, whose expression can be recovered by multiplying an overall $(-1)$ factor and replacing $\widehat b_{\mathbf p}, \widehat b_{\mathbf p}^\dagger$ with $\widehat d_{\mathbf p}$ and $\widehat d_{\mathbf p}^\dagger$, respectively.
Returning to the \(\pi^+\pi^-\) final state with momenta \(p_+\) and \(p_-\), we identify the positively (negatively) charged scalar $\pi^+$ ($\pi^-$) as the particle (antiparticle) whose dynamics is governed by $\widehat H_{bb}^{(1)}$ ($\widehat H_{dd}^{(1)}$).
Physically,
the expressions of $\widehat H_{bb}^{(1)}$ and $\widehat H_{dd}^{(1)}$ indicate that
the photon field is evaluated on the straight trajectories of \(\pi^\pm\), which reduces the mode expansion of $\widehat A_{\mu}$ into,
\begin{equation}
  \widehat A_\mu(t,\mathbf v_{\pm}t)
  =
  \int\dd^3x\,
  \delta^{(3)}(\mathbf x-\mathbf v_{\pm}t)
  \widehat A_\mu(t,\mathbf x),
  \qquad
  \mathbf v_{\pm}\equiv\frac{\mathbf p_{\pm}}{E_{\pm}},
  \qquad
  E_{\pm}\equiv E_{\mathbf p_{\pm}} .
  \label{sqed:eq:A-on-pion-trajectories}
\end{equation}
By our convention, the $bb$ ($dd$) sector inherits the momentum label $p_+$ ($p_-$),
allowing us to write,
\begin{align}
  \widehat H_{bb}^{(1)}(t)
  &=
  e\int\dd\Pi(p_+)
  \widehat b_{\mathbf p_+}^\dagger\widehat b_{\mathbf p_+}
  \frac{p_+^\mu}{E_+}
  \widehat A_\mu(t,\mathbf v_+t)
  +O(q^0)
  \notag\\
  &=
  \int\dd^3x\,
  \int\dd\Pi(p_+)
  e\frac{p_+^\mu}{E_+}
  \delta^{(3)}(\mathbf x-\mathbf v_+t)
  \widehat b_{\mathbf p_+}^\dagger\widehat b_{\mathbf p_+}
  \widehat A_\mu(t,\mathbf x)
  +O(q^0),
  \notag\\
  \widehat H_{dd}^{(1)}(t)
  &=
  -e\int\dd\Pi(p_-)
  \widehat d_{\mathbf p_-}^\dagger\widehat d_{\mathbf p_-}
  \frac{p_-^\mu}{E_-}
  \widehat A_\mu(t,\mathbf v_-t)
  +O(q^0)
  \notag\\
  &=
  -\int\dd^3x\,
  \int\dd\Pi(p_-)
  e\frac{p_-^\mu}{E_-}
  \delta^{(3)}(\mathbf x-\mathbf v_-t)
  \widehat d_{\mathbf p_-}^\dagger\widehat d_{\mathbf p_-}
  \widehat A_\mu(t,\mathbf x)
  +O(q^0).
  \label{sqed:eq:Hbb-Hdd-coordinate-eikonal}
\end{align}
The hard sector for the scattering process contains one \(b\)-particle and one \(d\)-antiparticle. We denote its momentum basis and the identity operator by
\begin{equation}
  |p_+,p_-\rangle
  =
  \widehat b_{\mathbf p_+}^\dagger
  \widehat d_{\mathbf p_-}^\dagger|0\rangle,
  \qquad
  \widehat I_{\pi\pi}
  =
  \int\dd\Pi(p_+)\dd\Pi(p_-)
  |p_+,p_-\rangle\langle p_+,p_-|.
  \label{sqed:eq:pion-pair-identity}
\end{equation}
The commutation relations in eq.~\eqref{sqed:eq:momentum-measure} imply that the action of the particle-number operators $\widehat N_{bb}$ and $\widehat N_{dd}$ on the momentum eigenstates of the hard particles takes the form,
\begin{align}
  \widehat N_{bb} |p_+,p_-\rangle &\equiv \int\dd\Pi(k)\,f(\mathbf k)
  \widehat b_{\mathbf k}^\dagger\widehat b_{\mathbf k}
  |p_+,p_-\rangle
  =f(\mathbf p_+)|p_+,p_-\rangle,
  \notag\\
  \widehat N_{dd} |p_+,p_-\rangle &\equiv
  \int\dd\Pi(k)\,f(\mathbf k)
  \widehat d_{\mathbf k}^\dagger\widehat d_{\mathbf k}
  |p_+,p_-\rangle
=
  f(\mathbf p_-)|p_+,p_-\rangle.
  \label{sqed:eq:number-operators-on-pion-pair}
\end{align}
Applying eq.~\eqref{sqed:eq:number-operators-on-pion-pair} to
eq.~\eqref{sqed:eq:Hbb-Hdd-coordinate-eikonal}, and hereafter restricting
the hard operator to the \(\pi^+\pi^-\) sector, gives
\begin{align}
  \widehat H^{(1)}(t)
  &\equiv
  \left[
  \widehat H_{bb}^{(1)}(t)+\widehat H_{dd}^{(1)}(t)
  \right]\widehat I_{\pi\pi}
  \notag\\
  &=
  \int\dd^3x
  \int\dd\Pi(p)
  |p\rangle\langle p|
  \otimes
  \left[
  e\frac{p_+^\mu}{E_+}\delta^{(3)}(\mathbf x-\mathbf v_+t)
  -e\frac{p_-^\mu}{E_-}\delta^{(3)}(\mathbf x-\mathbf v_-t)
  \right]
  \widehat A_\mu(t,\mathbf x)
  \notag\\
  &\equiv
  \int\dd^3x
  \sum_p
  |p\rangle\langle p|\otimes
  J_p^\mu(t,\mathbf x)\widehat A_\mu(t,\mathbf x)\, ,
  \label{sqed:eq:eikonal-H-projector-form}
\end{align}
where, from here on, \(p\equiv(p_+,p_-)\), \(|p\rangle\equiv |p_+,p_-\rangle\), and
\(\sum_p\equiv\int\dd\Pi(p)\). The pion-pair current selected by the hard momentum state \(p\) is
\begin{equation}
  J_p^\mu(t,\mathbf x)
  \equiv
  e\frac{p_+^\mu}{E_+}
  \delta^{(3)}(\mathbf x-\mathbf v_+t)
  -
  e\frac{p_-^\mu}{E_-}
  \delta^{(3)}(\mathbf x-\mathbf v_-t)\,,
  \label{sqed:eq:branch-current}
\end{equation}
which can also be formulated as a distribution on the
worldlines of the two hard scalars. Let
\begin{equation}
  z_{p_\pm}^\mu(s)\equiv p_\pm^\mu s,\qquad s\ge0,
  \label{sqed:eq:eikonal-trajectory}
\end{equation}
so that \(z_{p_\pm}^0(s)=E_\pm s\) and
\(\mathbf z_{p_\pm}(s)=\mathbf p_\pm s\). Then
\begin{equation}
  J_p^\mu(x)
  =
  e\int_0^\infty\dd s\,
  p_+^\mu\delta^{(4)}(x-z_{p_+}(s))
  -
  e\int_0^\infty\dd s\,
  p_-^\mu\delta^{(4)}(x-z_{p_-}(s))\, ,
  \label{sqed:eq:branch-current-worldline}
\end{equation}
gives us eq.~\eqref{eq:eikonal-EM-current}.
Indeed, each line separately yields
\begin{equation}
  \int_0^\infty\dd s\,
  p_\pm^\mu
  \delta(x^0-E_\pm s)
  \delta^{(3)}(\mathbf x-\mathbf p_\pm s)
  =
  \frac{p_\pm^\mu}{E_\pm}
  \delta^{(3)}(\mathbf x-\mathbf v_\pm x^0),
  \label{sqed:eq:worldline-current-localization}
\end{equation}
and setting \(x^0=t\) reproduces eq.~\eqref{sqed:eq:branch-current}.

The Fourier transform of the
current~\eqref{sqed:eq:branch-current-worldline} is
\begin{align}
  \int\dd^4x\,\e^{\ii q\cdot x}J_p^\mu(x)
  &=
  e\,p_+^\mu\int_0^\infty\dd s\,\e^{\ii s\,p_+\cdot q}
  -e\,p_-^\mu\int_0^\infty\dd s\,\e^{\ii s\,p_-\cdot q}
  \notag\\
  &=
  \frac{\ii\,e\,p_+^\mu}{p_+\cdot q+\ii0}
  -\frac{\ii\,e\,p_-^\mu}{p_-\cdot q+\ii0}
  \notag\\
  &=
  \ii\,J_p^\mu(q),
  \label{sqed:eq:current-fourier}
\end{align}
where the \(\ii0\) comes from the convergence factor \(\e^{-\varepsilon s}\). Finally, the \(S\)-matrix describing the final-state soft LP emission can therefore be written as,
\begin{align}
  \widehat S_{\rm soft}
  &=
  \mathcal T\exp\left[-\ii\int\dd t\,\widehat H^{(1)}(t)\right]
  \notag\\
  &=
  \mathcal T\exp\left[-\ii\int\dd^4x\sum_p
  |p\rangle\langle p|\otimes J_p^\mu(x)\widehat A_\mu(x)\right]\, ,
  \label{eq:widehatS}
\end{align}
where the second line follows from eq.~\eqref{sqed:eq:eikonal-H-projector-form} together with the worldline current~\eqref{sqed:eq:branch-current-worldline}, and the result is identical to eq.~\eqref{wl:eq:Ssoft-before-factorization}.

\section{Path ordering for Abelian Wilson lines}
\label{wl:app:path-ordering-commutator}

In this appendix we justify the step used in
eq.~\eqref{wl:eq:Wilson-line-soft-factor-form}, where the path-ordered exponential
along the eikonal trajectory is replaced by an ordinary exponential, i.e., without path-ordering, with an additional $c$-valued phase from the commutator.
Denote the unresolved photon resolution window by \((\xi_2,\xi_1)\)
and define,
\begin{equation}
  \widehat B(s)
  \equiv
  -\ii e\,p^\mu
  \widehat A_\mu^{(\xi_2,\xi_1)}(z_p(s)).
  \label{wl:eq:appendix-B-definition}
\end{equation}
The claim is that, when \([\widehat B(s),\widehat B(s')]\) is a $c$-number distribution,
\begin{align}
  \mathcal P
  \exp\!\left[
    \int_0^\infty\dd s\,\widehat B(s)
  \right]
  &=
  \exp\!\left[
    \widehat X+Y
  \right],
  \label{wl:eq:path-ordering-magnus-claim}
\end{align}
where the path-ordered exponential on the left-hand side is exactly the Wilson-line operator $\widehat W_p(\xi_2,\xi_1)$ in eq.~\eqref{wl:eq:Ssoft-wilson-factorization} with $\widehat A_\mu(z)$ constrained along the path of the hard scalars, and
\begin{align}
  \widehat X
  &\equiv
  \int_0^\infty\dd s\,\widehat B(s),
  \notag\\
  Y
  &\equiv
  \frac12
  \int_0^\infty\dd s
  \int_0^\infty\dd s'\,
  \Theta(s-s')\,[\widehat B(s),\widehat B(s')].
  \label{wl:eq:XY-definition}
\end{align}
For QED this condition is satisfied because the commutator of two photon field
operators is indeed a $c$-number distribution.  Thus \(Y\) commutes with \(\widehat X\), and all nested commutators such as \([\widehat B(s_1),[\widehat B(s_2),\widehat B(s_3)]]\) vanish.

We now check eq.~\eqref{wl:eq:path-ordering-magnus-claim} explicitly through third
order in \(\widehat B\).  With \(\widehat B_i\equiv \widehat B(s_i)\) and
\(\Theta_{ij}\equiv\Theta(s_i-s_j)\), the left-hand side expands as
\begin{align}
  \mathcal P
  \exp\!\left[
    \int_0^\infty\dd s\,\widehat B(s)
  \right]
  &=
  1
  +
  \int_0^\infty\dd s_1\,\widehat B_1
  +
  \int_0^\infty\dd s_1
  \int_0^\infty\dd s_2\,
  \Theta_{12}\,\widehat B_1\widehat B_2
  \notag\\
  &+
  \int_0^\infty\dd s_1
  \int_0^\infty\dd s_2
  \int_0^\infty\dd s_3\,
  \Theta_{12}\Theta_{23}\,\widehat B_1\widehat B_2\widehat B_3
  +
  O(\widehat B^4).
  \label{wl:eq:path-ordered-expansion-third-order}
\end{align}
The right-hand side of eq.~\eqref{wl:eq:path-ordering-magnus-claim} expands as an ordinary exponential.  Since \(Y=O(\widehat B^2)\)
and \(Y\) commutes with \(\widehat X\), terms through third order are
\begin{align}
  \exp(\widehat X+Y)
  &=
  1+\widehat X
  +
  \left(
    \frac12 \widehat X^2+Y
  \right)
  +
  \left(
    \frac16 \widehat X^3+\widehat XY
  \right)
  +
  O(\widehat B^4)\, ,
  \label{wl:eq:ordinary-expansion-third-order}
\end{align}
where we follow the counting $Y \sim \widehat B^2\sim \widehat X^2$ by its definition~\eqref{wl:eq:XY-definition}.
The first-order terms clearly match.  At second order, write
$\displaystyle\int_{12}\equiv\int_0^\infty\dd s_1 \dd s_2$.  Then,
\begin{align}
  \frac12 \widehat X^2+Y
  &=
  \frac12\int_{12}
  \widehat B_1\widehat B_2
  +
  \frac12\int_{12}
  \Theta_{12}\,[\widehat B_1,\widehat B_2]
  \notag\\
  &=
  \frac12\int_{12}
  (\Theta_{12}+\Theta_{21})\widehat B_1\widehat B_2
  +
  \frac12\int_{12}
  \Theta_{12}\widehat B_1\widehat B_2
  -
  \frac12\int_{12}
  \Theta_{12}\widehat B_2\widehat B_1
  \notag\\
  &=
  \frac12\int_{12}
  (\Theta_{12}+\Theta_{21})\widehat B_1\widehat B_2
  +
  \frac12\int_{12}
  \Theta_{12}\widehat B_1\widehat B_2
  -
  \frac12\int_{12}
  \Theta_{21}\widehat B_1\widehat B_2
  \notag\\
  &=
  \int_{12}
  \Theta_{12}\,\widehat B_1\widehat B_2\, ,
  \label{wl:eq:second-order-match}
\end{align}
which reproduces the second-order term in eq.~\eqref{wl:eq:path-ordered-expansion-third-order}.

The same derivation applies to the third order.
Adopt the short-hand notation,
$\displaystyle\int_{123}\equiv \int_0^\infty\dd s_1\int_0^\infty\dd s_2\int_0^\infty\dd s_3$ and
substituting the definitions of \(\widehat X\) and \(Y\) in eq.~\eqref{wl:eq:XY-definition}, we find,
\begin{align}
  \frac16\widehat X^3+\widehat XY
  &=
  \frac16\int_{123}\widehat B_1\widehat B_2\widehat B_3
  +
  \frac12\int_{123}\Theta_{23}\widehat B_1[\widehat B_2,\widehat B_3]
  \notag\\
  &=
  \frac16\int_{123}\Theta_{12}\Theta_{23}
  \left(
    \widehat B_1\widehat B_2\widehat B_3+\widehat B_1\widehat B_3\widehat B_2+\widehat B_2\widehat B_1\widehat B_3
  \right)
  \notag\\
  &+
  \frac16\int_{123}\Theta_{12}\Theta_{23}
  \left(
    \widehat B_3\widehat B_1\widehat B_2+\widehat B_2\widehat B_3\widehat B_1+\widehat B_3\widehat B_2\widehat B_1
  \right)
  \notag\\
  &+
  \frac12\int_{123}\Theta_{12}\Theta_{23}
  \left(
    \widehat B_1[\widehat B_2,\widehat B_3]+\widehat B_2[\widehat B_1,\widehat B_3]+\widehat B_3[\widehat B_1,\widehat B_2]
  \right)
  \notag\\
  &=
  \int_{123}\Theta_{12}\Theta_{23}\widehat B_1\widehat B_2\widehat B_3
  -
  \frac12\int_{123}\Theta_{12}\Theta_{23}
  \left(
    \widehat B_1[\widehat B_2,\widehat B_3]+\widehat B_2[\widehat B_1,\widehat B_3]+\widehat B_3[\widehat B_1,\widehat B_2]
  \right)
  \notag\\
  &+
  \frac12\int_{123}\Theta_{12}\Theta_{23}
  \left(
    \widehat B_1[\widehat B_2,\widehat B_3]+\widehat B_2[\widehat B_1,\widehat B_3]+\widehat B_3[\widehat B_1,\widehat B_2]
  \right)
  \notag\\
  &=
  \int_{123}
  \Theta_{12}\Theta_{23}\,
  \widehat B_1\widehat B_2\widehat B_3\, ,
  \label{wl:eq:third-order-match}
\end{align}
which is exactly the third-order path-ordered term in
eq.~\eqref{wl:eq:path-ordered-expansion-third-order}.  Thus the two exponentials agree
through third order. In the derivation we have used,
\begin{align}
  1
  &=
  \Theta_{12}\Theta_{23}
  +\Theta_{13}\Theta_{32}
  +\Theta_{21}\Theta_{13}
  +\Theta_{23}\Theta_{31}
  +\Theta_{31}\Theta_{12}
  +\Theta_{32}\Theta_{21},\label{wl:eq:third-order-theta-partitions}
  \\
  \Theta_{23}
  &
 =
  \Theta_{23}(\Theta_{12}+\Theta_{21})
=
  \Theta_{12}\Theta_{23}
  +\Theta_{21}\Theta_{23}(\Theta_{13}+\Theta_{31})
=
  \Theta_{12}\Theta_{23}
  +\Theta_{21}\Theta_{13}
  +\Theta_{23}\Theta_{31}\, ,
  \notag
\end{align}
and relabeled the dummy indices as in the third line of eq.~\eqref{wl:eq:second-order-match}.

There are two ways to promote these checks to an all-order proof.  The
first one is closest to the physical definition of a Wilson line.  It is sufficient to consider
the integration with a finite upper limit \(S\), divide the interval into
\(N\) ordered subintervals, and write \(\widehat B_i\equiv \widehat B(s_i)\) with
\(s_N>\cdots>s_2>s_1\).  The path-ordered exponential is then the continuum
limit of the ordered product
\begin{align}
  \mathcal P\exp\left[\int_0^S\dd s\,\widehat B(s)\right]
  &=
  \lim_{N\to\infty}
  \exp(\widehat B_N\Delta s)\cdots\exp(\widehat B_2\Delta s)\exp(\widehat B_1\Delta s).
  \label{wl:eq:discrete-path-ordered-product}
\end{align}
For two adjacent factors, the Baker--Campbell--Hausdorff formula gives
\begin{align}
  \e^{\widehat A}\e^{\widehat B}
  &=
  \exp\left(\widehat A+\widehat B+\frac12[\widehat A,\widehat B]\right),
  \label{wl:eq:bch-cnumber-truncation}
\end{align}
if and only if \([\widehat A,\widehat B]\) is a $c$-number, ensuring all nested commutators of higher orders such as \([\widehat A,[\widehat A,\widehat B]]\) and \([\widehat B,[\widehat A,\widehat B]]\) vanish.  Applying
eq.~\eqref{wl:eq:bch-cnumber-truncation} repeatedly to
eq.~\eqref{wl:eq:discrete-path-ordered-product}, with \(\widehat A_i=\widehat B_i\Delta s\), gives
\begin{align}
  \exp(\widehat B_N\Delta s)\cdots\exp(\widehat B_1\Delta s)
  &=
  \exp\left[
    \sum_{i=1}^N \widehat B_i\Delta s
    +\frac12\sum_{i>j}[\widehat B_i,\widehat B_j](\Delta s)^2
  \right].
  \label{wl:eq:discrete-bch-result}
\end{align}
The first term in the exponent becomes \(\displaystyle\int_0^S \dd s\,\widehat B(s)\).  The second
term becomes the ordered double integral because \(i>j\) means \(s_i>s_j\):
\begin{align}
  \frac12\sum_{i>j}[\widehat B_i,\widehat B_j](\Delta s)^2
  &\longrightarrow
  \frac12\int_0^S\dd s\int_0^S\dd s'\,
  \Theta(s-s')\,[\widehat B(s),\widehat B(s')].
  \label{wl:eq:discrete-commutator-continuum-limit}
\end{align}
Thus the discretized Wilson line gives the all-order identity
\begin{align}
  \mathcal P\exp\left[\int_0^S\dd s\,\widehat B(s)\right]
  &=
  \exp\left[
    \int_0^S\dd s\,\widehat B(s)
    +\frac12\int_0^S\dd s\int_0^S\dd s'\,
    \Theta(s-s')\,[\widehat B(s),\widehat B(s')]
  \right],
  \label{wl:eq:bch-all-order-path-ordering-removal}
\end{align}
and the formula used in the Wilson line follows by taking \(S\to\infty\).

The second method directly uses the Magnus expansion.  Define
\begin{align}
  \widehat U(S)
  &\equiv
  \mathcal P\exp\left[\int_0^S\dd s\,\widehat B(s)\right].
  \label{wl:eq:finite-path-ordered-exponential}
\end{align}
By definition, \(\widehat U(S)\) is the solution of the first-order operator equation
\begin{align}
  \frac{\dd \widehat U(S)}{\dd S}
  &=
  \widehat B(S)\widehat U(S),
  &
  \widehat U(0)&=1.
  \label{wl:eq:path-ordered-initial-value-problem}
\end{align}
The Magnus theorem states that the solution can be written exactly in an exponential form,
\begin{align}
  \widehat U(S)
  &=
  \e^{\widehat\Omega_{\rm M}(S)},
  \notag\\
  \widehat\Omega_{\rm M}(S)
  &\equiv
  \widehat\Omega_1(S)+\widehat\Omega_2(S)+\widehat\Omega_3(S)+\cdots ,
  \label{wl:eq:magnus-series-definition}
\end{align}
where the first terms are
\begin{align}
  \widehat\Omega_1(S)
  &=
  \int_0^S\dd s_1\,\widehat B(s_1),
  \notag\\
  \widehat\Omega_2(S)
  &=
  \frac12\int_0^S\dd s_1\int_0^{s_1}\dd s_2\,
  [\widehat B(s_1),\widehat B(s_2)],
  \notag\\
  \widehat\Omega_3(S)
  &=
  \frac16\int_0^S\dd s_1\int_0^{s_1}\dd s_2\int_0^{s_2}\dd s_3
  \left(
    [\widehat B(s_1),[\widehat B(s_2),\widehat B(s_3)]]
    +[\widehat B(s_3),[\widehat B(s_2),\widehat B(s_1)]]
  \right)\,,
  \notag\\
  &\cdots
  \label{wl:eq:first-magnus-terms}
\end{align}
All higher terms in the Magnus expansion are built from higher nested commutators.  In the
present Abelian Wilson-line problem, \([\widehat B(s),\widehat B(s')]\) is a $c$-number
distribution, hence
\begin{align}
  [\widehat B(s_1),[\widehat B(s_2),\widehat B(s_3)]]
  &=
  0,
  &
  \widehat\Omega_n(S)&=0\quad(n\ge3).
  \label{wl:eq:magnus-nested-commutators-vanish}
\end{align}
Therefore the Magnus series truncates exactly after \(\widehat\Omega_2\):
\begin{align}
  \widehat\Omega_{\rm M}(S)
  &=
  \int_0^S\dd s\,\widehat B(s)
  +\frac12\int_0^S\dd s\int_0^s\dd s'\,[\widehat B(s),\widehat B(s')]
  \notag\\
  &=
  \int_0^S\dd s\,\widehat B(s)
  +\frac12\int_0^S\dd s\int_0^S\dd s'\,
  \Theta(s-s')\,[\widehat B(s),\widehat B(s')]\, .
  \label{wl:eq:magnus-truncated-omega}
\end{align}
This is exactly the exponent in
eq.~\eqref{wl:eq:bch-all-order-path-ordering-removal}.  For completeness, we can also
check directly that the truncated Magnus exponent solves the same
initial-value problem.  Let \(\widehat V(S)\equiv \e^{\widehat\Omega_{\rm M}(S)}\) and consider $\left(\frac{\dd\widehat V(S)}{\dd S}\right)\widehat V^{-1}(S)$, we have,
\begin{align}
  \left(\frac{\dd}{\dd S}\e^{\widehat\Omega_{\rm M}(S)}\right)\e^{-\widehat\Omega_{\rm M}(S)}
  &=
  \frac{\dd\widehat\Omega_{\rm M}(S)}{\dd S}
  +\frac12\left[\widehat\Omega_{\rm M}(S),\frac{\dd\widehat\Omega_{\rm M}(S)}{\dd S}\right]\,,
  \label{wl:eq:exponential-derivative-truncated}
\end{align}
which can be easily verified by expanding both sides and noticing
\begin{align}
\frac{\dd}{\dd S}\widehat\Omega_{\rm M}^n(S)= \sum_{k=0}^{n-1} \widehat\Omega_{\rm M}^k(S)\frac{\dd\widehat\Omega_{\rm M}(S)}{\dd S}\widehat\Omega_{\rm M}^{n-k-1}(S) \, .%
\end{align}
The two terms on the right-hand side of eq.~\eqref{wl:eq:exponential-derivative-truncated} are
\begin{align}
  \frac{\dd\widehat\Omega_{\rm M}(S)}{\dd S}
  &=
  \widehat B(S)
  +\frac12\int_0^S\dd s\,[\widehat B(S),\widehat B(s)],
  \notag\\
  \left[\widehat\Omega_{\rm M}(S),\frac{\dd\widehat\Omega_{\rm M}(S)}{\dd S}\right]
  &=
  \int_0^S\dd s\,[\widehat B(s),\widehat B(S)]
=
  -\int_0^S\dd s\,[\widehat B(S),\widehat B(s)].
  \label{wl:eq:omega-derivative-and-commutator}
\end{align}
Substituting eq.~\eqref{wl:eq:omega-derivative-and-commutator} into
eq.~\eqref{wl:eq:exponential-derivative-truncated} gives
\begin{align}
  \left(\frac{\dd}{\dd S}\e^{\widehat\Omega_{\rm M}(S)}\right)\e^{-\widehat\Omega_{\rm M}(S)}
  &=
  \widehat B(S)
  +\frac12\int_0^S\dd s\,[\widehat B(S),\widehat B(s)]
  -\frac12\int_0^S\dd s\,[\widehat B(S),\widehat B(s)]
  =
  \widehat B(S).
  \label{wl:eq:magnus-initial-value-check}
\end{align}
Therefore \(\dd \widehat V(S)/\dd S=\widehat B(S)\widehat V(S)\), and \(\widehat V(0)=1\).  By uniqueness of the
solution to the first-order differential (operator) equation with a definite initial-value, \(\widehat V(S)=\widehat U(S)\), which is exactly
eq.~\eqref{wl:eq:bch-all-order-path-ordering-removal}.  Therefore the path ordering
in eq.~\eqref{wl:eq:Wilson-line-soft-factor-form} may be replaced by an ordinary
exponential containing the commutator phase.

\section{Mass renormalization of the Wilson line}
\label{app:wilson-matrix-element}

In this appendix, we show that the vacuum matrix element of the Wilson line
reproduces the LSZ result~\eqref{pion:eq:virtual-eikonal-current} once the
mass counterterm is included.  Since the path ordering along the outgoing worldlines coincides with time
ordering, the vacuum matrix element of the unresolved Wilson line \(\widehat W_p(\xi,0)\) of
eq.~\eqref{wl:eq:unresolved-wilson-line} follows from its time-ordered form.  Wick's theorem
then gives in Feynman gauge,
\begin{align}
  \langle0|\widehat W_p(\xi,0)|0\rangle
  &=
  \langle0|\,\mathcal T\exp\Big[-\ii\int\dd^4x\,J_p^\mu(x)\widehat A_\mu^{(\xi,0)}(x)\Big]|0\rangle
  \notag\\
  &=
  \exp\Big[-\frac12\int\dd^4x\,\dd^4y\,J_p^\mu(x)\,
  \langle0|\mathcal T\widehat A_\mu^{(\xi,0)}(x)\widehat A_\nu^{(\xi,0)}(y)|0\rangle\,J_p^\nu(y)\Big]
  \notag\\
  &=
  \exp\Big[-\frac12\int\dd^4x\,\dd^4y\,J_p^\mu(x)J_p^\nu(y)
  \int_{q,\xi}\e^{-\ii q\cdot(x-y)}\,\frac{-\ii\,g_{\mu\nu}}{q^2+\ii0}\Big]
  \notag\\
  &=
  \exp\Big[\frac12\int_{q,\xi}\frac{-\ii}{q^2+\ii0}\,
  J_p(q)\cdot J_p(-q)\Big]\, ,
  \label{sqed:eq:wilson-vev-wick}
\end{align}
which can be rewritten as,
\begin{align}
  &\langle0|\widehat W_p(\xi,0)|0\rangle
  \notag\\
  &=
  \lim_{\varepsilon\to0^+}\exp\bigg[
  \frac12\int_{q,\xi}\frac{-\ii}{q^2+\ii0}\,
  \frac{e^2p_+^2}{(p_+\cdot q+\ii\varepsilon)(-p_+\cdot q
  \mathbin{\textcolor{red}{+}}\textcolor{red}{\ii\varepsilon})}
  \notag\\
  &\qquad
  +\frac12\int_{q,\xi}\frac{-\ii}{q^2+\ii0}\,
  \frac{e^2p_-^2}{(p_-\cdot q+\ii\varepsilon)(-p_-\cdot q
  \mathbin{\textcolor{red}{+}}\textcolor{red}{\ii\varepsilon})}
  \notag\\
  &\qquad
  -\frac12\int_{q,\xi}\frac{-\ii}{q^2+\ii0}
  \bigg(\frac{e^2p_+\cdot p_-}{(p_+\cdot q+\ii\varepsilon)(-p_-\cdot q+\ii\varepsilon)}
  +\frac{e^2p_-\cdot p_+}{(p_-\cdot q+\ii\varepsilon)(-p_+\cdot q+\ii\varepsilon)}\bigg)
  \bigg]\, .
  \label{sqed:eq:wilson-vev-explicit}
\end{align}
The above expression differs from the YFS/BN exponentiation of the LSZ result~\eqref{pion:eq:virtual-eikonal-current} only
in the sign of the \(\ii0\) prescription marked in red.  By the Sokhotski--Plemelj formula, this
difference is a pure phase.  It arises because the LSZ reduction in
Section~\ref{pion:sec:reduced-density-soft-exponentiation} uses the physical
mass \(m\), while the Wilson line is built from
eq.~\eqref{sqed:eq:sqed-lagrangian} with the bare mass \(m_0\),
which is related to its physical counterpart $m$ by
\begin{equation}
  -m_0^2\phi^\dagger\phi
  =
  -m^2\phi^\dagger\phi
  +\delta m^2\phi^\dagger\phi\, ,
  \label{sqed:eq:mass-counterterm}
\end{equation}
where the counterterm \(\delta m^2\phi^\dagger\phi\) results in an
additional term in the interaction Hamiltonian,
\begin{equation}
  \widehat H_{\delta m}(t)
  \equiv
  -\delta m^2\int\dd^3x\,
  \widehat\phi^\dagger(t,\mathbf x)\widehat\phi(t,\mathbf x)\, .
  \label{sqed:eq:counterterm-hamiltonian-def}
\end{equation}
Following the same leading-power simplification procedure as in
Appendix~\ref{app:scalarQED}, we have
\begin{align}
  \widehat H_{\delta m}(t)\,\widehat I_{\pi\pi}
  &=
  -\delta m^2\int\dd^3x\int\dd\Pi(k)\,\dd\Pi(k')
  \Big[\widehat b_{\mathbf k}^\dagger\widehat b_{\mathbf k'}\,\e^{\ii(k-k')\cdot x}
  +\widehat d_{\mathbf k'}^\dagger\widehat d_{\mathbf k}\,\e^{-\ii(k-k')\cdot x}\Big]
  \widehat I_{\pi\pi}
  \notag\\
  &=
  -\delta m^2\int\dd\Pi(k)\,\frac{1}{2E_{\mathbf k}}
  \Big[\widehat b_{\mathbf k}^\dagger\widehat b_{\mathbf k}
  +\widehat d_{\mathbf k}^\dagger\widehat d_{\mathbf k}\Big]\widehat I_{\pi\pi}
  \notag\\
  &=
  -\sum_p|p\rangle\langle p|
  \Big(\frac{\delta m^2}{2E_+}+\frac{\delta m^2}{2E_-}\Big)\, ,
  \label{sqed:eq:counterterm-hamiltonian}
\end{align}
where the last line follows from eq.~\eqref{sqed:eq:number-operators-on-pion-pair}.  Including this term, the renormalized Wilson line reads
\begin{equation}
  \widehat W_p^{\rm ren}(\xi,0)
  \equiv
  \lim_{\varepsilon\to0^+}\mathcal T\exp\Big[-\ii\int_0^\infty\dd t\Big(
  \int\dd^3x\,J_p^\mu(t,\mathbf x)\widehat A_\mu^{(\xi,0)}(t,\mathbf x)
  -\sum_\pm\frac{\delta m^2}{2E_\pm}\,\e^{-2\varepsilon t/E_\pm}\Big)\Big]\, ,
  \label{sqed:eq:wilson-line-renormalized}
\end{equation}
and its vacuum matrix element is
\begin{align}
  &\langle0|\widehat W_p^{\rm ren}(\xi,0)|0\rangle
  \notag\\
  &=
  \lim_{\varepsilon\to0^+}\langle0|\,\mathcal T\exp\Big[-\ii\int_0^\infty\dd t\Big(
  \int\dd^3x\,J_p^\mu(t,\mathbf x)\widehat A_\mu^{(\xi,0)}(t,\mathbf x)
  -\sum_\pm\frac{\delta m^2}{2E_\pm}\,\e^{-2\varepsilon t/E_\pm}\Big)\Big]|0\rangle
  \notag\\
  &=
  \lim_{\varepsilon\to0^+}\exp\Big[\sum_\pm\frac{\ii\,\delta m^2}{4\varepsilon}\Big]\,
  \langle0|\widehat W_p(\xi,0)|0\rangle
  \notag\\
  &=
  \lim_{\varepsilon\to0^+}\exp\Big[\sum_\pm\frac{\ii}{4\varepsilon}\Big(-4\ii e^2\int_{q,\xi}
  \frac{m^2}{(q^2+\ii0)(2p_\pm\cdot q+\ii0)}\Big)\Big]\,
  \langle0|\widehat W_p(\xi,0)|0\rangle
  \notag\\
  &=
  \lim_{\varepsilon\to0^+}\exp\Big[\sum_\pm\frac{e^2m^2}{2\varepsilon}\int_{q,\xi}\frac{1}{q^2+\ii0}
  \!\Big(\mathrm{PV}\frac{1}{p_\pm\cdot q}-\ii\pi\delta(p_\pm\cdot q)\Big)
  +\frac12\int_{q,\xi}\frac{-\ii}{q^2+\ii0}J_p(q)\cdot J_p(-q)\Big]
  \notag\\
  &=
  \lim_{\varepsilon\to0^+}\exp\bigg[
  \frac12\int_{q,\xi}\frac{-\ii}{q^2+\ii0}\,e^2p_+^2
  \Big(\frac{1}{(p_+\cdot q+\ii\varepsilon)(-p_+\cdot q\mathbin{\textcolor{red}{+}}\textcolor{red}{\ii\varepsilon})}
  +\frac{\pi}{\varepsilon}\,\delta(p_+\cdot q)\Big)
  \notag\\
  &\qquad
  +\frac12\int_{q,\xi}\frac{-\ii}{q^2+\ii0}\,e^2p_-^2
  \Big(\frac{1}{(p_-\cdot q+\ii\varepsilon)(-p_-\cdot q\mathbin{\textcolor{red}{+}}\textcolor{red}{\ii\varepsilon})}
  +\frac{\pi}{\varepsilon}\,\delta(p_-\cdot q)\Big)
  \notag\\
  &\qquad
  -\frac12\int_{q,\xi}\frac{-\ii}{q^2+\ii0}
  \bigg(\frac{e^2p_+\cdot p_-}{(p_+\cdot q+\ii\varepsilon)(-p_-\cdot q+\ii\varepsilon)}
  +\frac{e^2p_-\cdot p_+}{(p_-\cdot q+\ii\varepsilon)(-p_+\cdot q+\ii\varepsilon)}\bigg)
  \bigg]
  \notag\\
  &=
  \exp\bigg[
  \frac12\int_{q,\xi}\frac{-\ii}{q^2+\ii0}\,
  \frac{e^2p_+^2}{(p_+\cdot q+\ii0)(-p_+\cdot q\mathbin{\textcolor{blue}{-}}\textcolor{blue}{\ii0})}
  \notag\\
  &\qquad
  +\frac12\int_{q,\xi}\frac{-\ii}{q^2+\ii0}\,
  \frac{e^2p_-^2}{(p_-\cdot q+\ii0)(-p_-\cdot q\mathbin{\textcolor{blue}{-}}\textcolor{blue}{\ii0})}
  \notag\\
  &\qquad
  -\frac12\int_{q,\xi}\frac{-\ii}{q^2+\ii0}
  \bigg(\frac{e^2p_+\cdot p_-}{(p_+\cdot q+\ii0)(-p_-\cdot q+\ii0)}
  +\frac{e^2p_-\cdot p_+}{(p_-\cdot q+\ii0)(-p_+\cdot q+\ii0)}\bigg)
  \bigg]
  \notag\\
  &=
  \exp\bigg[
  \frac12\int_{q,\xi}\frac{\ii}{q^2+\ii0}
  \bigg(\frac{e\,p_+}{p_+\cdot q+\ii0}-\frac{e\,p_-}{p_-\cdot q-\ii0}\bigg)^2\bigg]\, ,
  \label{sqed:eq:wilson-vev-renormalized}
\end{align}
where the PV-terms do not contribute because of the symmetric $q$-integral. Also, in the second-to-last line we have used,
\begin{equation}
  \frac{1}{(p_\pm\cdot q+\ii\varepsilon)(-p_\pm\cdot q\mathbin{\textcolor{red}{+}}\textcolor{red}{\ii\varepsilon})}
  =
  \frac{1}{(p_\pm\cdot q+\ii\varepsilon)(-p_\pm\cdot q\mathbin{\textcolor{blue}{-}}\textcolor{blue}{\ii\varepsilon})}
  -\frac{\pi}{\varepsilon}\,\delta(p_\pm\cdot q)\, .
  \label{sqed:eq:pinch-difference}
\end{equation}
We have thus reproduced the LSZ result~\eqref{pion:eq:virtual-eikonal-current},
and the Wilson line \(\widehat W_p(\xi,0)\) in the main text should be understood as
\(\widehat W_p^{\rm ren}(\xi,0)\).

\section{Gauge invariance}
\label{pion:app:gauge-invariance}

This appendix verifies the gauge invariance of the resolution-dependent soft factor $\langle X|\widehat W_p|0\rangle$
multiplying the hard amplitude, and of the coherence factor it induces on the hard density matrix.  We distinguish two boundary
conditions.  For gauge transformations that vanish at the asymptotic
boundary, current conservation makes the Wilson line induced by the hard pion-pair invariant.  For large gauge transformations with nonzero endpoint values, the unresolved Wilson line remains invariant, because the non-vanishing phase induced by a large gauge transformation is due to zero-energy photons, which we include in the infrared dressing of the pion pair, and the dressed neutral pion pair is itself invariant under large gauge transformations~\cite{Gabai:2016kuf}.  The hard density matrix \(\rho_{pp'}(0)\) is generated by
the complete scalar-QED hard \(S\)-matrix and therefore satisfies the
corresponding hard Ward identities.  In either case, the same conserved
pion-pair eikonal current removes the photon's longitudinal degree of freedom.

We first return to the eikonal current in
eq.~\eqref{sqed:eq:branch-current-worldline}.  Allowing the two outgoing
worldlines to begin at a common production point \(x_0\), the current selected
by the pion momentum configuration \(p=(p_+,p_-)\) with $p_\pm = (\sqrt{m^2+\mathbf p^2},\pm\mathbf p)$ is
\begin{equation}
  J_p^\mu(x)
  =
  e\int_0^\infty\dd s\,p_+^\mu
  \delta^{(4)}(x-x_0-p_+s)
  -
  e\int_0^\infty\dd s\,p_-^\mu
  \delta^{(4)}(x-x_0-p_-s)\, ,
  \label{gauge:eq:pair-current-with-production-point}
\end{equation}
where the two currents share the endpoint \(x_0\), at which the two pion currents cancel each other.
For either trajectory,
\begin{equation}
  p_\pm^\mu\partial_\mu
  \delta^{(4)}(x-x_0-p_\pm s)
  =
  -\frac{\dd}{\dd s}
  \delta^{(4)}(x-x_0-p_\pm s).
  \label{gauge:eq:worldline-delta-derivative}
\end{equation}
Using this identity in eq.~\eqref{gauge:eq:pair-current-with-production-point}, we find,
\begin{align}
  \partial_\mu J_p^\mu(x)
  &=
  -e\int_0^\infty\dd s\,
  \frac{\dd}{\dd s}
  \delta^{(4)}(x-x_0-p_+s)
  +e\int_0^\infty\dd s\,
  \frac{\dd}{\dd s}
  \delta^{(4)}(x-x_0-p_-s)
   =0.
  \label{gauge:eq:pair-current-conservation}
\end{align}
With the convention \(D_\mu=\partial_\mu+\ii eA_\mu\) used in
Appendix~\ref{app:scalarQED}, the Abelian gauge
transformation acts as
\begin{equation}
  \widehat A_\mu(x)\longrightarrow
  \widehat A_\mu(x)+\partial_\mu\alpha(x),
  \qquad
  \widehat\phi(x)\longrightarrow
  \e^{-\ii e\alpha(x)}\widehat\phi(x).
  \label{gauge:eq:abelian-gauge-transformation}
\end{equation}
The pion-pair induced Wilson line in eq.~\eqref{wl:eq:Ssoft-wilson-factorization} may
be written directly in terms of the conserved current as
\begin{equation}
  \widehat W_p[A]
  =
  \mathcal P\exp\!\left[
  -\ii\int\dd^4x\,
  J_p^\mu(x)\widehat A_\mu(x)
  \right].
  \label{gauge:eq:pair-wilson-current-form}
\end{equation}
Restricting the photon field in eq.~\eqref{gauge:eq:pair-wilson-current-form} to the window of eq.~\eqref{wl:eq:unresolved-photon-field} gives the unresolved Wilson line \(\widehat W_p(\xi,0)\) of eq.~\eqref{wl:eq:unresolved-wilson-line}.  The change of the gauge field in eq.~\eqref{gauge:eq:abelian-gauge-transformation} is a \(c\)-function and factors out of the path ordering, so that
\begin{align}
  \widehat W_p(\xi,0)[A+\partial\alpha]
  &=
  \widehat W_p(\xi,0)[A]\,\exp\Big[-\ii\int\dd^4x\,J_p^\mu(x)\,\partial_\mu\alpha(x)\Big]
  \notag\\
  &\equiv
  \widehat W_p(\xi,0)[A]\,\e^{-\ii\#}\, .
  \label{gauge:eq:unresolved-wilson-gauge-shift}
\end{align}
For a gauge transformation that vanishes at the boundary, integration by parts gives
\begin{align}
  \#
  &=
  \int\dd^4x\,J_p^\mu(x)\,\partial_\mu\alpha(x)
  \notag\\
  &=
  \int\dd^4x\,\partial_\mu\big[\alpha(x)\,J_p^\mu(x)\big]-\int\dd^4x\,\alpha(x)\,\partial_\mu J_p^\mu(x)
  \notag\\
  &=
  0\, ,
  \label{gauge:eq:ordinary-gauge-exponent}
\end{align}
where the total derivative drops out by this boundary condition and the second term vanishes by current conservation, eq.~\eqref{gauge:eq:pair-current-conservation}.  The unresolved Wilson line is therefore invariant,
\begin{equation}
  \widehat W_p(\xi,0)[A+\partial\alpha]
  =
  \widehat W_p(\xi,0)[A]\, .
  \label{gauge:eq:pair-wilson-gauge-invariance}
\end{equation}
For a large gauge transformation, to analyze its effect on the unresolved Wilson line with ease, we write \(\alpha\) in Feynman gauge in the form
\begin{equation}
  \alpha(x)
  =
  \int\dd\Phi(q)\Big[\tilde\alpha(q)\,\e^{-\ii q\cdot x}+\tilde\alpha^\ast(q)\,\e^{\ii q\cdot x}\Big]\, .
  \label{gauge:eq:feynman-gauge-alpha}
\end{equation}
Inserting it into eq.~\eqref{gauge:eq:unresolved-wilson-gauge-shift} gives
\begin{align}
  \#
  &=
  \int\dd^4x\,J_p^\mu(x)\int\dd\Phi_\xi(q)\Big[-\ii q_\mu\tilde\alpha(q)\,\e^{-\ii q\cdot x}+\ii q_\mu\tilde\alpha^\ast(q)\,\e^{\ii q\cdot x}\Big]
  \notag\\
  &=
  \int\dd\Phi_\xi(q)\Big[-\ii q_\mu\tilde\alpha(q)\int\dd^4x\,\e^{-\ii q\cdot x}J_p^\mu(x)+\ii q_\mu\tilde\alpha^\ast(q)\int\dd^4x\,\e^{\ii q\cdot x}J_p^\mu(x)\Big]
  \notag\\
  &=
  \int\dd\Phi_\xi(q)\Big[\tilde\alpha(q)\,q\cdot J_p(-q)-\tilde\alpha^\ast(q)\,q\cdot J_p(q)\Big]
  \notag\\
  &=
  e\int\dd\Phi_\xi(q)\bigg\{\tilde\alpha(q)\bigg[\frac{p_+\cdot q}{-p_+\cdot q+\ii\varepsilon}-\frac{p_-\cdot q}{-p_-\cdot q+\ii\varepsilon}\bigg]
  -\tilde\alpha^\ast(q)\bigg[\frac{p_+\cdot q}{p_+\cdot q+\ii\varepsilon}-\frac{p_-\cdot q}{p_-\cdot q+\ii\varepsilon}\bigg]\bigg\}
  \notag\\
  &=
  e\int\dd\Phi_\xi(q)\bigg\{\tilde\alpha(q)\bigg[\Big(-1+\frac{\varepsilon}{\varepsilon+\ii p_+\cdot q}\Big)-\Big(-1+\frac{\varepsilon}{\varepsilon+\ii p_-\cdot q}\Big)\bigg]
  \notag\\
  &\hspace{6.5em}
  -\tilde\alpha^\ast(q)\bigg[\Big(1-\frac{\varepsilon}{\varepsilon-\ii p_+\cdot q}\Big)-\Big(1-\frac{\varepsilon}{\varepsilon-\ii p_-\cdot q}\Big)\bigg]\bigg\}
  \notag\\
  &=
  e\int\dd\Phi_\xi(q)\bigg\{\tilde\alpha(q)\bigg[\frac{\varepsilon}{\varepsilon+\ii p_+\cdot q}-\frac{\varepsilon}{\varepsilon+\ii p_-\cdot q}\bigg]
  +\tilde\alpha^\ast(q)\bigg[\frac{\varepsilon}{\varepsilon-\ii p_+\cdot q}-\frac{\varepsilon}{\varepsilon-\ii p_-\cdot q}\bigg]\bigg\}\, ,
  \label{gauge:eq:large-gauge-exponent}
\end{align}
where the fourth line inserts \(J_p(\pm q)\) of eq.~\eqref{eq:Jp-current} with its \(\ii0\) written as \(\ii\varepsilon\).

Since \(\omega_q>\lambda_\IR\), we have
\begin{align}
  |\#|
  &\le
  2e\sum_\pm\int\dd\Phi_\xi(q)\,|\tilde\alpha(q)|\,\bigg|\frac{\varepsilon}{\varepsilon+\ii p_\pm\cdot q}\bigg|
  \notag\\
  &\le
  2e\sum_\pm\int\dd\Phi_\xi(q)\,|\tilde\alpha(q)|\,\frac{\varepsilon}{p_\pm\cdot q}
  \notag\\
  &\le
  2e\sum_\pm\int\dd\Phi_\xi(q)\,|\tilde\alpha(q)|\,\frac{\varepsilon}{\omega_q\big(p_\pm^0-|\mathbf p_\pm|\big)}
  \notag\\
  &\le
  \frac{\varepsilon}{\lambda_\IR}\,2e\sum_\pm\frac{1}{p_\pm^0-|\mathbf p_\pm|}\int\dd\Phi_\xi(q)\,|\tilde\alpha(q)|\, ,
  \label{gauge:eq:large-gauge-bound}
\end{align}
where we used \(p_\pm\cdot q\ge\omega_q(p_\pm^0-|\mathbf p_\pm|)\) and \(\omega_q>\lambda_\IR\).  For \(\lambda_\IR\neq0\), we thus have
\begin{equation}
  \lim_{\varepsilon\to0^+}\#
  =
  0\, ,
  \label{gauge:eq:large-gauge-limit}
\end{equation}
so that the unresolved Wilson line is again invariant,
\begin{equation}
  \widehat W_p(\xi,0)[A+\partial\alpha]
  =
  \widehat W_p(\xi,0)[A]\, .
  \label{gauge:eq:unresolved-wilson-large-gauge-invariance}
\end{equation}
The phase due to a large gauge transformation is non-vanishing only for \(\lambda_\IR=0\), namely through the zero-energy photons.  In our construction, these modes belong to the infrared dressing of the pions, Section~\ref{pion:sec:reduced-density-soft-exponentiation}.  For the dressed neutral pion pair, the large gauge transformation of the dressing is cancelled by that of the pions, making the dressed pion state gauge invariant by itself~\cite{Gabai:2016kuf}.  Therefore, the reduced density matrix \(\widehat\rho(\xi)\) is invariant under both ordinary and large gauge transformations.

The momentum-space form of the same statement is the soft Ward identity.  For
the positive-energy on-shell photons entering the unresolved trace, the
\(\ii0\) prescription in eq.~\eqref{eq:Sp} does
not affect the contraction with \(q_\mu\), and
\begin{align}
  q_\mu J_p^\mu(q)
  &=
  e\frac{q\cdot p_+}{p_+\cdot q}
  -e\frac{q\cdot p_-}{p_-\cdot q}
    =0.
  \label{gauge:eq:soft-current-ward-identity}
\end{align}
Under a change of polarization convention equivalent to a gauge transformation,
\(\epsilon^\mu(q,\lambda)\to
\epsilon^\mu(q,\lambda)+c(q,\lambda)q^\mu\), the one-photon soft factor
therefore transforms as,
\begin{align}
  S_p(q,\lambda)
  &\longrightarrow
  \left[\epsilon^\ast(q,\lambda)+c^\ast(q,\lambda)q\right]\cdot J_p(q)
  =
  S_p(q,\lambda)
  +c^\ast(q,\lambda)q\cdot J_p(q)
  =
  S_p(q,\lambda).
  \label{gauge:eq:soft-factor-gauge-invariance}
\end{align}
Consequently both soft factors
$S_p(q,\lambda)$ and $S_{p'}(q,\lambda)$ are separately gauge invariant.  The
zero-temperature and thermal decoherence exponents may be displayed together
as
\begin{align}
  \Gamma_{pp'}(\xi)
  &=
  \frac12\sum_\lambda\int\dd\Phi_\xi(q)\,
  |S_p(q,\lambda)-S_{p'}(q,\lambda)|^2,
  \notag\\
  \Gamma^{(\rm th)}_{pp'}(\xi)
  &=
  \frac12\sum_\lambda\int\dd\Phi_\xi(q)\,
  \coth\!\left(\frac{\omega_q}{2k_BT}\right)
  |S_p(q,\lambda)-S_{p'}(q,\lambda)|^2.
  \label{gauge:eq:vacuum-thermal-gauge-invariant-kernels}
\end{align}
The phase-space restriction and the Bose factor are scalar functions of the
photon momentum hence invariant under gauge transformations, so eq.~\eqref{gauge:eq:soft-factor-gauge-invariance} proves
the gauge invariance of both kernels.

The virtual correction is independently insensitive to the covariant-gauge
parameter.  Denoting this parameter by $\zeta$, the photon propagator is
\begin{equation}
  D_{\mu\nu}^{(\zeta)}(q)
  =
  \frac{-\ii}{q^2+\ii0}
  \left[
  g_{\mu\nu}
  -(1-\zeta)\frac{q_\mu q_\nu}{q^2+\ii0}
  \right].
  \label{gauge:eq:covariant-gauge-propagator}
\end{equation}
Repeating the derivation of eq.~\eqref{sqed:eq:wilson-vev-wick} with \(D^{(\zeta)}_{\mu\nu}\) in place of the Feynman propagator, the one-loop virtual correction~\eqref{pion:eq:virtual-eikonal-current} becomes
\begin{equation}
  V_p^{(\zeta)}(\xi)
  =
  \frac12\int_{q,\xi}
  J_p^\mu(q)D_{\mu\nu}^{(\zeta)}(q)J_p^\nu(-q).
  \label{gauge:eq:virtual-exponent-general-gauge}
\end{equation}
Subtracting the Feynman-gauge result \(V_p^{(1)}\) and using eq.~\eqref{gauge:eq:soft-current-ward-identity} then gives,
\begin{align}
  V_p^{(\zeta)}(\xi)-V_p^{(1)}(\xi)
  &=
  \frac{\ii(1-\zeta)}{2}
  \int_{q,\xi}
  \frac{
  [q\cdot J_p(q)][q\cdot J_p(-q)]
  }{(q^2+\ii0)^2}
  =0.
  \label{gauge:eq:virtual-gauge-parameter-cancellation}
\end{align}
The same contraction appears for the primed current appearing in $\widehat W_{p'}$ and for the cross term $\widehat W_{p'}^\dagger\widehat W_p$ in
the influence functional of eq.~\eqref{if:eq:influence-functional-form}.  Hence the virtual contribution and the eikonal
phase \(\chi_{pp'}(\xi)\) are independent of the parameter \(\zeta\), in agreement with the operator proof in eq.~\eqref{gauge:eq:pair-wilson-gauge-invariance}.

Finally, the transverse gauge used in
Section~\ref{pion:sec:angular-decoherence} is only a convenient representation
of the same conserved-current contraction.  For the back-to-back kinematics
in eq.~\eqref{pion:eq:nearby-pion-kinematics}, with
\(q=\omega(1,\mathbf n_q)\), direct substitution into the eikonal current gives
\begin{equation}
  J_p^\mu(q)
  =
  \frac{2ev}{
  \omega[1-v^2(\mathbf n\cdot\mathbf n_q)^2]
  }
  \left(\mathbf n\cdot\mathbf n_q,\mathbf n\right)\, ,
  \label{gauge:eq:angular-four-current}
\end{equation}
and its contraction with the photon momentum
is identically zero
\begin{align}
  q\cdot J_p(q)
  &=
  \frac{2ev}{1-v^2(\mathbf n\cdot\mathbf n_q)^2}
  \left[
  \mathbf n\cdot\mathbf n_q-\mathbf n_q\cdot\mathbf n
  \right]
  =0\, .
  \label{gauge:eq:angular-current-conservation}
\end{align}
Since the polarization vector $\epsilon^\mu(q,\lambda)$ is always contracted with the eikonal currents and respects $q\cdot \epsilon(q,\lambda) = 0\, \implies \epsilon^\mu(q,\lambda) = c(q,\lambda)\, q^\mu + (0,\boldsymbol\epsilon)$ where $\mathbf q\cdot\boldsymbol\epsilon=0$,
choosing the polarization vector which satisfies \(\epsilon^0(q,\lambda)=0\) and \(\mathbf n_q\cdot\epsilon(q,\lambda)=0\) is therefore equivalent to fixing $c(q,\lambda)=0$ and it only removes components that
vanish anyway by eq.~\eqref{gauge:eq:angular-current-conservation}.  The angular
decoherence rate in eq.~\eqref{pion:eq:angular-decoherence-main} is consequently
independent of this gauge choice.

Combining the real and virtual contributions in
eq.~\eqref{pion:eq:combined-coherence-factor}, the soft
factor, with the unresolved photons traced over,
satisfies,
\begin{equation}
  \delta_\alpha\left[
  \frac{\rho_{pp'}(\xi)}{\rho_{pp'}(0)}\right]
  =0,
  \qquad
  \frac{\partial}{\partial\zeta}\left[
  \frac{\rho_{pp'}^{(\zeta)}(\xi)}{\rho_{pp'}(0)}\right]
  =0.
  \label{gauge:eq:coherence-factor-gauge-invariance}
\end{equation}
where \(\delta_\alpha\) denotes the variation under
eq.~\eqref{gauge:eq:abelian-gauge-transformation} and $\rho_{pp'}(0)$ is the hard density matrix at the scalar production point $t=0$.
For gauge transformations that vanish at the asymptotic boundary,
eq.~\eqref{gauge:eq:coherence-factor-gauge-invariance} shows directly that the
soft factor generated by the charged hard scalar particles, and hence the reduced density matrix, is explicitly gauge invariant within the leading-power
approximation used throughout this paper.
For large gauge transformations, the same conclusion follows because the unresolved Wilson line, eq.~\eqref{gauge:eq:unresolved-wilson-large-gauge-invariance}, and the dressed pion pair are separately gauge invariant.

\bibliographystyle{JHEP}
\bibliography{ref}

\end{document}